\documentclass[]{Outils_latex/aa}

\usepackage[utf8]{inputenc}
\usepackage{orcidlink}

\usepackage{natbib}
\usepackage{graphicx}
\usepackage{longtable}
\usepackage{xcolor}
\usepackage{amssymb}
\usepackage[switch]{lineno}
\usepackage{hyperref}

\bibpunct{(}{)}{;}{a}{}{,}

\hypersetup{
colorlinks = true,
breaklinks = true,
urlcolor = blue,
linkcolor = blue,
citecolor = blue,
}

\newcommand{\ind}[1]{_{\mathrm{#1}}}
\newcommand{\diff}{\mathrm{d}}

\newcommand\Kepler{\emph{Kepler}}
\newcommand\ktwo{K2}
\newcommand\tess{TESS}
\newcommand\corot{CoRoT}

\newcommand\adipls{ADIPLS}
\newcommand\mesa{MESA}

\newcommand\nunl{\nu\ind{n,\ell}}
\newcommand\nunlUP{\nu_{n,\ell}^{\mathrm{UP}}}
\newcommand\numax{\nu\ind{max}}
\newcommand\nmax{n\ind{max}}

\newcommand\Dnu{\Delta\nu}
\newcommand\DPi{\Delta\Pi_{1}}
\newcommand\Teff{T\ind{eff}}

\newcommand\eps{\varepsilon}
\newcommand\dol{d\ind{0\ell}}

\newcommand\alphaovHe{\alpha_{\mathrm{ov, He}}}
\newcommand\alphaovH{\alpha_{\mathrm{ov, H}}}
\newcommand\alphaov{\alpha_{\mathrm{ov}}}
\newcommand\alphaMLT{\alpha_{\mathrm{MLT}}}
\newcommand\alphaundersh{\alpha_{\mathrm{ov, env}}}
\newcommand\alphath{\alpha_{\mathrm{th}}}

\newcommand\gradad{\nabla_{\mathrm{ad}}}
\newcommand\gradrad{\nabla_{\mathrm{rad}}}
\newcommand\gradT{\nabla_{T}}

\newcommand{\NBV}{N\ind{BV}}

\newcommand{\Amplgl}{\mathcal{A}\ind{gl}}
\newcommand{\Ggl}{\mathcal{G}\ind{gl}}
\newcommand{\Phigl}{\Phi\ind{gl}}

\newcommand\tHeII{t_{\mathrm{HeII}}}
\newcommand\HHeII{H_{\mathrm{HeII}}}
\newcommand\DHeII{b_{\mathrm{HeII}}}
\newcommand\THeII{T\ind{HeII}}

\newcommand\Omegacrit{\Omega_{\mathrm{crit}}}

\usepackage{amstext}

\begin{document}

\title{Physical interpretation of the oscillation spectrum on the RGB and AGB}
\titlerunning{Oscillation spectrum of evolved RGB and AGB stars}

\author{
 G. Dr\'eau\inst{1,2}\orcidlink{0000-0002-0135-8720},
 Y. Lebreton\inst{1,3}\orcidlink{0000-0002-4834-2144},
 B. Mosser\inst{1}\orcidlink{0000-0002-7547-1208},
  D. Stello\inst{2}\orcidlink{0000-0002-4879-3519}
}

\institute{
\inst{1} LIRA, Observatoire de Paris, Universit\'e PSL, Sorbonne Universit\'e, Universit\'e Paris Cit\'e, CY Cergy Paris Universit\'e, CNRS 5 Place Jules Janssen, 92190 Meudon, FRANCE \\
 \inst{2} School of Physics, University of New South Wales, Sydney, NSW 2052, Australia \\
 email: \texttt{g.dreau@unsw.edu.au}\\
\inst{3} Univ Rennes, CNRS, IPR (Institut de Physique de Rennes) - UMR 6251, F-35000 Rennes, France\\
 }

\abstract{The high frequency resolution of the four-year time series collected by the space-borne telescope \Kepler\ gives us an opportunity to study the seismic mode structure of highly luminous giants in great detail. Seismic observables can be used as to infer the interior structure through comparisons with stellar models. However, we still need to extend the physical interpretation of previously observed seismic differences between hydrogen-shell burning (Red-Giant Branch; RGB) and helium-burning (red clump and Asymptotic-Giant Branch; AGB) stars towards high luminosity stages.}
{Here we aim to investigate which physical conditions differ between H-shell and He-burning stars in the helium-second ionisation zone, through the signature this zone leaves in mode frequencies. In addition, we explore the sensitivity of seismic parameters to the physics implemented in models.}
{We used a grid of stellar models with mass between $0.8\, M_{\odot}$ and $2.5\, M_{\odot}$ and metallicity between $-1.0\, $dex and $0.25\,$dex. Transfer mechanisms are implemented such as mass loss, core and envelope overshooting, and thermohaline mixing. We infer the p-mode frequencies of the models by artificially suppressing the gravity modes in the core. 
}
{In accordance with observations, we find that the main stellar properties affecting the seismic observables in the models are the stellar mass and metallicity. Mass loss on the RGB and rotation-induced mixing from the main sequence to the early-AGB cause a phase difference of the helium ionisation zone glitch signature between H-shell and He-burning stars. The amplitude of the glitch signature in the local large separation, $\Dnu$, is correlated with the density in the helium ionisation zone, which explains the different glitch amplitudes observed between H-shell and He-burning stars. The amplitude exceeds 10\% of the observed value of $\Dnu$ in high-luminosity red giants, which makes the asymptotic expansion less accurate when $\Dnu \leq 0.5\, \mu$Hz.}
{An efficient mass loss on the RGB, typically encountered when $M \leq 1.5\, M_{\odot}$, can explain the classification of H-shell and He-burning stars based on the p-mode pattern. When $M \geq 1.5\, M_{\odot}$, efficient mixing mechanisms might leave an important detectable signature in the p-mode frequencies, permitting a potential classification of these stars.
}

\maketitle

\section{Introduction}

A breakthrough in our understanding of stellar structure and evolution has been provided by the advent of the space missions \corot\ \citep{2006ESASP1306...33B}, \Kepler\ \citep{2010Sci...327..977B, 2010PASP..122..131G}, \ktwo\ \citep{2014PASP..126..398H}, and now \tess\ \citep{2015JATIS...1a4003R}. The ultra-high precision photometric data collected by these space-borne telescopes give access to the frequencies of stellar oscillation modes. The observables describing the global frequency patterns of stars can be used to determine their masses, and hence mass loss, through the use of seismic scaling relations \citep{2012MNRAS.419.2077M, 2018A&A...616A.104K, 2021A&A...645A..85M, 2021MNRAS.501.5135Y}. The more detailed fine structure of the frequency patterns can be used to estimate the evolutionary stage of red-giant stars \citep{2011Natur.471..608B, 2012A&A...541A..51K, 2016A&A...588A..87V, 2017MNRAS.469.4578H, 2018MNRAS.476.3233H, 2019A&A...622A..76M}. \citet{2011A&A...525L...9M} have shown that the asymptotic p-mode frequencies $\nunlUP$ of red giants, which are derived under the assumption that $n \gg \ell$, can be expressed as

\begin{equation}
\label{eq:nunl_asympt}
\nunlUP = \left(n + \frac{\ell}{2} + \eps - \dol + \frac{\alpha}{2}[n - \nmax]^{2}\right)\Dnu,
\end{equation}
where UP stands for Universal Pattern, $n$ is the mode radial order, $\ell$ is its degree, $\eps$ is the acoustic offset that allows us to locate the radial modes in the spectrum, $\Dnu$ is the observed mean frequency spacing between consecutive radial modes (the large frequency separation), $\dol$ is a reduced small frequency separation defined as $\dol = \delta\nu_{0\ell}/\Dnu$ where $\delta\nu_{0\ell}$ is the small frequency separation between a mode of degree $\ell$ and its neighbouring radial mode, $\alpha = (\diff \log{\Dnu}/\diff n)$ is a curvature term that accounts for the linear dependence of the large frequency separation on the radial order, and $\nmax = \numax/\Dnu - \varepsilon$ is the equivalent radial order that the frequency of the maximum oscillation power $\numax$ would have. 

The observed frequencies $\nunl$ are of major importance as they are sensitive to the stellar internal structure \citep{1980ApJS...43..469T, 1983SoPh...82...75S}, and can be used to probe interior stellar physics such as overshooting \citep[e.g.,][]{2012A&A...538A..73B, 2015MNRAS.453.2290B, 2018ApJ...859..156K, 2022A&A...668A.115D}. Specifically in main-sequence stars, the small separations $\dol$ are sensitive to regions with strong gradients of sound speed \citep{1986HiA.....7..283G} and can be used to probe the stellar structure at localised depths  \citep{2003A&A...411..215R}. Observations of giant stars has highlighted that these small separations depend on the stellar mass \citep{2010ApJ...723.1607H}. The resulting distribution of small separations $\dol$, at fixed $\Dnu$, can be theoretically explained by a difference in the distance between the location of the bottom of the convective envelope and the inner turning point of the mode cavity for stars of different masses \citep{2010ApJ...721L.182M}, especially for dipole modes ($\ell = 1$). 

It has been shown both observationally and theoretically that the frequency pattern of the acoustic modes changes as stars evolve up the RGB. The dipole modes move closer to their neighbouring radial mode while the quadrupole modes move slightly away, resulting in a triplet, fork-like, pattern in the frequency spectrum \citep{2014ApJ...788L..10S, 2021A&A...650A.115D}. As this happens, the observational range of modes is also shifted to a narrower frequency range towards lower frequencies, meaning fewer modes and of lower radial orders \citep{2013A&A...559A.137M, 2017ApJ...847..139T, 2020MNRAS.493.1388Y}. 

The so-called semi-regular variables, which are solar-like pulsators at high luminosity stages \citep{2001MNRAS.328..601D, 2001ApJ...562L.141C}, show only a small number of stochastically excited oscillation modes, limiting the power of seismology to probe their interiors. Moreover, the assumption that $n \gg \ell$ is no longer satisfied, and hence the modes are no longer expected to follow the asymptotic pattern given by Eq.~\ref{eq:nunl_asympt}. Here, we explore the limits of the asymptotic approach and assess its potential to describe the oscillation spectrum of high-luminosity red giants.

As presented here, Eq.~\ref{eq:nunl_asympt} assumes that the Jeffreys, Wentzel, Kramers and Brillouin (JWKB) approximation is valid, which states that the physical parameters inside the star vary on a scale much greater than the wavelength of the oscillations. However, clear signatures of sharp structural variations have been predicted \citep{1988IAUS..123..151V}, then confirmed for the Sun \citep{2007MNRAS.375..861H}, for main-sequence stars \citep{2012A&A...544L..13L,2012A&A...540A..31M, 2014ApJ...782...18M, 2014ApJ...790..138V, 2016A&A...589A..93D}, and for red giants \citep{2010A&A...520L...6M, 2014MNRAS.440.1828B, 2015A&A...579A..84V, 2015A&A...578A..76C}. These structural variations are called glitches \citep{2002ESASP.485...65G} and are known to introduce a smooth frequency-dependent modulation to the otherwise regular asymptotic mode frequencies $\nunl$ given by Eq.~\ref{eq:nunl_asympt} \citep{1990LNP...367..283G}. Three main structural variations have been identified, which are the base of the convective envelope, the helium ionisation zones, and the boundary of the convective core \citep{1994A&A...283..247M, 2005MNRAS.361.1187M, 2007MNRAS.375..861H, 2007ApJ...666..413C, 2016A&A...589A..93D}. In red giants, the helium second-ionisation (HeII) zone creates the dominant glitch \citep{2010A&A...520L...6M}. The glitch signature is expected to depend on stellar properties such as the helium abundance \citep{2007MNRAS.375..861H, 2021A&A...655A..85H, 2022A&A...663A..60H}, which paves the way for estimating the helium abundance in cool stars \citep[e.g.,][]{1992MNRAS.257...32V, 1997ApJ...480..794L, 2014MNRAS.440.1828B, 2014ApJ...790..138V, 2019MNRAS.483.4678V, 2019A&A...622A..98F}. Also, the study of the HeII zone signature in intermediate- and high-luminosity red giants has shown that the glitch morphology (such as the amplitude, phase and frequency dependence of the glitched-induced frequency modulation) depend on the evolutionary stage \citep{2015A&A...579A..84V, 2021A&A...650A.115D}. Particularly interesting is that the glitch modulation is stronger and phase-shifted during He-burning phases compared to the H-shell burning phase, which makes a classification between Red-Giant Branch (RGB) and He-burning stars possible, including clump and Asymptotic-Giant Branch (AGB) stars. In the following, we aim to investigate the physical origin of these differences caused by stellar evolution, which are attributed to a change of the temperature and the density in the HeII ionisation zone \citep{2014MNRAS.445.3685C}.

In this work, we focus on the analysis of the oscillation spectra of evolved red giants. By means of the Aarhus adiabatic oscillation package \citep[\adipls,][]{2008Ap&SS.316..113C}, we extract the p-mode frequencies of stellar models derived with the evolution code Modules for Experiments in Stellar Astrophysics \citep[\mesa,][]{2011ApJS..192....3P, 2013ApJS..208....4P, 2015ApJS..220...15P, 2018ApJS..234...34P, 2019ApJS..243...10P}. We investigate the impact of input physics on seismic parameters, in comparison with observations. We look into the physical differences between RGB and AGB stars that we can deduce from their different glitch signatures and explore the limits of the asymptotic expansion (Eq.~\ref{eq:nunl_asympt}) at high-luminosity stages. In Sect.~\ref{sec:models}, we describe the set of input physics that we adopted in the \mesa\ calculations. The procedure we use to extract the p-mode pattern of evolved red giants, including RGB and clump/AGB stars, is developed in Sect.~\ref{sec:method}. We examine how the seismic parameters vary as a function of both the evolutionary stage and stellar parameters in Sect.~\ref{sec:results}. Then, we discuss the validity of the asymptotic expansion and the differences in the oscillation spectrum between RGB and AGB stars in Sect.~\ref{sec:discussion}. We finally conclude in Sect.~\ref{sec:conclusion}.

\section{Stellar models}
\label{sec:models}

A grid of stellar models with initial mass $M = [0.8,\ 0.9,\ 1.0,\ 1.1,\ 1.2,\ 1.5,\ 1.75,\ 2.0,\ 2.5]\,M_{\odot}$, initial metallicity $\mathrm{[Fe/H]} = [-1.0, -0.5, -0.25, 0.0, 0.25]\,$dex, and different input physics has been computed with the release 12778 of the stellar evolution code Modules for Experiments in Stellar Astrophysics \citep[\mesa, ][]{2011ApJS..192....3P, 2013ApJS..208....4P, 2015ApJS..220...15P, 2018ApJS..234...34P, 2019ApJS..243...10P}. Modelling is exhaustively described in \citet{2022A&A...668A.115D}. Here, we summarise the main input physics that are likely to affect seismic observables. We used the \texttt{1M\_pre\_ms\_to\_wd}\footnote{Located at \$MESA\_DIR/star/test\_suite/1M\_pre\_ms\_to\_wd after dowloading from \href{https://zenodo.org/records/3698354}{https://zenodo.org/records/3698354}} test suite case and defined a reference model from which the input physics are modified (Table~\ref{Table:reference_model}). Convection is treated following the mixing-length theory formalism, where the convective efficiency is taken to vary with the opacity of the convective element, especially in the outer stellar layers \citep{1965ApJ...142..841H}. Three sets of opacity tables are used to take the changing internal temperature and chemical composition into account. At low temperature ($\log T < 3.95$), we used the AESOPUS tables \citep{2009A&A...508.1539M} that allow us to take continuum and discrete sources into account such as molecular absorption bands and collision-induced absorption in the photosphere of AGB stars. At high temperature ($\log T > 4.05$), we took OPAL1 tables for regions that do not experience metal enrichment, applicable before He burning, and OPAL2 tables for those that undergo C and O enhancements due to He burning \citep{1996ApJ...464..943I}. 
The transition from one opacity table to another is treated as explained in \citet{2011ApJS..192....3P}, according to their Eq. 1. AESOPUS and OPAL1 tables are blended in the interval $\log T = 4.00 \pm 0.05$, while the transition from OPAL1 to OPAL2 is done in the region where the metal mass fraction is increased by an amount $\mathrm{d}Z$ where $\mathrm{d}Z \in [0.001, 0.01]$.\\

\begin{table*}[htbp]
\caption{Specifications of the \mesa\ reference stellar model}
\begin{center}
\begin{tabular}{rrrrrrrrrr}
\hline
\hline
$Y_{0}$\ \ & $\alphaMLT$ & $\eta\ind{R}$ & $\eta\ind{B}$ & $\alphaovH$ & $\alphaovHe$ & $\alphaundersh$ & $\alphath$ & $\alpha\ind{sc}$ & $\Omega_{\mathrm{ZAMS}}/\Omegacrit$ \\
& & & & \\
\hline
 0.253 & 1.92\ \ & 0.3 & 0.1 & 0.2\ \  & 0\ \ \ \  & 0\ \ \ \ \  & 0\ \  & 0$^{*}$\  & 0\ \ \ \ \ \ \ \\
  \hline
  \label{Table:reference_model}
\end{tabular}
\\
\end{center}
\textbf{Notes:} $Y_{0}$ is the initial helium mass fraction; $\alphaMLT$ is the mixing length parameter; $\eta\ind{R}$ and $\eta\ind{B}$ are the Reimers's and Blöcker's scaling factors for the mass loss prescriptions on the RGB and AGB, respectively; $\alphaovH$, $\alphaovHe$, and $\alphaundersh$ are the H-core overshooting, He-core overshooting, and envelope undershooting parameters, respectively, in units of the pressure scale height $H_{P}$; $\alpha\ind{sc}$ and $\alphath$ are the efficiencies of semiconvection and thermohaline convection following Eq.~12 and Eq.~14 of \citet{2013ApJS..208....4P}, respectively; $\Omega_{\mathrm{ZAMS}}/\Omegacrit$ is the ratio between the angular velocity at the ZAMS and the surface critical angular velocity for the star to be dislocated. These physical ingredients are described in Sect~\ref{sec:models}. (*) $\alpha\ind{sc}$ is always $0$ except when considering core-He overshooting. In that case, we set $\alpha\ind{sc} = 0.1$.
\end{table*}

Mixing processes in stellar interiors affect the chemical composition profile and the physical evolution of layers near the boundary of convective zones. Mixing therefore has non-negligible effects on the stellar oscillation modes that probe these layers. For instance, extending the mixing beyond the convective core instability boundary defined by the Schwarzschild criterion significantly modifies the period spacing $\DPi$ of dipole modes \citep{2015MNRAS.453.2290B}. Here, we considered several mixing processes, both in the core and the envelope. Among them, we take convective core overshooting and envelope undershooting into account to extend the mixing beyond the Schwarzschild boundary (above the core and below the envelope). In \mesa, we assume that the temperature gradient $\gradT$ is equal to the radiative gradient $\gradrad$ in the extra mixing region \citep{1991A&A...252..179Z} and we follow a step scheme \citep[e.g. ][]{1975A&A....40..303M}, in which the additional mixing region spreads over a distance 

\begin{equation}
    \label{eq:overshooting}
    d_{\mathrm{ov}} = 
\left\lbrace
\begin{array}{ccc}
\alphaov H_{P} \qquad \mathrm{if}\ H_{P} \leq R_{\mathrm{cz}} \\ 
\alphaov R_{\mathrm{cz}}\ \ \ \ \ \ \mathrm{otherwise}, \ \ \
\end{array}
\right.
\end{equation}
where $H_{P}$ is the pressure scale height at the boundary of the convective zone and $R_{\mathrm{cz}}$ is the radial thickness of the convective zone. In particular, we apply core overshooting during core nuclear burning phases, where a convective core grows with time. Including overshooting modifies the radiative gradient $\gradrad$ profile in the extra mixing region through an opacity increase. This may bring $\gradrad$ above $\gradad$, then leave the boundary of the convective core ambiguous \citep{2017MNRAS.469.4718B}. To avoid misidentifying the convective border in \mesa, we follow the treatment presented in \citet{2017MNRAS.469.4718B}, which consists in defining the convective border either at the position of the local minimum of $\gradrad$ if the latter increases over $\gradad$ in the extra mixing region, or at the usual border of the convective instability where $\gradrad = \gradad$ otherwise. We also included envelope undershooting from the main sequence up to the AGB following \citet{2019A&A...628A..35K}. When He-core overshooting is applied, we included a partially mixed He-semiconvection region between the convective core border and the outer radiative zone, according to the diffusion scheme presented in \citet{1985A&A...145..179L} where the efficiency factor $\alpha\ind{sc}$ is $0.1$.

Once the hydrogen-burning shell reaches the homogeneous zone of the envelope after the first dredge-up, nuclear reactions in the shell, such as $^{3}\mathrm{He}(^{3}\mathrm{He},2p)^{4}\mathrm{He}$, create an inversion of molecular weight \citep{1972ApJ...172..165U, 2006Sci...314.1580E, 2007A&A...467L..15C, 2008ApJ...677..581E}. Thermohaline convection sets in when the molecular weight gradient becomes negative (i.e. $\nabla_{\mu} = \mathrm{d}\ln \mu / \mathrm{d}\ln P < 0$) in regions that are stable against convection (according to the Ledoux criterion). As a result, this mixing process occurs between the convective envelope and the H-burning shell surrounding the degenerate core and affects the surface composition, especially for stars of mass below $1.5\, M_{\odot}$ \citep{2010A&A...521A...9C, 2012A&A...543A.108L}. In \mesa, we treat thermohaline convection as a diffusive process \citep{1980A&A....91..175K}, taking the diffusion coefficient presented in Sect.~4.2 of \citet{2013ApJS..208....4P} with the efficiency parameter $\alphath = 2$ when effective. In our models, we did not include microscopic diffusion, such as atomic diffusion and radiative accelerations. Although this process has been shown to be non-negligible around the hydrogen-burning shell, significantly affecting the location of the RGB bump, it only weakly alters the helium core mass at the helium flash \citep{2010A&A...510A.104M}. As an additional test, we applied the atomic diffusion settings of the \texttt{1.5M\_with\_diffusion}\footnote{Located at \$MESA\_DIR/star/test\_suite/1.5M\_with\_diffusion after dowloading from \href{https://zenodo.org/records/3698354}{https://zenodo.org/records/3698354}} test case to the reference model listed in Table~\ref{Table:reference_model}, from the main sequence to the luminosity tip of the RGB, and found that they had a negligible impact on the seismic parameters described in Eq.~\ref{eq:nunl_asympt}, as measured on the RGB. \\

Finally, we consider a simple grey atmosphere with an Eddington $T(\tau)$-law where $\tau$ is the optical depth. In \mesa, the interior is connected to the atmosphere at the meshpoint corresponding to $\tau = 2/3$, which lies at the photospheric boundary where $T = \Teff$ \citep{2011ApJS..192....3P}. At high luminosity stages, stars experience significant mass loss due to the radiative pressure pushing the envelope outwards. On the RGB, we used Reimers's prescription \citeyearpar{1975MSRSL...8..369R}, which reads

\begin{equation}
\label{eq:Reimers_mass_loss}
\dot{M}_{\mathrm{R}} = -4\times 10^{-13}\ \eta_{\mathrm{R}} \frac{L}{L_{\odot}} \frac{R}{R_{\odot}}\left(\frac{M}{M_{\odot}}\right)^{-1} M_{\odot}.\mathrm{yr}^{-1},
\end{equation}
while on the AGB, we adopt Blöcker's prescription \citeyearpar{1995A&A...297..727B}, that is
\small
\begin{equation}
\label{eq:Blocker_mass_loss}
\dot{M}_{\mathrm{B}} = - 1.93\times 10^{-21} \eta_{\mathrm{B}}\ \left(\frac{M}{M_{\odot}}\right)^{-3.1}\frac{R}{R_{\odot}} \left(\frac{L}{L_{\odot}}\right)^{3.7} M_{\odot}.\mathrm{yr}^{-1}.
\end{equation}
\normalsize
In the previous equations, the scaling factors are chosen to be $\eta_{\mathrm{R}} = 0.3$, which is the maximum mass loss rate reported among several stellar-cluster populations with different age and chemical composition \citep{2021A&A...645A..85M}, and $\eta_{\mathrm{B}} = 0.1$, which allows to reproduce both the initial-final mass relation across evolution and the AGB luminosity function \citep{2016ApJ...823..102C}. \\

The mode frequencies associated to the \mesa\ models are computed with the stellar oscillation code \adipls\ \citep{2008Ap&SS.316..113C}. In the \adipls\ settings, we do not consider the Cowling approximation and solve the full set of fourth-order system of oscillation equations. The first trial frequency is taken to be low enough so that the first radial order is found, then the next trial frequencies are taken just above the mode frequencies computed at the previous radial order. The differential equations are integrated following the shooting method with centred difference equations, where the differential equations are replaced by difference equations. In this case, the solutions are integrated separately from the inner and outer boundaries while satisfying the boundary conditions, and the eigenvalue connected to the frequency is found by matching these solutions at the specific point $r/R = 0.5$, where $R$ is the stellar radius. If the frequency is below the acoustic cut-off frequency, the surface boundary condition is imposed by matching the interior solution to the exponentially decaying form that arises in an isothermal atmosphere at the matching point. We emphasize that the isothermal atmosphere is applied only for the boundary condition in the mode frequency calculations; in the \mesa\ structure models, the atmosphere is not isothermal. For higher frequencies, the surface pressure boundary condition is instead given by $\delta p = 0$, where $\delta p$ is the Lagrangian perturbation in pressure. Then, the computed frequencies are improved by using the Richardson extrapolation, which allows us to reduce the numerical errors due to the finite number of meshpoints \citep{1981PASJ...33..713S}.

\begin{figure*}[htbp]
    \centering
	\begin{minipage}{1.0\linewidth}  
		\rotatebox{0}{\includegraphics[width=0.50\linewidth]{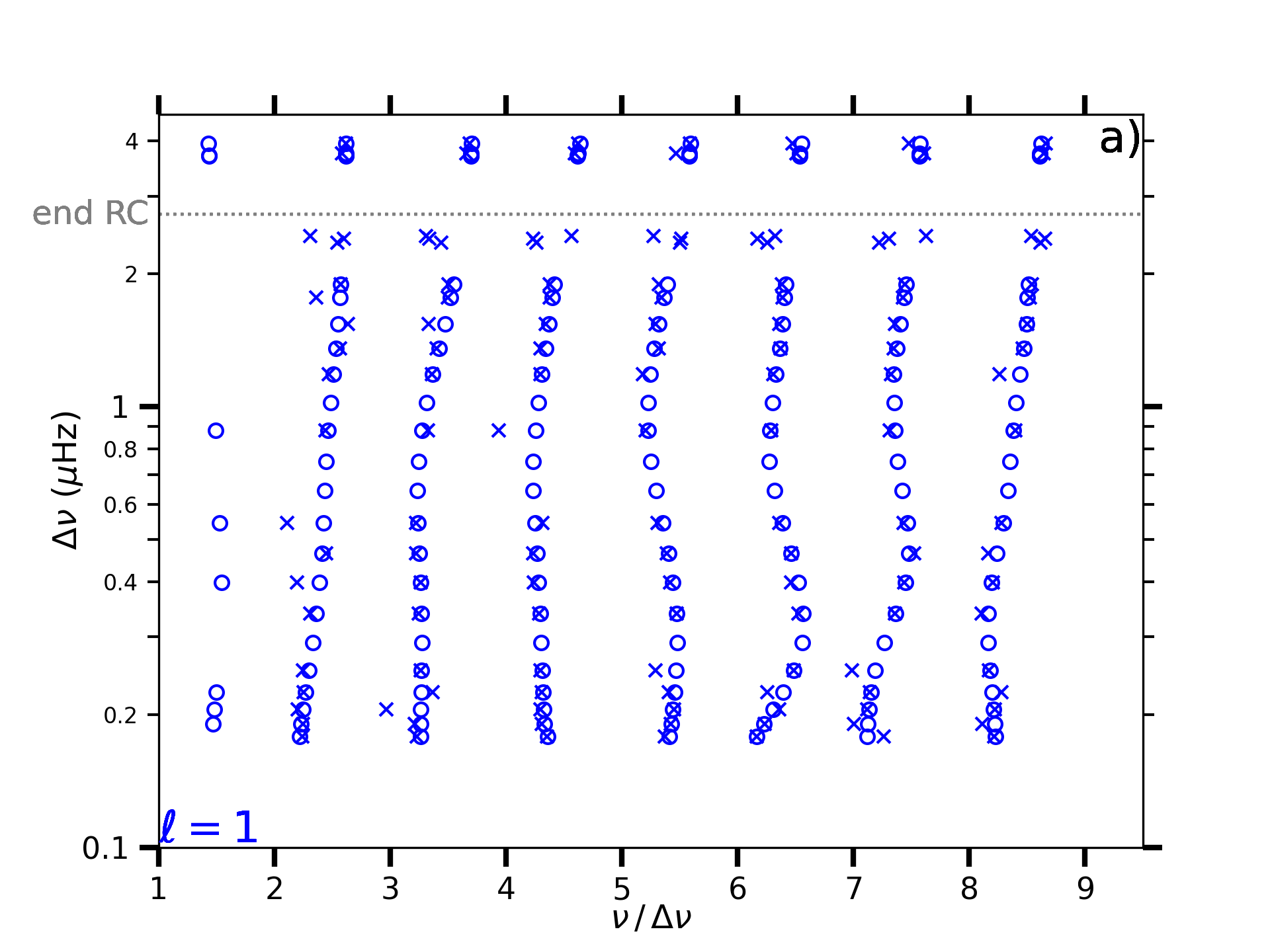}
		\rotatebox{0}{\includegraphics[width=0.50\linewidth]{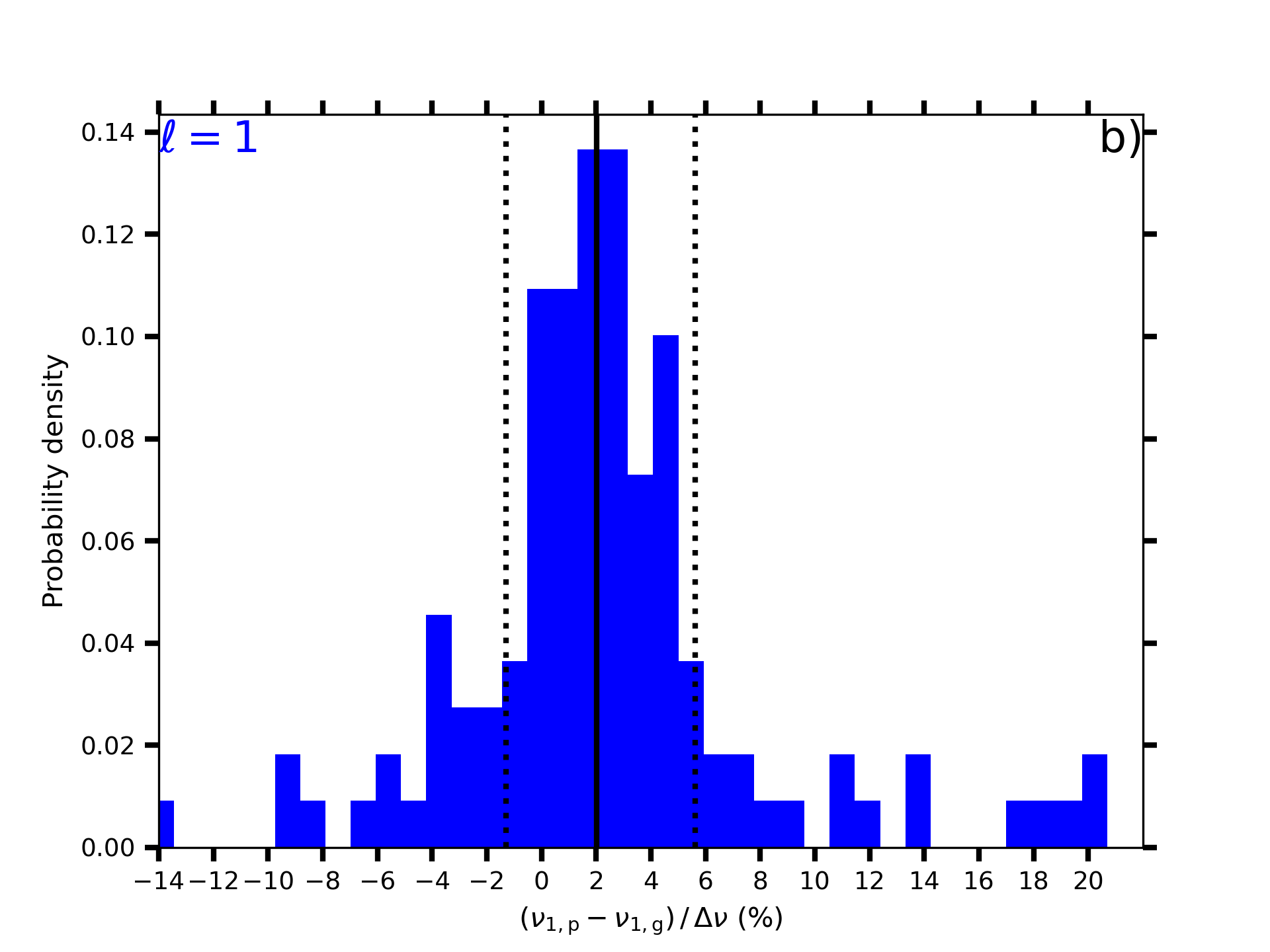}}
		}
	\end{minipage}
	\begin{minipage}{1.\linewidth}  
		\rotatebox{0}{\includegraphics[width=0.50\linewidth]{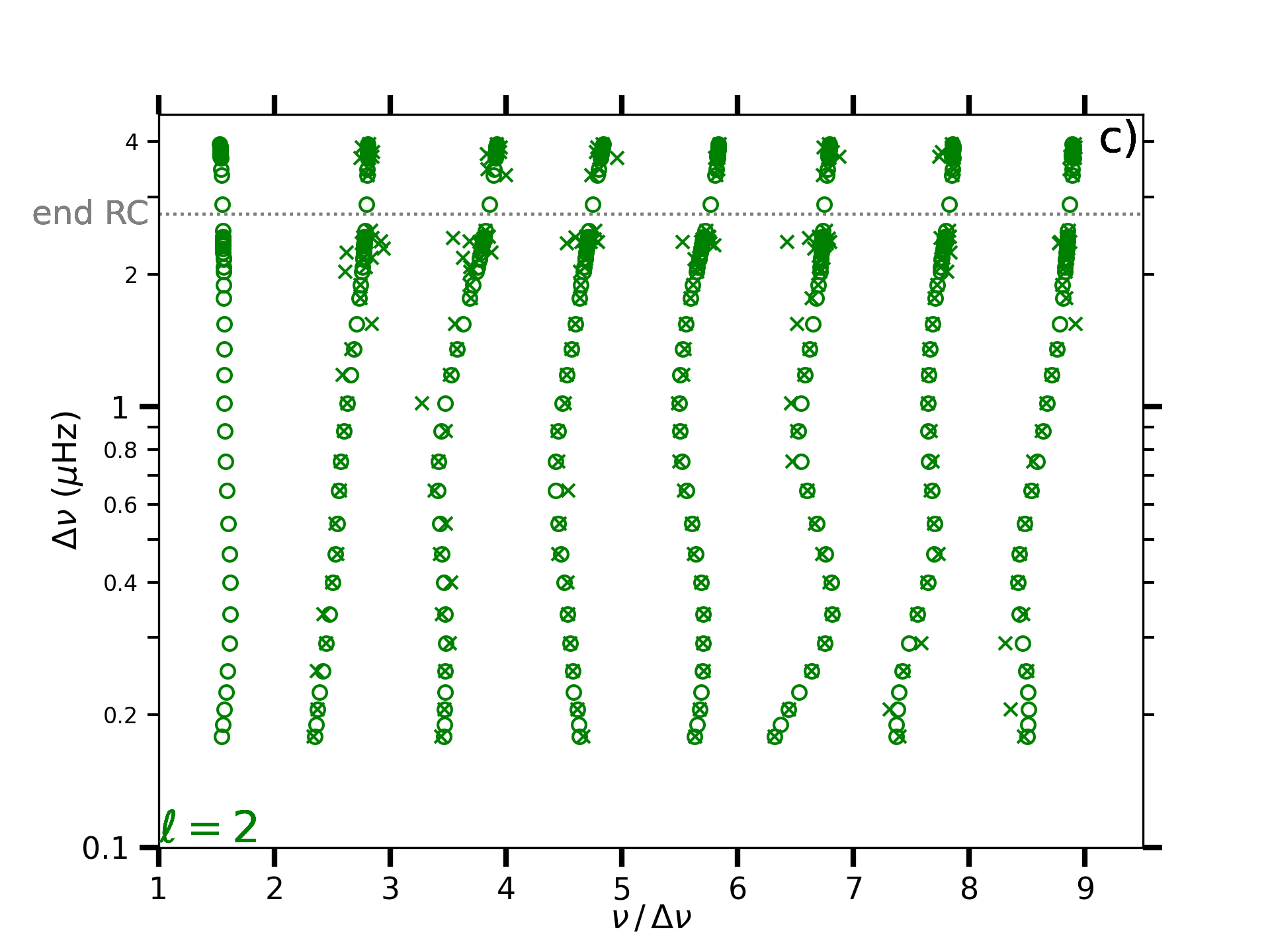}
		\rotatebox{0}{\includegraphics[width=0.50\linewidth]{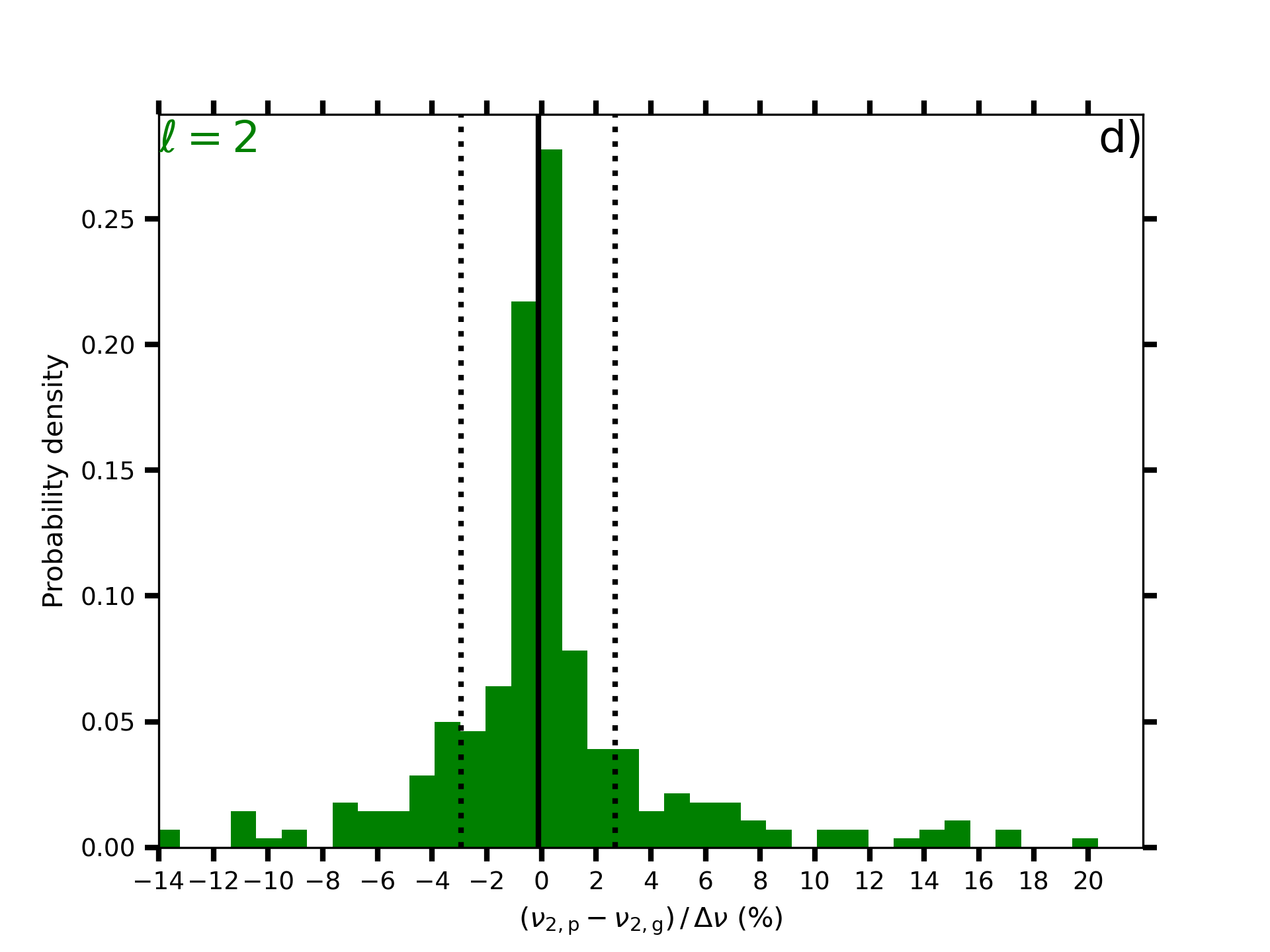}}
		}
	\end{minipage}
	
	\caption[Comparison of the mode frequencies computed by suppressing g modes in the core to the frequencies of the mixed modes of lowest inertia]{Comparison of the mode frequencies computed with the method $\NBV^{2} = 0$ to the frequencies of the mixed modes of lowest inertia in each $\Dnu$ interval, without modifying the outputs of \mesa. \emph{a)} Model frequencies of $\ell = 1$ modes in units of the large frequency separation $\Dnu$ for the 8 first radial orders and for a $1\,M_{\odot}$ track during the He-core burning and the early He-shell burning phase (input physics are summarised in Table~\ref{Table:reference_model}). The dipole modes computed with the method $\NBV^{2} = 0$ are shown in blue circles while the mixed dipole modes of lowest inertia are indicated by blue crosses. The grey dotted line shows the end of the clump phase, taken as when the central helium mass fraction goes below 0.01. \emph{b)} histogram of the differences between the dipole mode frequencies $\nu_{1,\, \mathrm{p}}$ computed with the method $\NBV^{2} = 0$ (blue circles) and the frequencies $\nu_{1,\, \mathrm{g}}$ of the dipole modes of lowest inertia (blue crosses). This histogram is computed for models in the whole He-burning phase and all radial orders up to $\nu/\Dnu = 9$. These differences are expressed as a percentile of $\Dnu$. The black solid line localises the median of the distribution, while the dotted lines show the $16^{\mathrm{th}}$ and $84^{\mathrm{th}}$ percentiles of the distribution. \emph{c)} and \emph{d)} same label as the upper panels, but for the $\ell = 2$ modes. Some modes could not be computed because they were missing or because they were inconsistent. This explains why some symbols are missing, especially the mixed non-radial modes of lowest inertia at $\nu/\Dnu \simeq 1$.
	}
	\label{fig:test_computation_pure_p_modes_with_NBV0}
\end{figure*}

\section{P-mode pattern extraction}
\label{sec:method}

\subsection{Computing the model pressure mode frequencies}
\label{sec:method_freq_computation}

In red giants, the p- and g-mode cavities are coupled \citep{2011Natur.471..608B}. As a consequence, non-radial modes are mixed, behaving like p-modes in the envelope and g-modes in the core. As a result of the coupling, mixed mode frequencies deviate from the expected pure p-mode (and pure g-mode) frequencies. The observed mixed modes with the largest amplitudes are those with the lowest inertia, as they need less energy input to be excited to observable amplitudes. These modes are also the most p-like, meaning that their frequencies are the closest to the acoustic resonances of the envelope (the frequencies that pure p-modes would have in the absence of the core). For evolved giants with $\numax \leq 40\, \mu$Hz, or equivalently $\Dnu \lessapprox 4\, \mu$Hz, it has been shown that the frequency of the mixed mode of lowest inertia, near each acoustic resonance, could be used as an approximate estimate for the frequency of the acoustic resonance \citep{2014MNRAS.440.1828B}. However, the computations required to first find this mode and to obtain an accurate estimate of its frequency makes this a time consuming approach. To address this difficulty, we followed an approach inspired by \citet{2018MNRAS.478.4697B}, who set the squared Brunt–Väisälä frequency $\NBV^{2}$ to zero in non-convective regions (i.e. where $\NBV^{2} > 0$). In this scenario, the cavity of g modes is removed, and hence we have no mixed modes, just pure p modes. This method is efficient for extracting the p-mode frequencies of dipole and quadrupole modes in RGB models with $\Dnu \leq 4.5\, \mu$Hz, showing a difference within 0.02$\Dnu$ relatively to the frequencies of the lowest inertia modes \citep{2018MNRAS.478.4697B}. While our treatment is based on the same principle, it differs in its implementation. In their analysis, \citet{2018MNRAS.478.4697B} defined an artificial $\Gamma_{1}$ (their Eq. 1) to maintain consistency within the oscillation equations. In our implementation, we retained the original $\Gamma_{1}$ profile. This choice has a negligible effect on the quadrupole mode frequencies but introduces a bias of 0.01$\Dnu$ in the dipole mode frequencies at low $\Dnu$ ($\Dnu \leq 4.5\,\mu$Hz). The magnitude of this bias is comparable to the precision with which the pressure-mode frequencies were extracted in \citet{2018MNRAS.478.4697B}. This bias has only a limited effect on the glitch signature in the large frequency separation, which is computed from the difference between consecutive radial orders at fixed degree. In this case, the bias is largely mitigated. The parameter most sensitive to this effect is the reduced small separation $d_{01}$ between dipole ($\ell = 1$) and radial ($\ell = 0$) modes. Nevertheless, we show in Sect.~\ref{subsec:result_seismic_param_asympt} that our interpretations are not affected by the presence of this bias. Similarly, we examine the applicability of this method for clump/AGB stars in the following section.

Finally, we performed a mesh redistribution to optimise the computation of p modes. This step is necessary as \adipls\ may otherwise fail to compute mode parameters at low-radial orders. This is particularly important for high-luminosity stages because the maximum oscillation power is located at low radial order \citep{2013A&A...559A.137M, 2014ApJ...788L..10S, 2020MNRAS.493.1388Y}. To further ensure we did not miss any non-radial modes, we ran \adipls\ with a variety of mesh redistributions. This mesh redistribution also guarantees the displacement eigenfunctions are resolved, regardless the evolutionary status. Our analysis rests on a fixed number of modes, associated to the seven radial orders that are the closest to $\numax$, which is estimated by the scaling relation

\begin{equation}
\label{eq:scaling_relation_nu_max_Teff}
\frac{\numax}{\nu\ind{max,\odot}} \simeq \left(\frac{\Dnu}{\Dnu\ind{\odot}}\right)^{4/3} \left( \frac{M}{M\ind{\odot}} \right)^{1/3} \left( \frac{\Teff}{T\ind{eff,\odot}} \right)^{-1/2},
\end{equation}
where $\nu\ind{max,\odot} = 3050\, \mu$Hz, $\Dnu\ind{\odot} = 135.5\, \mu$Hz and $T\ind{eff,\odot} = 5780\,$K \citep{1995A&A...293...87K, 2013A&A...550A.126M}.
This number of modes is representative of the number of observed modes for the least evolved stars in our sample (near $\Dnu \sim 4\,\mu$Hz) but in excess relative to the most luminous RGB and AGB stars. We chose a fixed number of modes to ensure the consistency of the determination of the glitch signature.

\begin{figure*}[htbp]
    \centering
	\begin{minipage}{1.0\linewidth}  
		\rotatebox{0}{\includegraphics[width=0.50\linewidth]{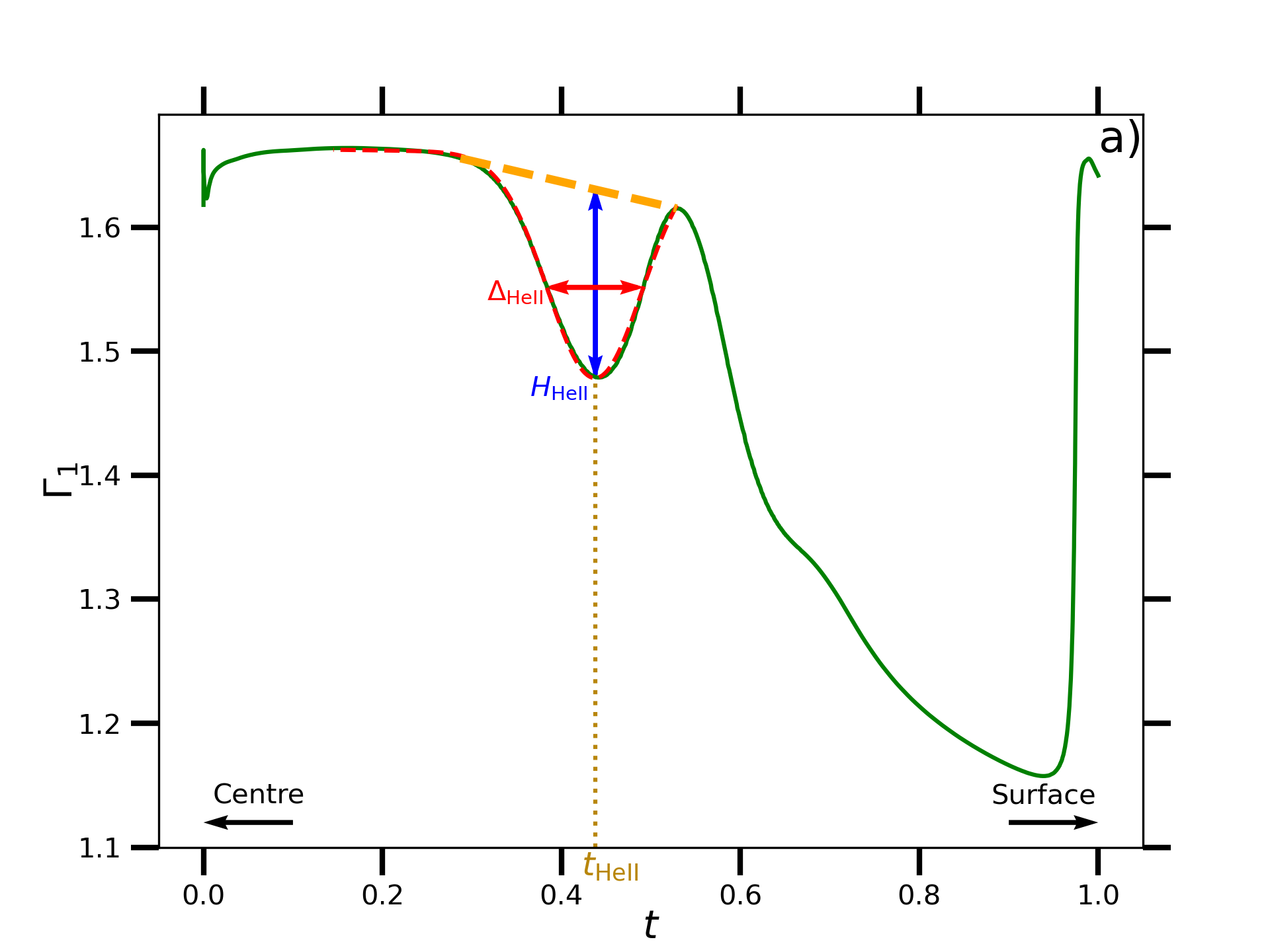}
		\rotatebox{0}{\includegraphics[width=0.50\linewidth]{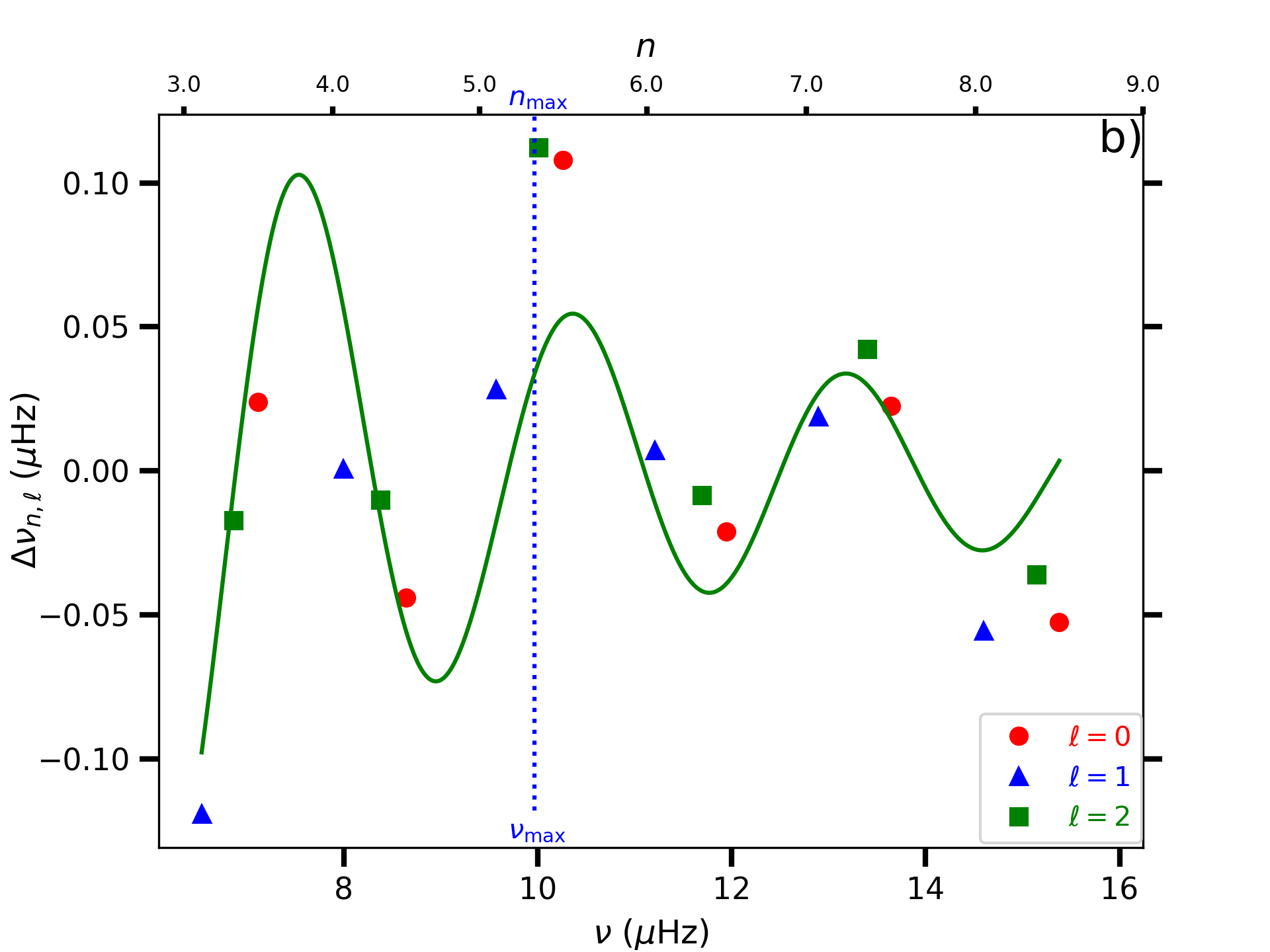}}
		}
	\end{minipage}
	
	\caption{\emph{a)} The profile of the first adiabatic exponent $\Gamma_{1}$ as a function of the normalised acoustic radius in a $1\,M_{\odot}$ RGB model computed with \mesa\ at $\Dnu = 1.65\, \mu$Hz. The parameters of the $\Gamma_{1}$ variations ($\HHeII$, $\tHeII$, and $\DHeII$) are directly shown in the figure. The green solid line is the $\Gamma_{1}$ profile throughout the star. The thick orange dashed line indicates the baseline that connects the local maximum after the dip caused by the second He-ionisation with the $\Gamma_{1}$ profile before the dip. The thin red dashed line gives the fit of the $\Gamma_{1}$ profile with Eq.~\ref{eq:fit_Gamma_1_around_HeII} around the dip. \emph{b)} Glitch modulation induced by the second He-ionisation zone in the same model as in the left panel. The local large separation $\Delta\nu_{n,\ell}$ is shown in red circles, blue triangles and green squares for radial, dipole, and quadrupole modes, respectively. The green solid line is the damped oscillator model given by Eq.~\ref{eq:fit_glitch_signature_Dreau_2021} fitted to the data points. We point out that the data points are plotted at the mean frequencies $(\nu_{n,\ell} + \nu_{n+1,\ell} )/ 2$. The upper x-axis indicates the radial order of $\ell = 0$ modes, and the blue dotted line locates the maximum oscillation power.
	}
	\label{fig:Gamma_1_profile_fitting}
\end{figure*}

\subsection{Validating the computation of pressure modes}

In order to check if our method for calculating the pure p-modes would give satisfactory results, we compared their frequencies with those of the mixed modes with the lowest inertias.
The comparison was done for our reference models in the He-burning phase defined in Table~\ref{Table:reference_model}. We used a fixed number of about 20,000 mesh points for this as it was sufficient to resolve the eigenfunctions of the mixed modes, and hence extract their frequencies reliably.
We obtained a set of several mixed modes per radial order $n$, or equivalently per $\Dnu$ interval. We took the frequencies of the modes of lowest inertia per radial order $n$ as a reference for the expected pure p-mode frequencies. For each reference, we derived the difference to the pure p-mode frequency (the one derived by setting $\NBV^{2} = 0$). The results are shown in Fig.~\ref{fig:test_computation_pure_p_modes_with_NBV0}\emph{a} and \emph{c}. Because most of the frequency differences are within $0.05\,\Dnu$ (\emph{b} and \emph{d} panels), we conclude that the pure p-mode frequencies are precise representations of the acoustic resonances of the envelope. However, for the dipole modes we do notice a mean difference of $\sim 0.02\Dnu$ between the two ways of estimating the acoustic resonant frequencies (see vertical solid line in panel \emph{b}). This bias has limited consequences for the study of the glitch signature because the signature is extracted from the large frequency separation $\Dnu_{n,\ell}$ (Eq.~\ref{eq:glitch_contribution_Vrard_2015}), which is defined as the difference of frequencies at consecutive radial order $n$ at the same degree $\ell$ (Eq.~\ref{eq:local_Dnu}). However, the spread of $\sim 0.05\,\Dnu$, may impact the amplitude of the glitch modulation, which is typically lower than $\Dnu/10$. In Sect.~\ref{sec:results}, we show that these differences do not prevent us from accurately reproducing the shape of the modulation. Quadrupole modes are better derived compared to dipole modes, with an unbiased measurement of their frequencies. In fact, the inner turning point of the dipole p-mode cavity is located deeper in the interior compared to the quadrupole p-mode cavity, resulting in a stronger coupling with the g-mode cavity where $\NBV^{2} > 0$, hence a more pronounced deviation for the dipole mode frequencies.

The reason for the above frequency differences could be one of the following. First, some mixed modes are poorly estimated because the \adipls\ settings we chose are not adapted for any $\Dnu$. For instance, the mesh redistribution could be improved at low $\Dnu \leq 1\,\mu$Hz to better resolve the displacement eigenfunction of mixed modes. Second, although the modes of lowest inertia are the mixed modes closest to the expected pure pressure modes, they still deviate a bit from the latter. To estimate this deviation we fitted a Gaussian profile to the mode inertia profiles in each $\Dnu$ interval, hence locating the acoustic resonance, and found that its difference to the mode of lowest inertia was in average $0.01\Dnu$. Finally, by keeping the inconsistent $\Gamma_{1}$ profile as presented in Sect.~\ref{sec:method_freq_computation}, a bias of approximately $0.01\Dnu$ is introduced in the dipole mode frequencies at low $\Dnu$ ($\Dnu \leq 4.5\,\mu$Hz).\\ 

The errors introduced in the dipole and quadrupole mode frequencies by setting $\NBV^{2} = 0$ while retaining inconsistent $\Gamma_{1}$ profiles in the core (see Fig.~\ref{fig:test_computation_pure_p_modes_with_NBV0}\emph{b},\emph{d}) are, on average, of the same order as the largest errors reported by \citet{2018MNRAS.478.4697B}, who applied the same technique to RGB models but with consistent $\Gamma_{1}$ profiles. In the following we use the dipole and quadrupole mode frequencies obtained by setting $\NBV^{2} = 0$ in the core without changing the $\Gamma_{1}$ profiles as the reference frequencies for the pure pressure dipole and quadrupole modes, both for RGB and clump/AGB stars. In parallel, we compute the radial modes with the unmodified \mesa\ models. These radial, dipole and quadrupole modes constitute the set of modes used in our study. 

\begin{figure*}[htbp]
    \centering
	\begin{minipage}{0.71\linewidth}  
		\rotatebox{0}{\includegraphics[width=1.0\linewidth]{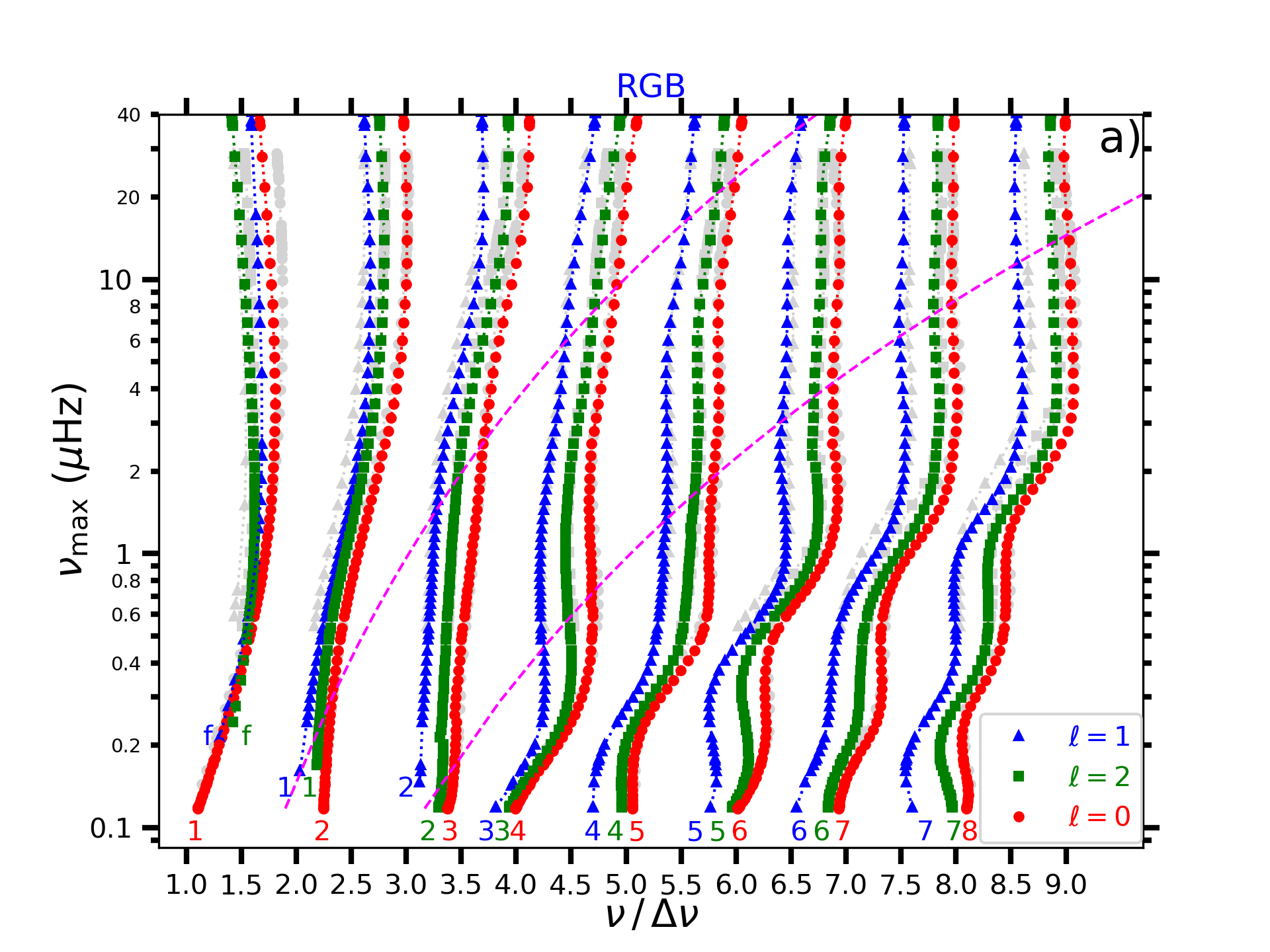}
		}
	\end{minipage}
	\begin{minipage}{0.71\linewidth}{
		\rotatebox{0}{\includegraphics[width=1.0\linewidth]{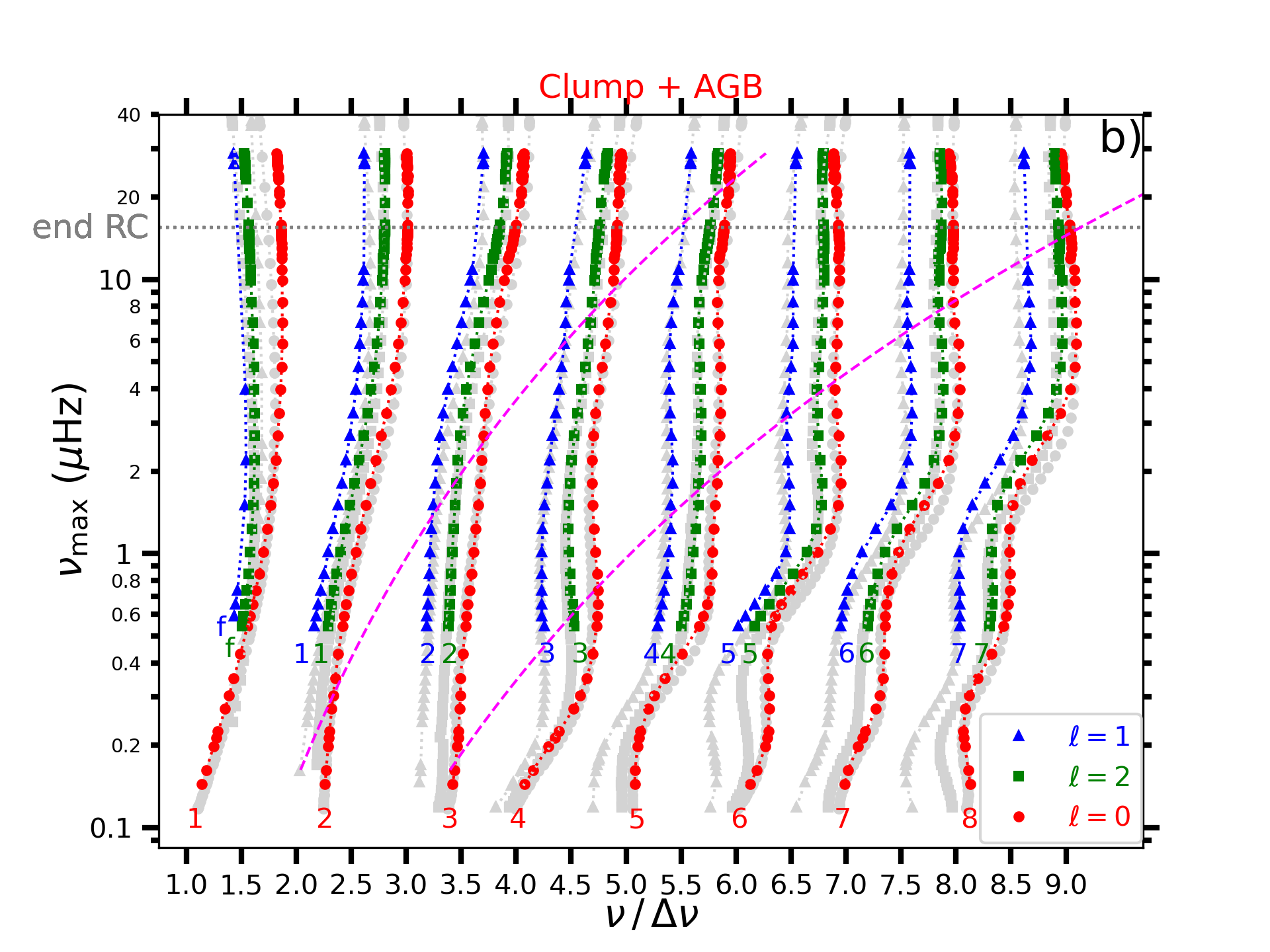}}
		}
	\end{minipage}
	
	\caption{Model frequencies from radial order $n = 1$ up to $n = 8$ computed with \adipls\ for a $1\,M_{\odot}$ track at solar metallicity. The \mesa\ models are computed with the reference input physics listed in Table~\ref{Table:reference_model}.  Radial, dipole and quadrupole modes are shown in red circles, blue triangles and green squares, respectively. For the RGB (panel \emph{a}) the grey frequencies are Clump+AGB models for comparison, and vice versa for the Clump+AGB (panel \emph{b}). The non-radial modes ($\ell = 1,\, 2$) are computed by setting the squared Brunt-V\"ais\"al\"a frequency $\NBV^{2} = 0$ in the core, but retaining the original $\Gamma_{1}$ profile, as described in Sect.~\ref{sec:method}. Modes of the same degree $\ell$ and same radial order $n$ are connected by dotted lines, in red for $\ell = 0$, in blue for $\ell = 1$, and in green for $\ell = 2$. The radial orders are indicated at the lower edge of each branch, with the same colour code as the mode degree $\ell$. The presence of non-radial modes with frequencies below that of the fundamental radial mode are labelled by an ``f". The magenta dashed lines delimit the typical frequency range [$\numax - 0.25\, \numax$, $\numax + 0.25\, \numax$] between which model frequencies are likely to be observed in the oscillation spectrum \citep{2011ApJ...743..161W}. In panel \emph{b)}, the grey dotted line shows the end of the clump phase, taken as when the central helium mass fraction goes below 0.01. We warn that the code could not find out the $\ell = 1$ modes during the clump phase and the early-AGB ($\numax \in [10, 25]\,\mu$Hz) when setting $\NBV^{2} = 0$ in the core. Because the code had troubles to return the non-radial modes for high-luminosity AGB models, we stopped the computation of non-radial modes of $\numax \leq 0.5\,\mu$Hz after the He-core burning phase.
	}
	\label{fig:model_frequencies_ADIPLS}
\end{figure*}

\subsection{Deriving the seismic parameters}
\label{subsec:method_seismic_param_asympt}

In this work, we aim to interpret the observational results of \citet{2021A&A...650A.115D} in terms of internal structure differences between RGB and AGB stars. With the set of frequencies described in Sect.~\ref{sec:method_freq_computation}, we computed the seismic parameters that are the acoustic offset $\varepsilon$ and the reduced small frequency separation $\dol$. The large frequency separation, $\Dnu$, is taken as the slope of the unweighted linear fit to the radial-mode frequencies versus their radial order $n$.

The local acoustic offset $\varepsilon(n)$, which represents the frequency spacing at radial order $n$ between the radial mode at frequency $\nu_{n,0}$ and $n\Dnu$, is computed as 
\begin{equation}
    \label{eq:eps_ADIPLS}
    \varepsilon(n) = \frac{\nu_{n,0} - n \Dnu}{\Dnu}.
\end{equation}
The dimensionless local small frequency separations $\dol(n)$ in fraction of $\Dnu$ are defined as 
\begin{equation}
\label{eq:dnu0l_definition}
\left\lbrace
\begin{array}{ccc}
d_{01}(n) = \frac{1}{2\Dnu}\left(\nu_{n,0} - 2 \ \nu_{n,1} + \nu_{n+1,0}\right) \\
d_{02}(n) = \frac{\nu_{n, 0} - \nu_{n-1, 2}}{\Dnu}, \qquad \qquad \qquad \ \ \ 
\end{array}
\right.
\end{equation}
and the global seismic parameters $\varepsilon$ and $\dol$ are taken as the average of the local values computed with our set of frequencies while the uncertainties are obtained as the standard deviation of the local values.\\

The method we used to extract the glitch parameters is in many aspects similar to that adopted in \citet{2021A&A...650A.115D}. We inferred the glitch signature $\delta_{n,\ell}^{\mathrm{gl}}$ by evaluating the difference between the measured local large frequency separation $\Dnu_{n,\ell}$ defined by 
\begin{equation}
    \label{eq:local_Dnu}
    \Dnu_{n,\ell} = \nu_{n+1,\ell} - \nu_{n,\ell},
\end{equation}
and the expected value $\Dnu_{n}^{\mathrm{UP}}$ without glitch derived from Eq.~\ref{eq:nunl_asympt}, which gives 
\begin{equation}
    \label{eq:glitch_contribution_Vrard_2015}
    \delta_{n,\ell}^{\mathrm{gl}} = \Dnu_{n,\ell} - \Dnu_{n}^{\mathrm{UP}},
\end{equation}
where $\Dnu_{n}^{\mathrm{UP}} = \left( 1 + \alpha \left(n - \nmax + \frac{1}{2}\right)  \right) \Dnu$. We then fitted the glitch signature by a damped oscillator function 
\begin{equation}
\label{eq:fit_glitch_signature_Dreau_2021}
\delta_{n,\ell}^{\mathrm{gl}} = \Amplgl\, \left( \frac{\numax}{\nu} \right)^{2} \Dnu\, \cos \left( 2\pi\frac{ \nu - \numax }{\Ggl\Dnu}  + \Phigl \right),
\end{equation}
where $\Amplgl$ and $\Ggl$ are, respectively, the dimensionless amplitude and period of the glitch modulation in units of $\Dnu$. $\Phigl$ is the phase of the modulation centred on $\numax$ \citep{2015A&A...579A..84V}. The uncertainties of the glitch parameters are extracted from the covariance matrix resulting from the Levenberg-Marquardt algorithm to fit the glitch signature. The main difference between this method and that adopted in \citet{2021A&A...650A.115D} lies in the choice of the initial guess for the glitch period $\Ggl$. The range of all possible values taken by the glitch parameters is large throughout the evolution from the RGB up to the AGB, and the optimisation function to be minimised has several local extrema according to the modulation period $\Ggl$. This complication is even more relevant at low $\Dnu$ because there the asymptotic expansion (Eq.~\ref{eq:nunl_asympt}) is less accurate, which makes it hard to extract the glitch signature. In order to mitigate the bias induced by the choice of initial conditions, the optimised glitch parameters of one stellar model are given as initial guesses for the glitch parameters at the next model. This allows us to extract glitch parameters that smoothly evolve between consecutive stellar models. 

Finally, we investigated the correlations between the physical properties of the HeII ionisation zone and the glitch parameters. As illustrated in Fig.~\ref{fig:Gamma_1_profile_fitting}, we characterise the HeII ionisation zone by three parameters, which are the amplitude, $\HHeII$, of the associated dip in the first adiabatic exponent $\Gamma_{1}$, the acoustic radius, $\tHeII$, of the dip (meaning the time for a sound wave to travel from the stellar centre to that location), and the width, $\DHeII$, of the dip. Hereafter, both $\tHeII$ and $\DHeII$ are normalised by the total acoustic radius of the star $T_{0} = 1/(2\Dnu\ind{as})$, which is the total time it takes a sound wave to travel from the centre to the surface. Here, $\Dnu\ind{as}$ is the asymptotic large frequency separation:

\begin{equation}
    \label{eq:Dnu_as}
    \Dnu\ind{as} = \left(2 \int_{0}^{R}{\frac{dr}{c\ind{s}}} \right)^{-1},
\end{equation}
where $c\ind{s}$ is the sound speed.
We fit the $\Gamma_{1}$ profile in the vicinity of the dip by a Gaussian on top of a linear baseline as follows

\begin{equation}
    \label{eq:fit_Gamma_1_around_HeII}
    \mathcal{L}\ind{HeII}(t) = - \HHeII\, \mathrm{e}^{-\frac{(t - \tHeII)^{2}}{2\DHeII^{2}}} + a_{0} + a_{1} t,
\end{equation}
where $t$ is the normalised acoustic radius, $a_{0}$ and $a_{1}$ are the coefficients of the linear baseline of the $\Gamma_{1}$ profile (orange dashed line). In the fitting process, both $\HHeII$, $\tHeII$, and $\DHeII$ are left as free parameters, but $a_{0}$ and $a_{1}$ are fixed by connecting the local maximum below and above the $\Gamma_{1}$ dip.

\section{Seismic diagnostics for RGB and AGB stages}
\label{sec:results}

In \citet{2021A&A...650A.115D}, we were able to accurately extract the seismic parameters of stars with $\Dnu$ values down to $0.5\,\mu$Hz with the 1470-day time series of \Kepler. Hereafter, we examine the p-mode parameters obtained with \adipls, giving us the opportunity to compare the seismic parameters derived from observations to those from stellar models as well as to extend the analysis up to higher luminosity stages of the RGB and AGB (equivalent to $\Dnu \gtrsim 0.06\, \mu$Hz).

\begin{figure}[htbp]
    \centering
	\begin{minipage}{1.0\linewidth}  
		\rotatebox{0}{\includegraphics[width=1.0\linewidth]{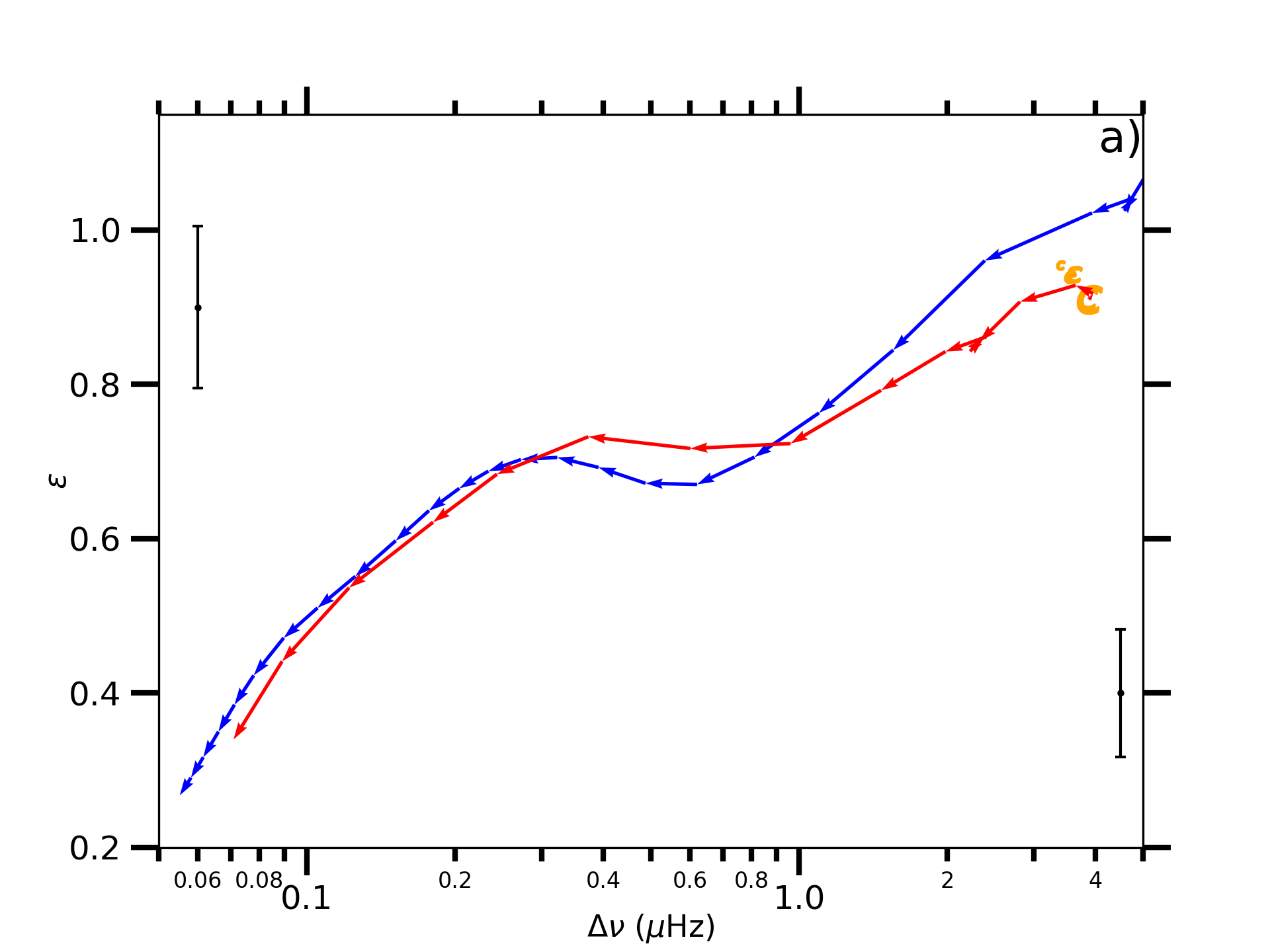}}
		\rotatebox{0}{\includegraphics[width=1.0\linewidth]{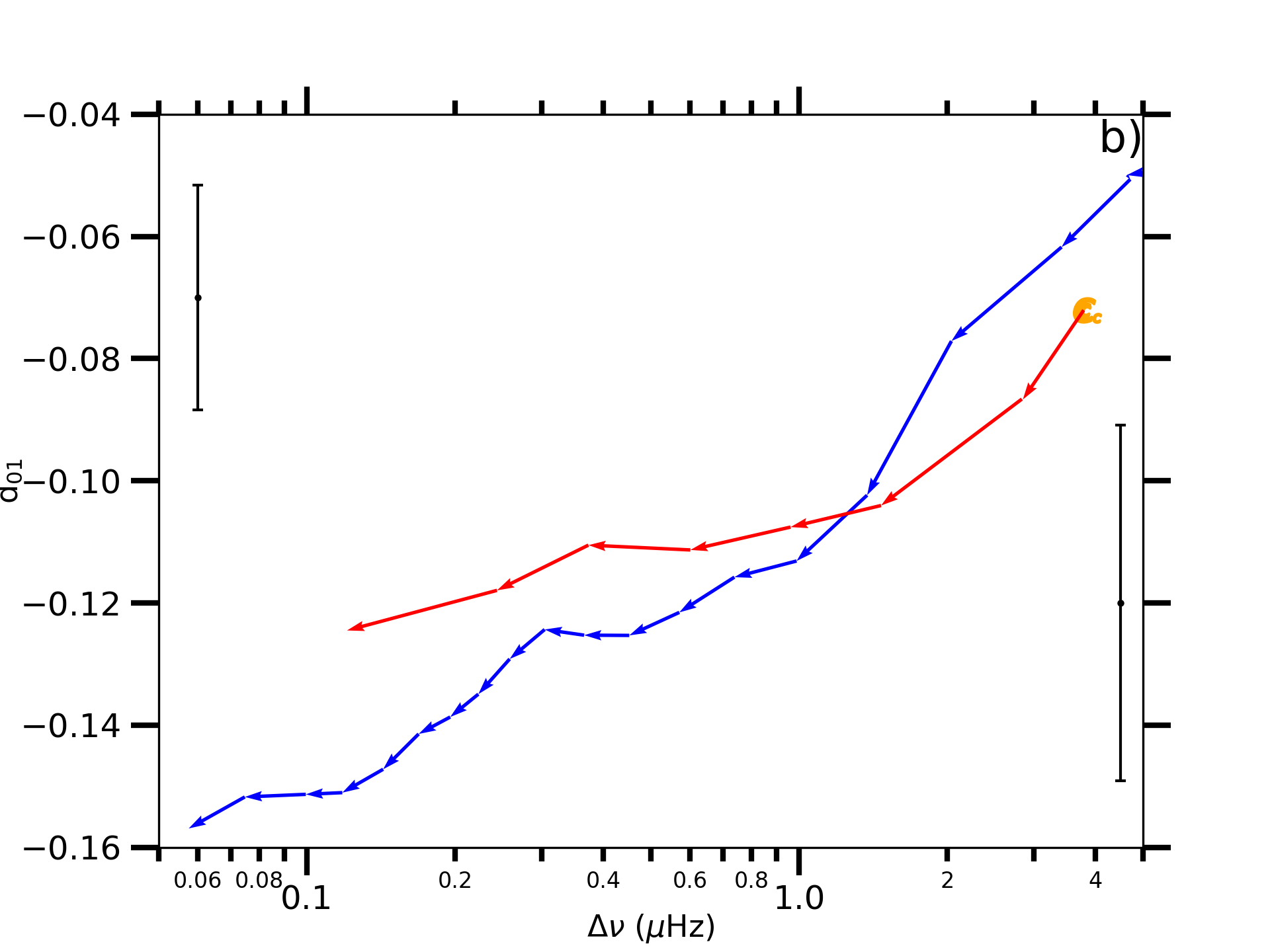}}
	\end{minipage}
	\begin{minipage}{1.0\linewidth}{
		\rotatebox{0}{\includegraphics[width=1.0\linewidth]{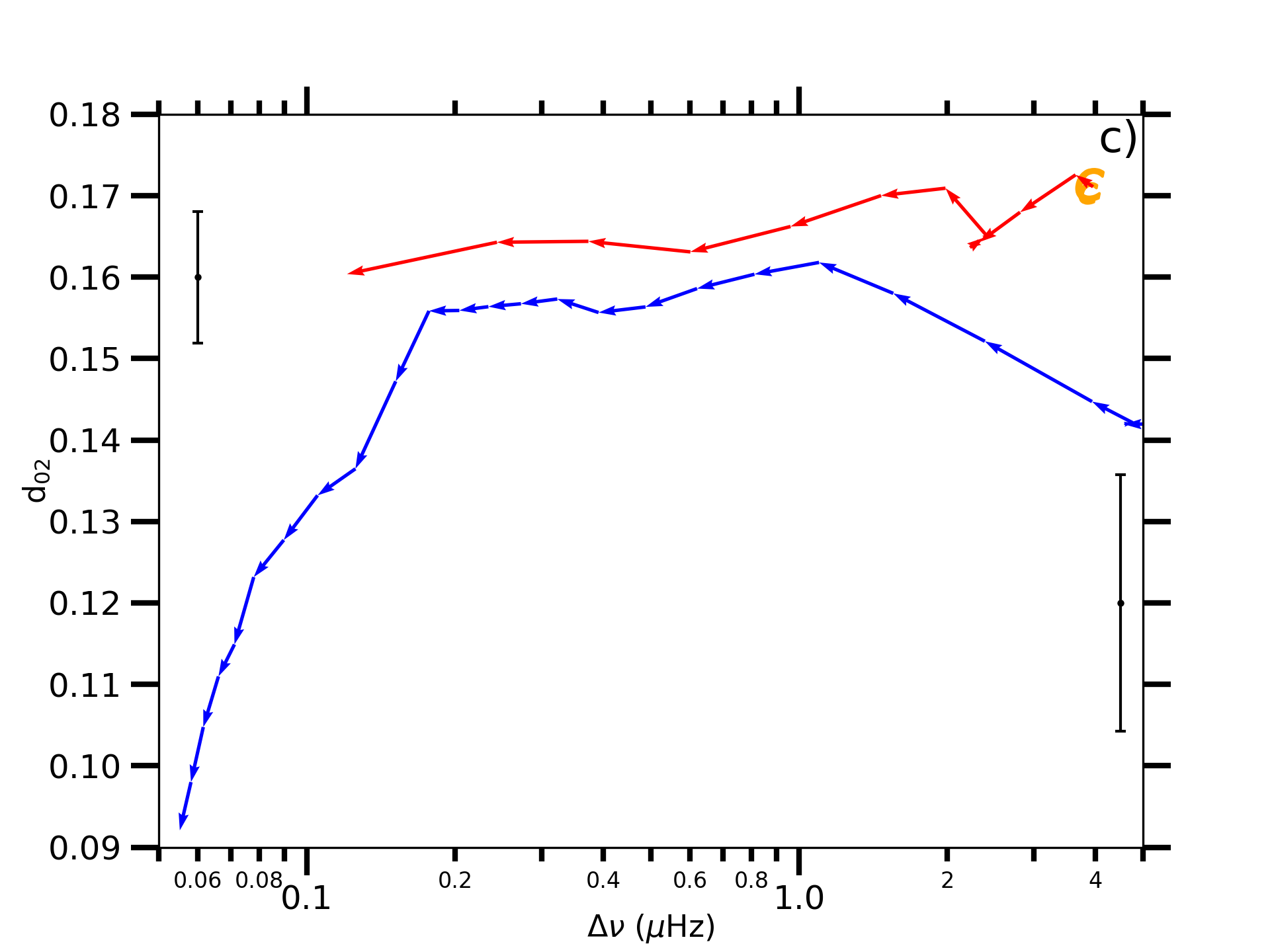}}
		}
	\end{minipage}
	
	\caption{Synthetic seismic parameters extracted from the p-mode frequencies computed with \adipls, as described in Sect.~\ref{subsec:method_seismic_param_asympt}. The \mesa\ models are computed with the reference input physics listed in Table~\ref{Table:reference_model}, with $M = 1.0\, M_{\odot}$ and solar metallicity. \emph{a)} Variation of the acoustic offset $\varepsilon$ as a function of $\Dnu$, with an emphasis on the evolutionary stage. Clump stars are shown by the orange ``C" symbols while the RGB and AGB are colour-coded in blue and red, respectively. ``C" symbols show the progress of the clump stage: small (large, respectively,) symbols correspond to the early (late, respectively,) clump phase. The arrows indicate the direction of evolution. \emph{b)} and \emph{c)} dimensionless small separations $\dol$ as a function of $\Dnu$. Mean error bars estimated for $\Dnu$ below or above $0.5\, \mu$Hz are represented on each panel.
	}
	\label{fig:seismic_parameters_asymptotic_pattern}
\end{figure}

\begin{figure*}[htbp]
    \centering
	\begin{minipage}{1.0\linewidth}  
		\rotatebox{0}{\includegraphics[width=0.50\linewidth]{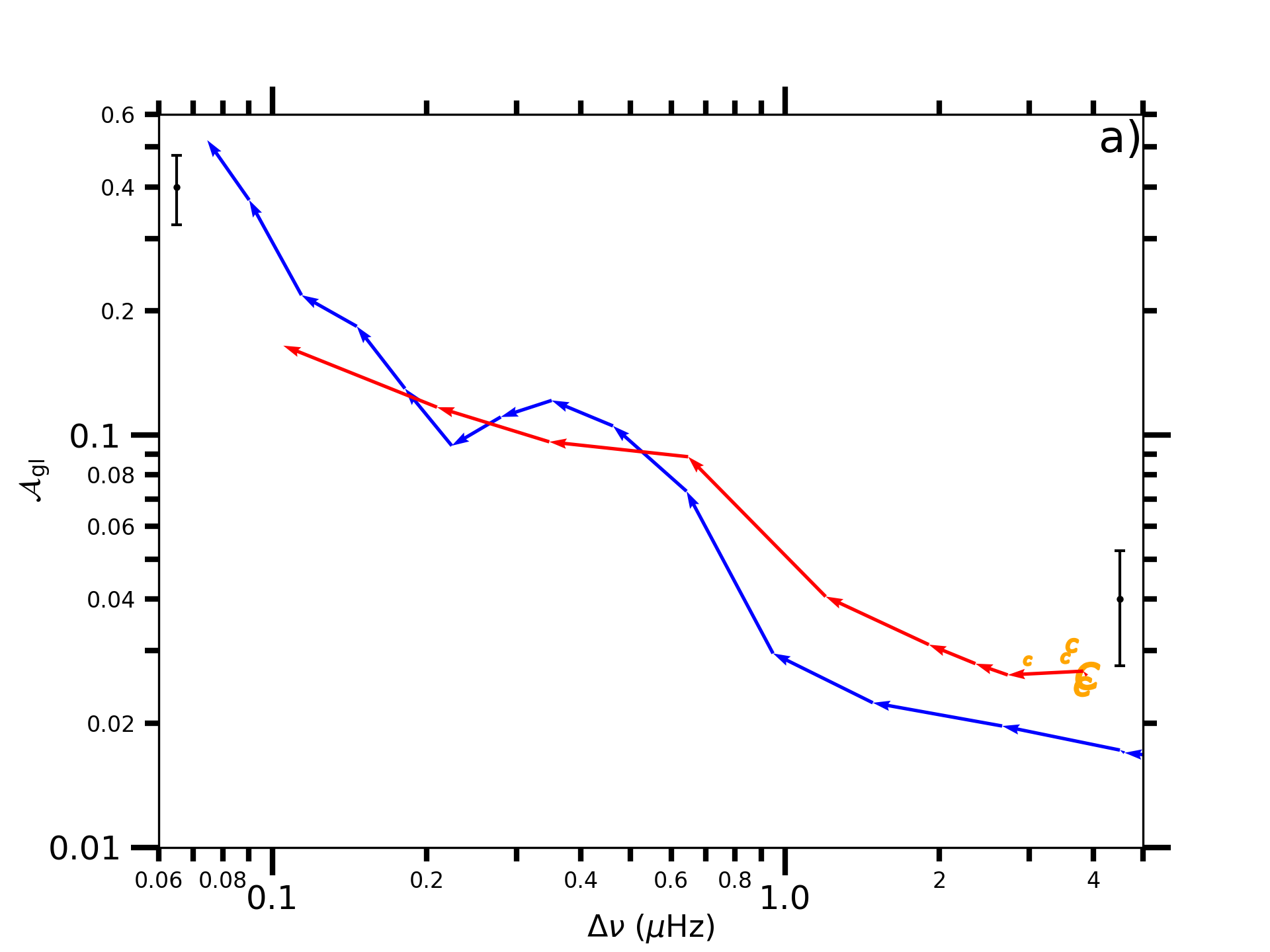}
		\rotatebox{0}{\includegraphics[width=0.50\linewidth]{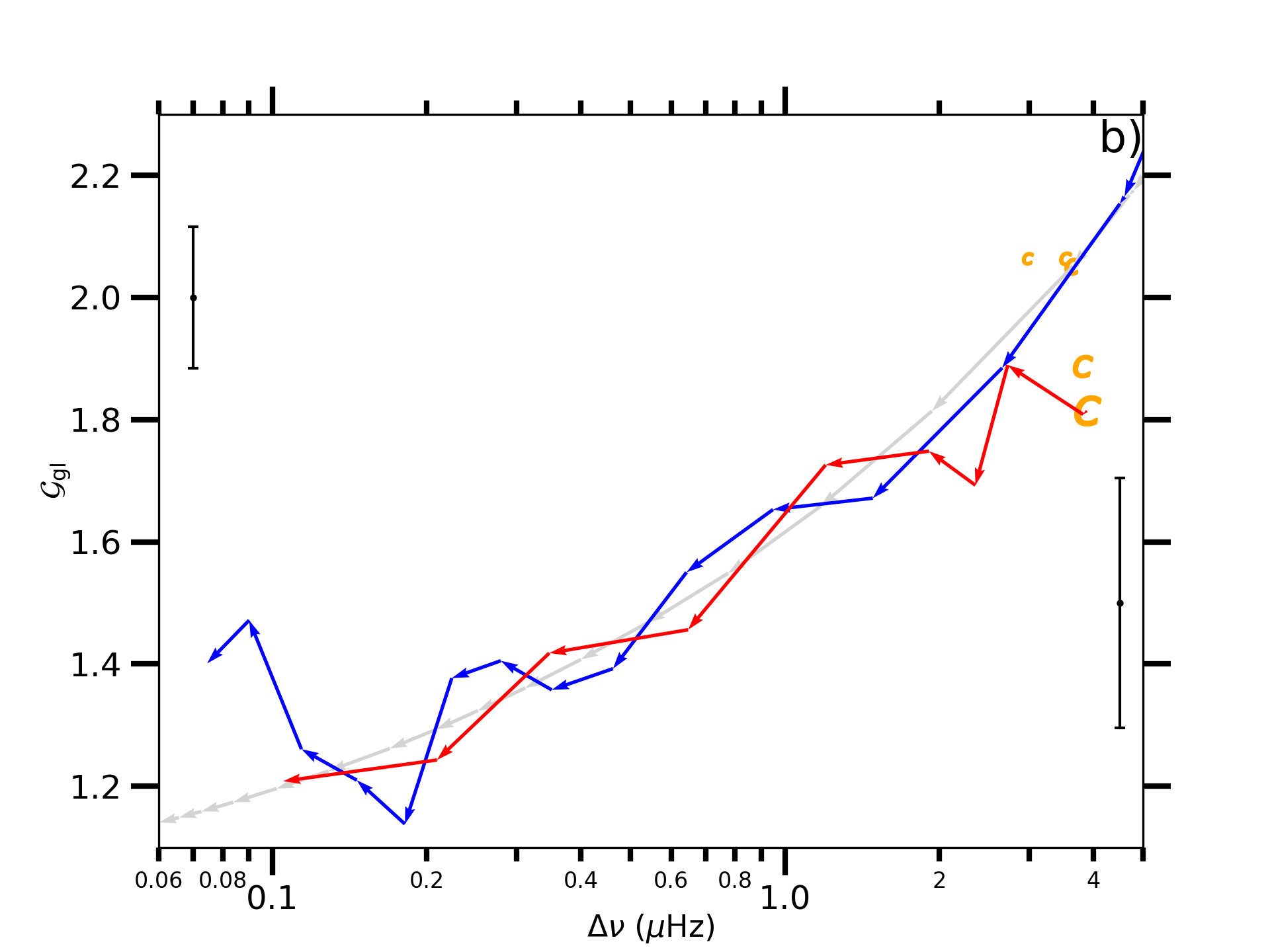}}
		}
	\end{minipage}
	\begin{minipage}{1.0\linewidth}  
		\rotatebox{0}{\includegraphics[width=0.50\linewidth]{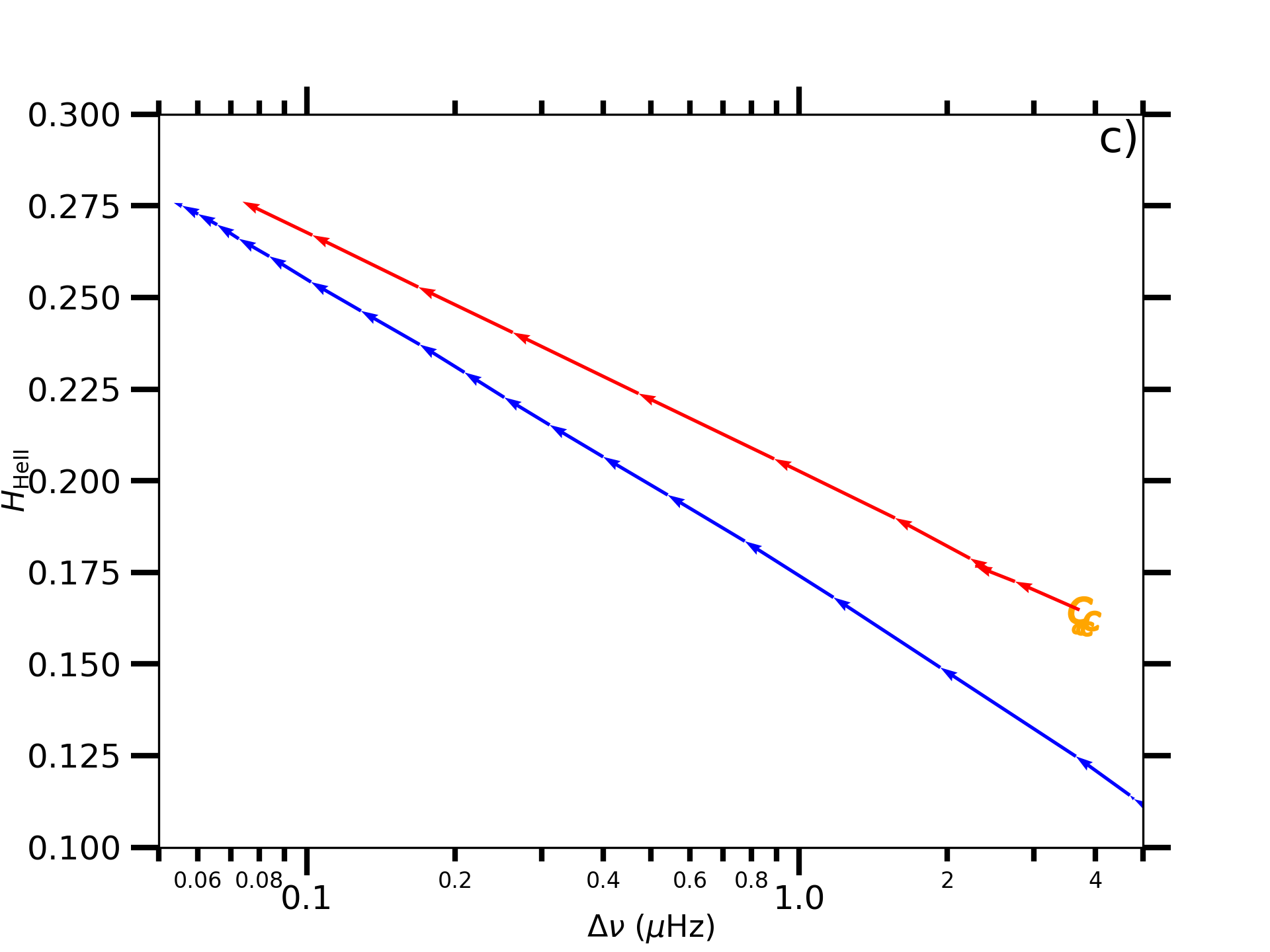}
		\rotatebox{0}{\includegraphics[width=0.50\linewidth]{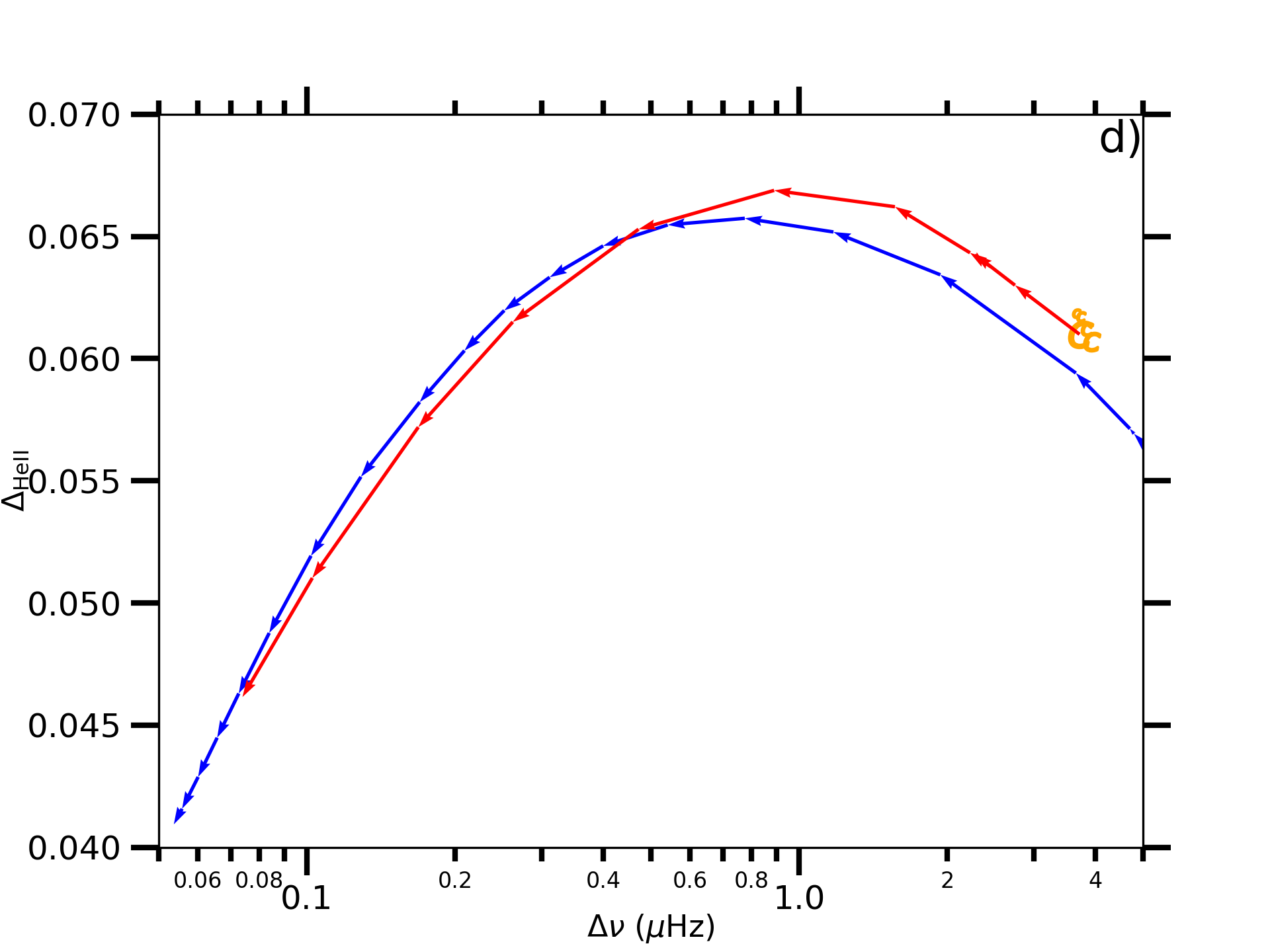}}
		}
	\end{minipage}
	\caption{Synthetic glitch and structure parameters computed with \mesa\, and \adipls. The \mesa\ models are the same as those shown in Fig.~\ref{fig:seismic_parameters_asymptotic_pattern}. The glitch amplitude $\Amplgl$ and period $\Ggl$ are shown on the \emph{a)} and \emph{b)} panels, respectively, while the amplitude $\HHeII$ and the width $\DHeII$ of the HeII zone in the $\Gamma_{1}$ profile are exhibited on the \emph{c)} and \emph{d)} panels, respectively. The label matches that of Fig.~\ref{fig:seismic_parameters_asymptotic_pattern}. In panel \emph{b)}, the additional light grey solid line is the modulation period expected from the location of the HeII ionisation zone $\tHeII$, which is computed according to Eq.~\ref{eq:period_modulation_as_function_acoustic_depth}. }
	
	\label{fig:glitch_structure_parameters}
\end{figure*}

\subsection{Oscillation spectrum across evolution}

Following the representation by \citet{2014ApJ...788L..10S}, we show in Fig.~\ref{fig:model_frequencies_ADIPLS} the structure of the oscillation spectrum at different evolutionary stages; each row being the frequencies of one model. From the top, they span from $\numax \sim 40\, \mu$Hz (or $\Dnu \sim 4\, \mu$Hz, equivalent to red clump luminosities) to more luminous stars with $\numax \sim 0.1\, \mu$Hz (or $\Dnu \sim 0.06\, \mu$Hz). RGB (panel \emph{a}) and clump/AGB (panel \emph{b}) stages are shown separately. As reported in \citet{2014ApJ...788L..10S}, the dipole modes are no longer located roughly halfway between adjacent radial modes as predicted by the asymptotic pattern \citep{1980ApJS...43..469T}. Rather, they get closer to the right-hand side radial and quadrupole modes, forming a triplet pattern. This behaviour is even more pronounced at low radial orders, which are detectable for low $\Dnu$ stars (as indicated by the modes between the dashed magenta lines). In addition, the whole oscillation spectrum narrows when $\Dnu$ decreases.
This explains why the observed frequency spacings between non-radial modes and the neighbouring radial mode narrows as stars become more luminous, as described in previous studies \citep[e.g.][]{2010ApJ...713L.176B, 2011A&A...525L...9M, 2011ApJ...743..143H, 2020MNRAS.493.1388Y, 2021A&A...650A.115D}. No clear difference can be seen between the radial and non-radial mode patterns of RGB and clump/AGB stars, except that the spacing between radial and non-radial modes differs slightly at fixed $\Dnu$ \citep{2010ApJ...721L.182M, 2012ASSP...26...23M}.\\
When $\numax \lesssim 10\, \mu$Hz, we see that the radial and non-radial ridges significantly deviate from vertical ridges, which we would not expect if the modes follow the asymptotic relation. The asymptotic relation assumes that physical properties vary smoothly in the interior. These assumptions may not be valid anymore at high-luminosity stages, in particular because of the occurrence of sharp variations in the structure such as the HeII ionisation zone that are not taken into account in the asymptotic expansion. This would explain the significant departure from the asymptotic ridges. The validity of the asymptotic approach is discussed in further details in Sect.~\ref{sec:discussion}. Finally, we note the presence of non-radial modes with frequencies below that of the fundamental radial mode. As discussed in \citet{2014ApJ...788L..10S}, these could be related to f modes \citep{1941MNRAS.101..367C, 1989nos..book.....U}.
F-modes are suspected to produce the sequence $F$ near the fundamental sequence C in the period-luminosity diagram of semi-regular variable stars \citep{2014ApJ...788L..10S}.

\subsection{Seismic parameters in the asymptotic relation}
\label{subsec:result_seismic_param_asympt}

The dependence of the acoustic offset $\varepsilon$ and the dimensionless small separations $\dol$ on stellar evolution is shown in Fig.~\ref{fig:seismic_parameters_asymptotic_pattern}. In agreement with \citet{2011ApJ...743..161W}, we see in Fig.~\ref{fig:seismic_parameters_asymptotic_pattern}\emph{a} that $\varepsilon$ decreases as RGB stars become more luminous (blue line). Here we show AGB stars follow a similar trend (red line). 
However, the differences seen between the RGB and clump/AGB stars visible for $\Dnu \gtrsim 0.3\, \mu$Hz are caused by the signature of the HeII ionisation zone. When computing $\varepsilon$ with Eq.~\ref{eq:eps_ADIPLS}, this signature is 'absorbed by' the $\varepsilon$ term and tends to vanish when averaged over a large number of modes. As proposed by \citet{2012A&A...541A..51K}, this local signature in $\varepsilon$ allows us to perform an efficient classification between RGB, clump and AGB stars based on the values of $\varepsilon$ and $\Dnu$. 

Fig.~\ref{fig:seismic_parameters_asymptotic_pattern}\emph{b}+\emph{c} shows that both on the RGB and AGB, $d_{01}$ increases in absolute value ($\ell = 1$ modes move closer to their higher-frequency radial mode neighbour) as $\Dnu$ decreases. In parallel, $d_{02}$ smoothly varies when $\Dnu \in [0.2, 4.0]\mu$Hz and decreases when $\Dnu \leq 0.2\,\mu$Hz on the RGB ($\ell = 2$ also moves towards the radial mode). This behaviour agrees with the observations of evolved stars \citep{2013A&A...559A.137M, 2014ApJ...788L..10S, 2020MNRAS.493.1388Y, 2021A&A...650A.115D}. 
No clear distinction can be made in the $d_{01}$ profile between RGB and clump/AGB models for $M = 1.0\, M_{\odot}$ as plotted here. However, we note that clump/AGB tend to have larger $\left| d_{01} \right|$ than RGB stars for $M \geq 1.5\, M_{\odot}$. In Fig.~\ref{fig:seismic_parameters_asymptotic_pattern}\emph{b}, we also note the uncertainties on $d_{01}$, which reflect the dispersion of its values across the seven radial orders closest to the order of maximum oscillation power. These uncertainties are about $0.01$ for $\Delta\nu \leq 0.5\,\mu$Hz and $0.015$ for $\Delta\nu \geq 0.5\,\mu$Hz, \emph{i.e.} of the same order of magnitude as the bias presented in Sect.~\ref{sec:method_freq_computation} by keeping an inconsistent $\Gamma_{1}$ profile while setting $\NBV^{2} = 0$ in the radiative core. This confirms that our interpretations are not affected by this bias.

However, we do see a clear difference in $d_{02}$ between RGB and clump/AGB stars (Fig.~\ref{fig:seismic_parameters_asymptotic_pattern}c). In addition, we noticed that the values of $d_{01}$ and $d_{02}$ are highly sensitive to the stellar mass and, to a lesser extent to the mass-loss rate, as already presented in \citet{2010ApJ...723.1607H, 2012ASSP...26...23M, 2021A&A...650A.115D} and depicted in Fig.~\ref{fig:relative_diff_params_aympt}\emph{c,d,e,f} of Appendix~\ref{appendix:effects_input_physics}. These differences between RGB and clump/AGB stars could be attributed to structure changes rather than mass loss processes, because the differences are also visible at high mass ($M \geq 1.5\, M_{\odot})$, where the mass loss rate is smaller. These structure changes could involve the distance between the base of the convective zone and the location of the turning point of the non-radial mode cavities \citep{2010ApJ...721L.182M}.
Aside from the initial stellar mass that is the main parameter affecting the seismic parameters, the metallicity is also expected to impact the measurement of these seismic parameters. These changes can exceed $10\%$ of the $d_{02}$ values, but remains $\sim 5\%$ of the $\varepsilon$ and $d_{01}$ values when switching [Fe/H] from $0\,$dex to $-1\,$dex (see Fig.~\ref{fig:relative_diff_params_aympt} in Appendix~\ref{appendix:effects_input_physics}). This could explain why we did not detect any clear metallicity effects in the sample of \Kepler\ targets studied in \citet{2021A&A...650A.115D}, with metallicities from $-0.75\,$dex to $0.25\,$dex.

\subsection{Signature of the HeII ionisation zone}

\subsubsection{The glitch amplitude $\Amplgl$}
\label{subsec:glitch_ampl}

The evolution of the modulation amplitude of the glitch signature is shown in Fig.~\ref{fig:glitch_structure_parameters}\emph{a}. We notice that the amplitude is larger on the clump stage/AGB than on the RGB, by 40\% on average. These results support the observations made for red giants \citep{2015A&A...579A..84V, 2021A&A...650A.115D}. This difference between RGB and red clump/AGB can be attributed to the strength of the $\Gamma_{1}$ variation at the HeII zone. The depth $\HHeII$ of the $\Gamma_{1}$ dip is larger in the clump/AGB phase than on the RGB (Fig.~\ref{fig:glitch_structure_parameters}\emph{c}), which demonstrates that the signature of the HeII zone on mode frequencies is stronger once He burning occurs. The physical origin of this difference is discussed in Sect.~\ref{sec:discussion}.

\subsubsection{The glitch period $\Ggl$}
\label{subsec:glitch_period_result}

The glitch period is directly related to the HeII ionisation location by\footnote{The computations used to derive this relation are presented in Appendix A of \citet{2021A&A...650A.115D}.}

\begin{equation}
\label{eq:period_modulation_as_function_acoustic_depth}
    \Ggl = \frac{1}{1 - \tHeII},
\end{equation}
where $\tHeII$ is the acoustic radius of the HeII ionisation zone normalised by the total acoustic radius of the stellar cavity $T_{0} = 1/(2\Dnu)$. In Fig.~\ref{fig:glitch_structure_parameters}\emph{b}, we superimpose the modulation period inferred from Eq.~\ref{eq:period_modulation_as_function_acoustic_depth} (grey curve) with that deduced from fitting the glitch modulation induced in the local large separation given by Eq.~\ref{eq:fit_glitch_signature_Dreau_2021} (blue and red curves). We obtain identical values with both methods, which confirms that the glitch signature is properly extracted.
The modulation period computed from the mode frequencies decreases when $\Dnu$ decreases, which reflects that the effective temperature, hence the internal temperature, decreases as stars evolve on the RGB and AGB, and the HeII ionisation zone moves closer to the centre. The decrease is in agreement with results from \Kepler\ high-luminosity stars \citep{2021A&A...650A.115D}, but we do find the trend to be steeper and the period values to be smaller in models than in \Kepler\ data (compare the values and variations from the models in Fig.~\ref{fig:glitch_structure_parameters}\emph{b} with the median values shown in Fig.~\ref{appendix:mass_dependence_Ggl}).

\begin{figure*}[htbp]
    \centering
	\begin{minipage}{1.0\linewidth}  
		\rotatebox{0}{\includegraphics[width=0.50\linewidth]{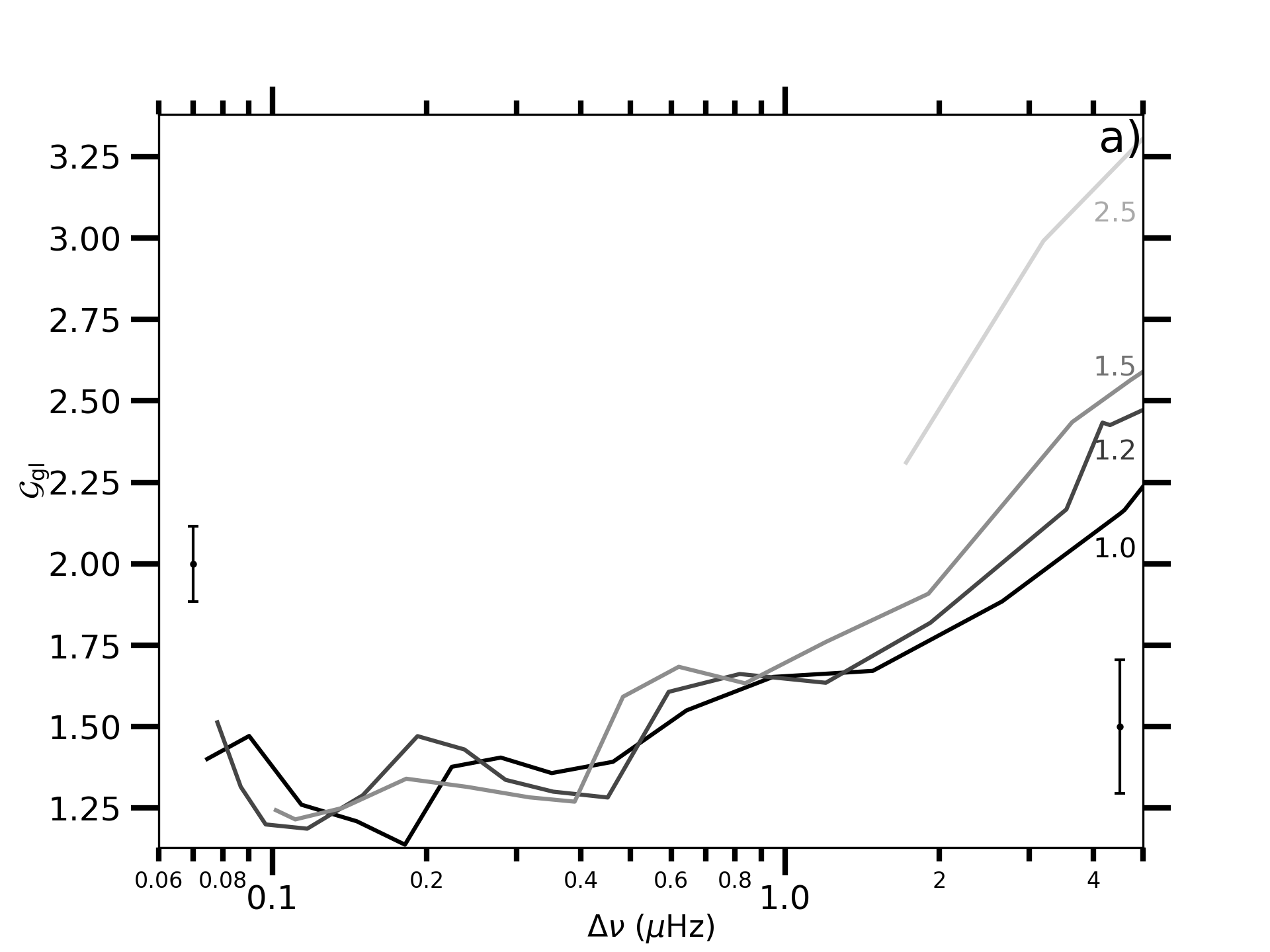}
		\rotatebox{0}{\includegraphics[width=0.50\linewidth]{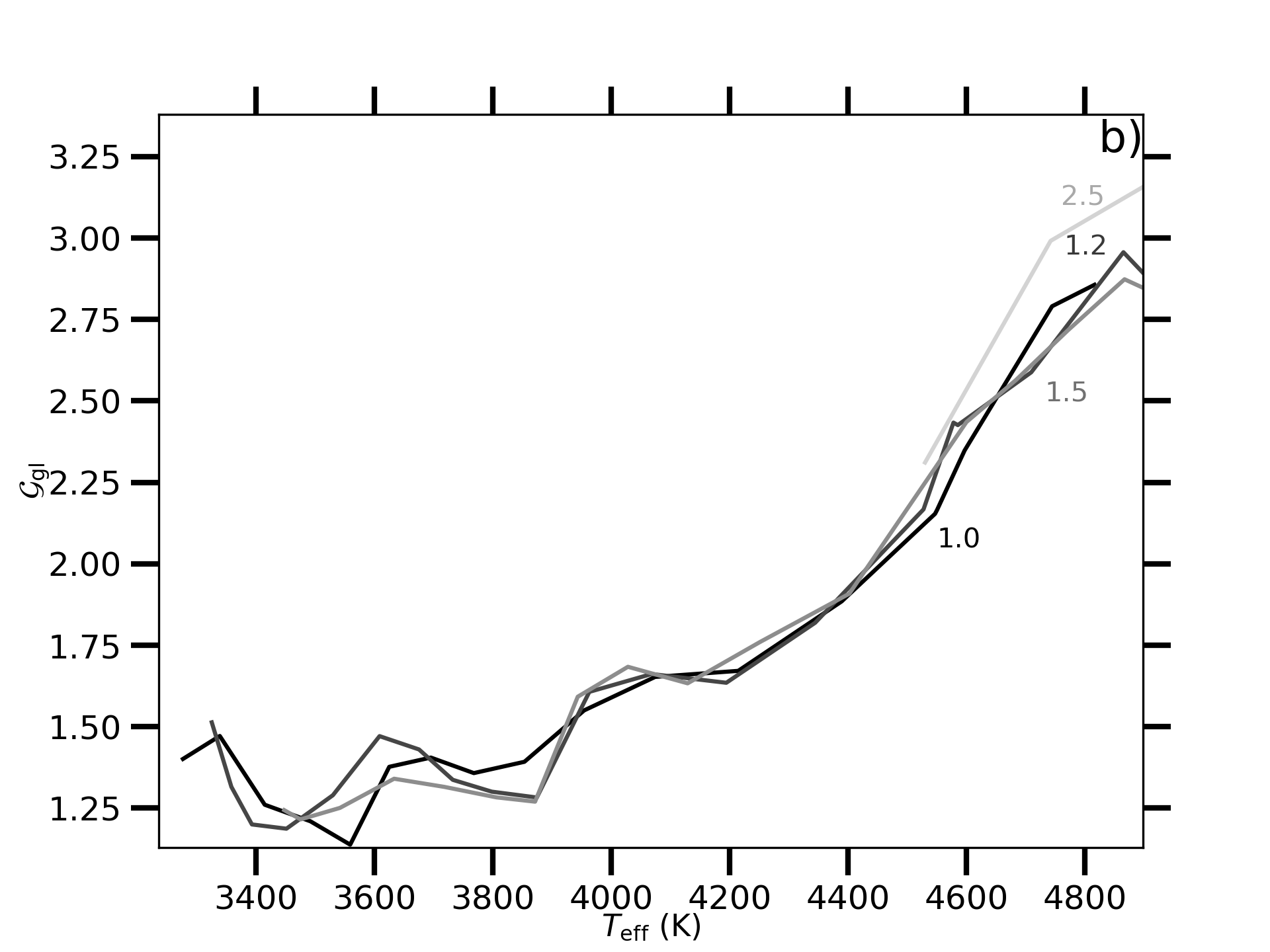}}
		}
	\end{minipage}
	\begin{minipage}{1.0\linewidth}  
		\rotatebox{0}{\includegraphics[width=0.50\linewidth]{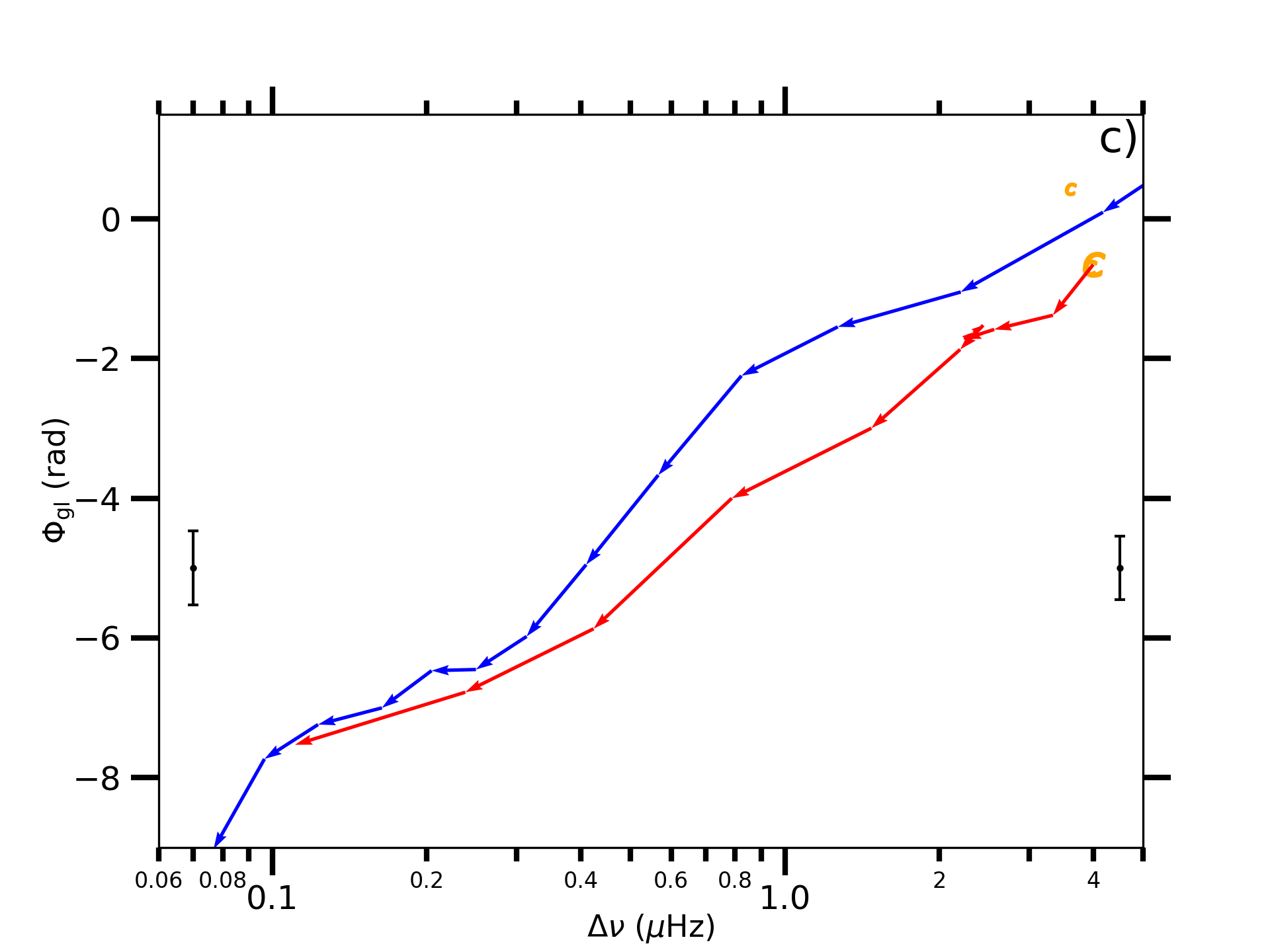}}
		\rotatebox{0}{\includegraphics[width=0.50\linewidth]{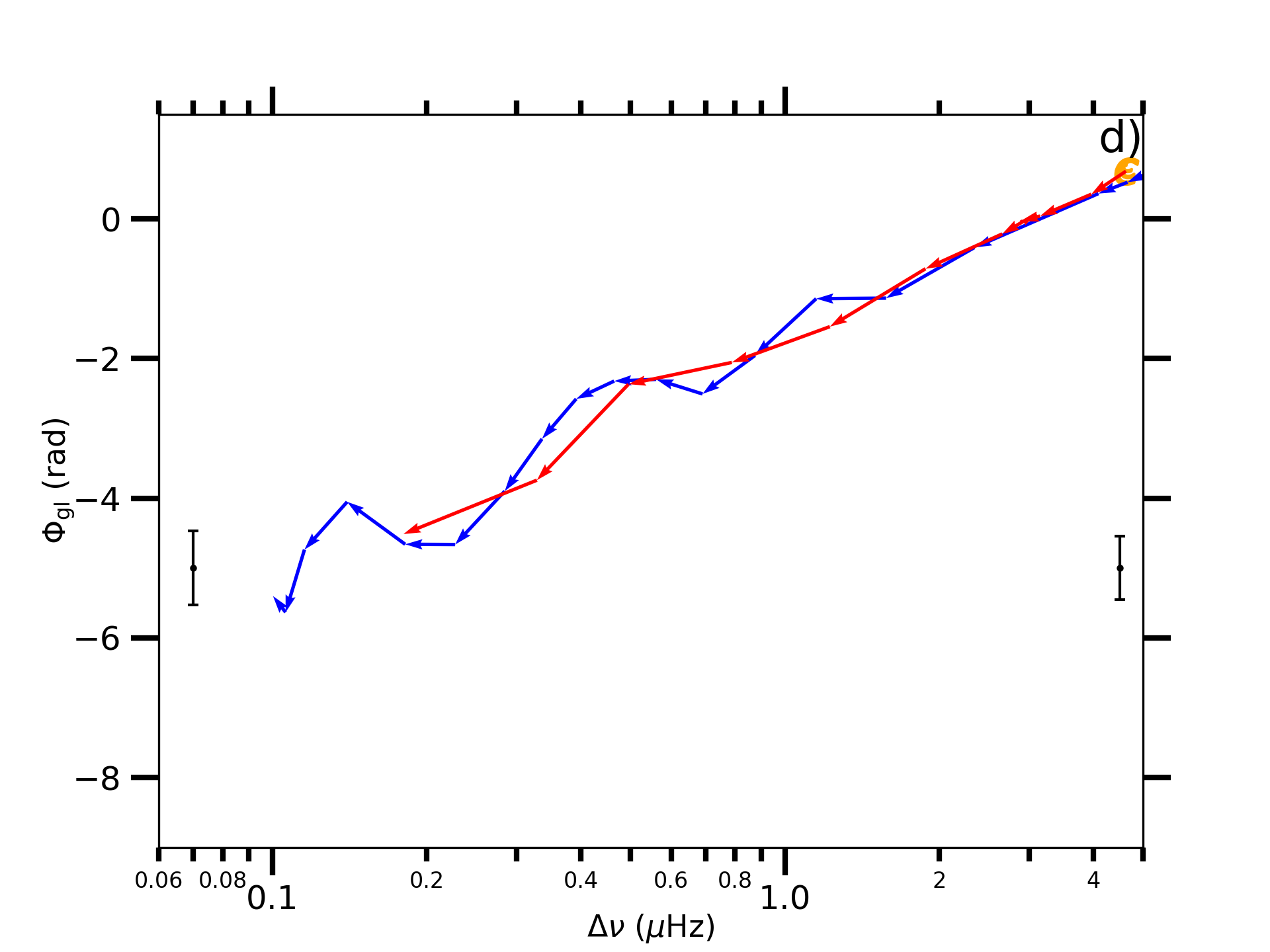}
		}
	\end{minipage}
	\caption{Synthetic glitch period $\Ggl$ and phase $\Phigl$ computed with \adipls. The \mesa\ models are computed with the reference input physics listed in Table~\ref{Table:reference_model}. \emph{a)} Dependence of the modulation period $\Ggl$ on the stellar mass as a function of $\Dnu$ during the RGB. The mass in solar units is shown with different shades of grey. \emph{b)} Same as \emph{a)} but for $\Ggl$ as a function of $\Teff$. \emph{c)} Evolution of the modulation phase $\Phigl$ with $\Dnu$ for models of initial mass $0.9\, M_{\odot}$ at solar metallicity, with an emphasis on the evolutionary stage. The labeling is identical to that in Fig.~\ref{fig:seismic_parameters_asymptotic_pattern}. \emph{d)} Same as \emph{c)}, but for models of initial mass $1.5\, M_{\odot}$. 
	}
	\label{fig:G_gl_Phi_gl_dependence_on_mass_models}
\end{figure*}

In the following, we investigate the mass dependence of the glitch period $\mathcal{G}\ind{gl}$. As a supplementary study of the \Kepler\ data analysis done in \citet{2021A&A...650A.115D}, in Appendix~\ref{appendix:mass_dependence_Ggl} we comment in more details on the dependence of the modulation period on the stellar mass obtained with the sample of \Kepler\ evolved stars used in \citet{2021A&A...650A.115D}. We notice that low-mass stars have a shorter modulation period than their high-mass counterparts at similar $\Dnu$, which is confirmed by stellar models (Fig.~\ref{fig:G_gl_Phi_gl_dependence_on_mass_models}\emph{a}.) In parallel, in Fig.~\ref{fig:G_gl_Phi_gl_dependence_on_mass_models}\emph{b}, we see that regardless of mass, the period is the same at fixed effective temperature. Such behaviour is also found in the acoustic radius of the HeII zone (not shown). The difference in the modulation period (hence in the acoustic radius of the HeII zone) between low- and high-mass stars can be explained by a difference of effective temperature $\Teff$, which is higher for high-mass stars at fixed $\Dnu$.

\subsubsection{The glitch phase $\Phigl$}
\label{subsec:glitch_phase_results}

In \Kepler\ observations, RGB and clump/AGB stars present a clear difference in the modulation phase, $\Phigl$, of the glitch signature \citep{2015A&A...579A..84V, 2021A&A...650A.115D}. This is also what we see from the frequencies of the model with mass $0.9\, M_{\odot}$ shown in Fig.~\ref{fig:G_gl_Phi_gl_dependence_on_mass_models}\emph{c}. There is a phase difference between clump/AGB and RGB stars of about $1\,$ radian all along the clump phase and the AGB, which results in a difference in the glitch contribution to $\Dnu$ (Eq.~\ref{eq:fit_glitch_signature_Dreau_2021}). As shown in Sect.~\ref{subsec:result_seismic_param_asympt}, differences in the measurements of the acoustic offset $\varepsilon$ can be observed between AGB and RGB stars. These differences would be even more pronounced if the $\varepsilon$ values were not averaged over several radial orders. They are in fact equivalent to those seen in the modulation phase $\Phigl$. Differentiating Eq.~\ref{eq:nunl_asympt} at fixed frequency highlights the connection between the small perturbations in $\eps$ and $\Dnu$, the latter being associated with the term $\delta_{n,0}^{\mathrm{gl}}$ \citep{2015A&A...579A..84V}:
\begin{equation}
\label{eq:link_deps_dDnu}
\delta\eps = - (n + \varepsilon_{0})\frac{\delta_{n,0}^{\mathrm{gl}}}{\Dnu},
\end{equation}
where $\delta\varepsilon$ is the contribution of the glitch signature to $\eps$ and $\eps_{0}$ is the acoustic offset in the absence of the glitch. This supports the observational results, showing that the physical basis, on which the classification of RGB and clump/AGB stars relies, is connected to the helium-second ionisation zone \citep{2012A&A...541A..51K, 2014MNRAS.445.3685C, 2015A&A...579A..84V, 2021A&A...650A.115D}. 

In the following we investigate how the modulation phase difference between the RGB and RC/AGB depend on stellar mass. Fig.~\ref{fig:G_gl_Phi_gl_dependence_on_mass_models}\emph{d} shows, for a $1.5\, M_{\odot}$ track, that $\Phigl$ behaves very similar for the two stages of evolution. This is in contrast to the reported phase difference from \Kepler\ observations at all masses (see lower right panel of Fig.~4 of \citet{2021A&A...650A.115D}). We discuss this further in Sect.~\ref{sec:discussion}.

\subsection{Influence of stellar model input physics on seismic parameters}

In order to investigate the sensitivity of seismic parameters of the asymptotic pattern (Eq.~\ref{eq:nunl_asympt}) and the glitch signature (Eq.~\ref{eq:fit_glitch_signature_Dreau_2021}) with respect to input physics, we modified the parameters of the reference model (Table~\ref{Table:reference_model}) step by step. The full scheme is detailed in Appendix~\ref{appendix:effects_input_physics} and the relative differences between the reference model and the modified models are shown in Figs.~\ref{fig:relative_diff_params_aympt};~\ref{fig:relative_diff_params_glitch} and \ref{fig:relative_diff_params_HeII}. To summarise, the major parameters that strongly affect the seismic parameters are the stellar mass $M$ and metallicity [Fe/H]. The initial helium abundance $Y_{0}$ also significantly impacts the glitch parameters, but does not alter the parameters of the asymptotic relation that much. The mixing-length parameter $\alphaMLT$ and mass loss $\eta\ind{R}$ mainly influence the modulation phase $\Phigl$ and the acoustic offset $\eps$. Finally, the inclusion of extra-mixing regions, such as core overshooting, envelope undershooting and thermohaline mixing, slightly affects the glitch parameters, while the impact on the seismic parameters from the asymptotic relation remains negligible. In Appendix~\ref{appendix:effects_input_physics}, we examine the effects of core overshooting during the main sequence for a $1\,M_{\odot}$ model, where they are expected to be negligible due to the small convective core. For comparison, the impact of such overshooting on the modulation phase $\Phigl$ is discussed in Sect.~\ref{subsec:phase_diff_physical_origin} for a $1.75\,M_{\odot}$ model, where the core convective zone is sufficiently developed to produce qualitatively noticeable effects.

\section{Discussion}
\label{sec:discussion}

\subsection{Understanding the strength of the glitch signal}

\begin{figure*}[htbp]
    \centering
	\begin{minipage}{1.0\linewidth}  
		\rotatebox{0}{\includegraphics[width=0.50\linewidth]{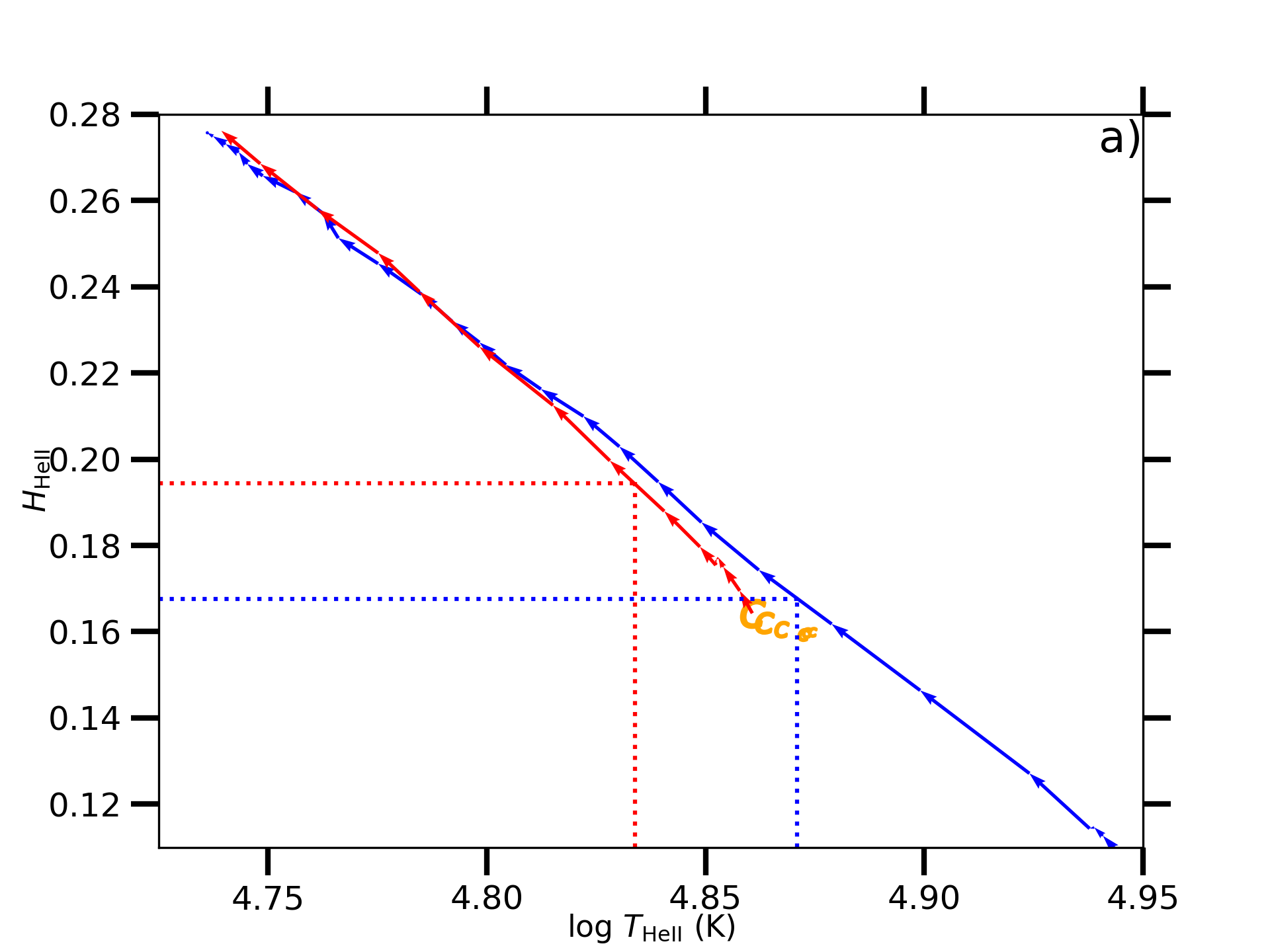}
		\rotatebox{0}{\includegraphics[width=0.50\linewidth]{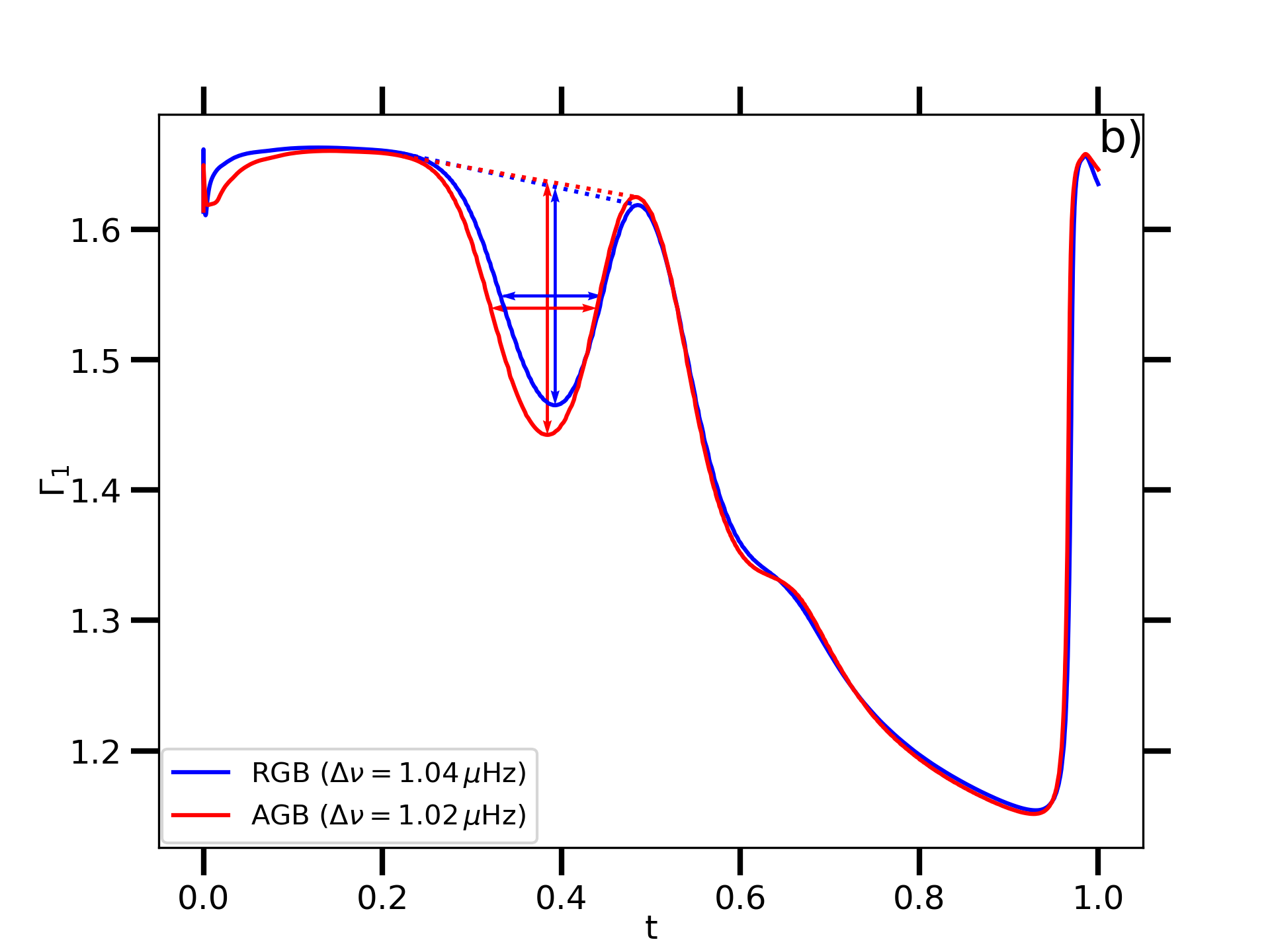}}
		}
    \end{minipage}
	\begin{minipage}{1.0\linewidth}  
		\rotatebox{0}{\includegraphics[width=0.50\linewidth]{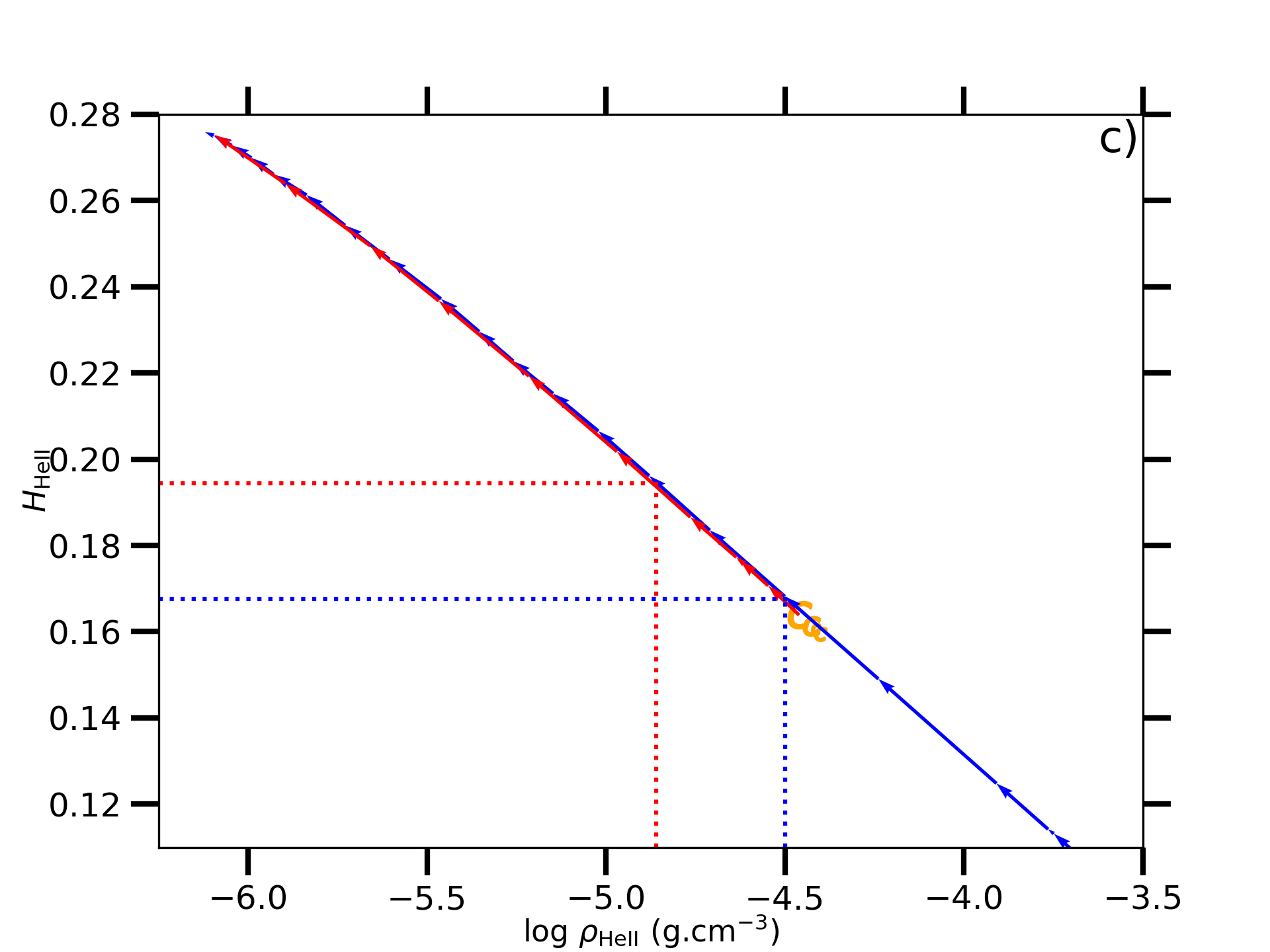}
		\rotatebox{0}{\includegraphics[width=0.50\linewidth]{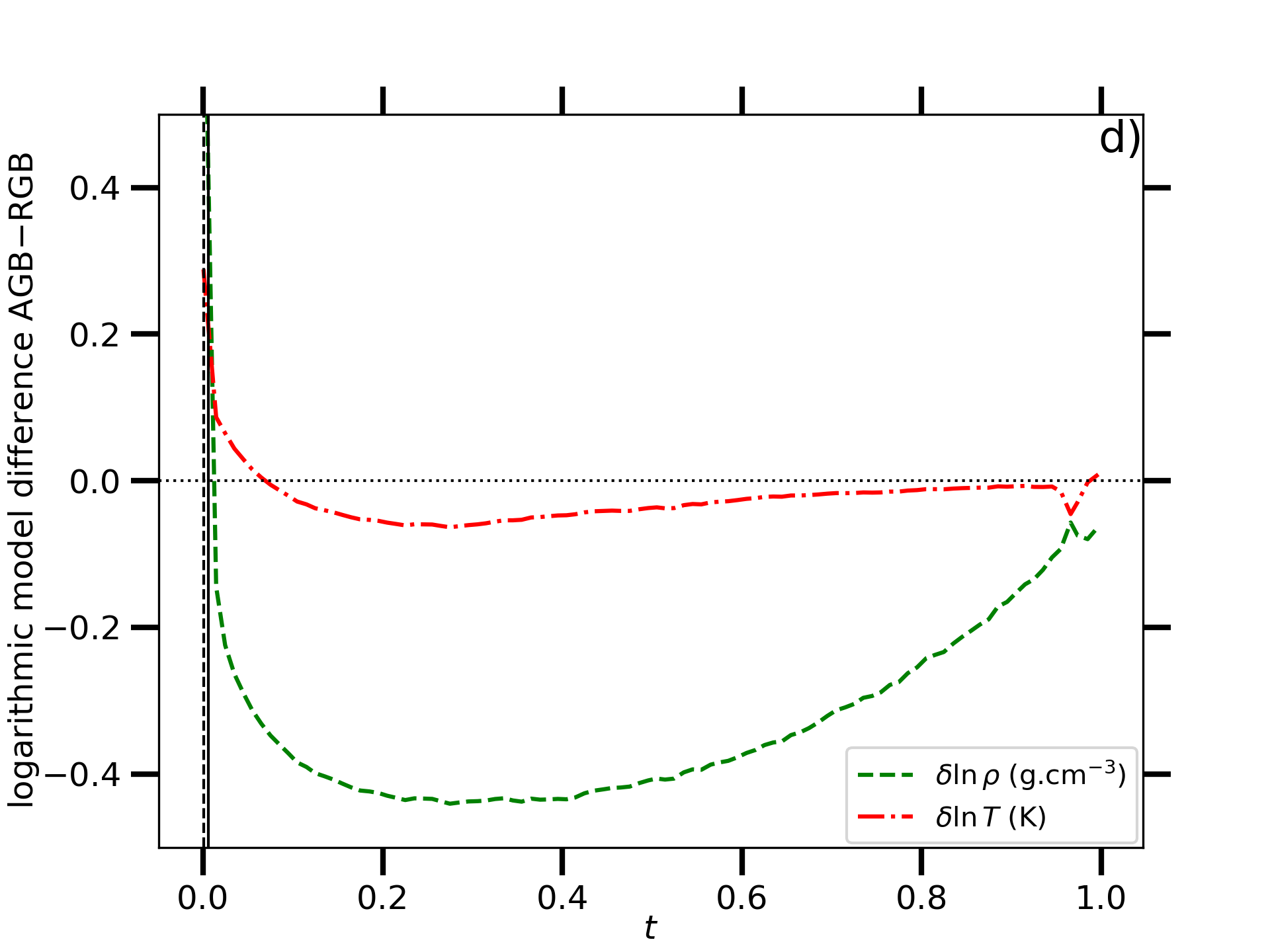}}
		}
    \end{minipage}
	\caption{\emph{a)} Dependence of the amplitude $\HHeII$ on the average temperature $\log \THeII$ in the HeII zone. The input physics and label are the same as in Fig.~\ref{fig:glitch_structure_parameters}, for models of mass $1\,M_{\odot}$ and solar metallicity. Dotted lines show the $\log \THeII-\HHeII$ coordinates of the RGB and AGB models presented in the right panel. \emph{b)} $\Gamma_{1}$ profiles on the RGB and AGB roughly at similar $\Dnu = 1.03\, \mu$Hz. The arrows indicate the amplitude and the width of the $\Gamma_{1}$ variation at the helium-second ionisation zone, in blue on the RGB and red on the AGB. \emph{c)} Same as panel \emph{a)}, but with the average density $\log \rho\ind{HeII}$ in the HeII zone instead of $\log \THeII$. \emph{d)} Logarithmic differences between the AGB and RGB models shown in \emph{b)} of the temperature (red dot-dashed line) and density (green dashed line) as a function of the normalised acoustic radius. The vertical solid and dashed lines indicate the location of the base of the convective envelope in the RGB and AGB models, respectively.
	}
	\label{fig:Amplitude_dependence_internal_structure}
\end{figure*}

The difference in the strength of the glitch signal reported in Sect.~\ref{subsec:glitch_ampl} between He-burning and H-shell burning phases reflects a difference of physical conditions and degree of ionisation in the envelope. The steeper the $\Gamma_{1}$ variation, the stronger the glitch signature. As illustrated in Fig.~\ref{fig:Amplitude_dependence_internal_structure}\emph{a}+\emph{c}, we observe a clear dependence of the amplitude of the $\Gamma_{1}$ variation, denoted $\HHeII$, on the average temperature $\THeII$ and density $\rho\ind{HeII}$ in the HeII ionisation region. Across all evolutionary stages considered, $\HHeII$ exhibits an approximately linear dependence on both $\log \THeII$ and $\rho\ind{HeII}$. This trend is consistent with theoretical predictions indicating a steeper $\Gamma_{1}$ profile in the HeII zone during the core He-burning phase \citep{2014MNRAS.445.3685C}. Such a steepening results from the lower temperature and density in the convective envelope, reflecting the reduced envelope mass in AGB stars at a given value of $\Dnu$.

\citet{2014MNRAS.445.3685C} further emphasised that differences in envelope density, rather than temperature, primarily account for the observed discrepancies in $\Gamma_{1}$ between RGB and clump stars. A similar inference can be drawn from our comparison of RGB and AGB models. In Fig.~\ref{fig:Amplitude_dependence_internal_structure}\emph{d}, we find that for models with $\Dnu \approx 1\,\mu$Hz, the density contrast throughout the convective envelope below the surface is more significant than the corresponding temperature difference. This enhanced density contrast affects the degree of helium ionisation and thus modifies the $\Gamma_{1}$ profile accordingly.

\subsection{Exploring the phase difference between RGB and clump/AGB stars}
\label{subsec:phase_diff_physical_origin}

Mass loss can contribute to the phase difference of the glitch modulation between RGB and clump/AGB stars highlighted in Sect.~\ref{subsec:glitch_phase_results} at low mass. For instance, taking a lower mass-loss rate $\eta\ind{R}$ from $0.3$ to $0.1$ (equivalent to a RC mass in the range $[0.85\, M_{\odot},0.95\, M_{\odot}]$ for an initial mass of $1\, M_{\odot}$) introduces an average phase shift between the modified and reference models $\Delta \Phigl = \Phi\ind{gl,\eta\ind{R} = 0.1} - \Phi\ind{gl,\eta\ind{R} = 0.3}$ of $0.3\,$rad on the clump stage/AGB (see Fig.~\ref{fig:relative_diff_params_glitch}\emph{f}). As a consequence, the absence of mass loss reduces the difference between the modulation phase of RGB and clump/AGB stars, which makes mass loss a solid candidate to explain this difference in low-mass stars. This would also explain why we do not notice any difference between RGB and clump/AGB stars in high-mass models with the same set of input physics because for those the mass loss is small, leading to similar RGB and RC masses. Indeed, with a Reimer's scaling factor $\eta\ind{R} = 0.3$ the RGB tip mass loss is $0.15\, M_{\odot}$ for a $M = 1.0\, M_{\odot}$ model while it is $0.03\, M_{\odot}$ for a $M = 1.75\, M_{\odot}$ model (at solar metallicity).

\begin{figure*}[htbp]
    \centering
	\begin{minipage}{1.0\linewidth}  
		\rotatebox{0}{\includegraphics[width=0.50\linewidth]{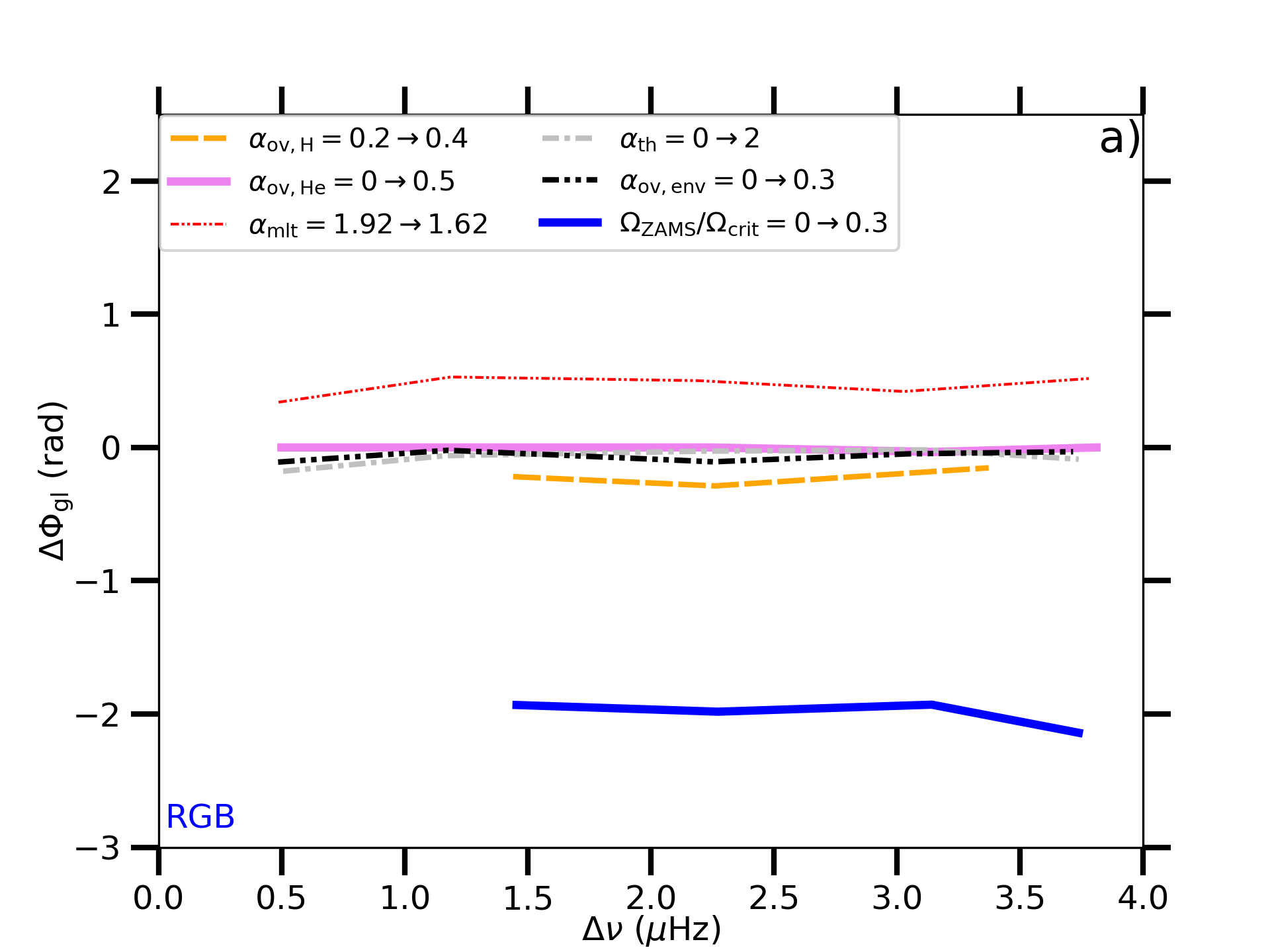}
		\rotatebox{0}{\includegraphics[width=0.50\linewidth]{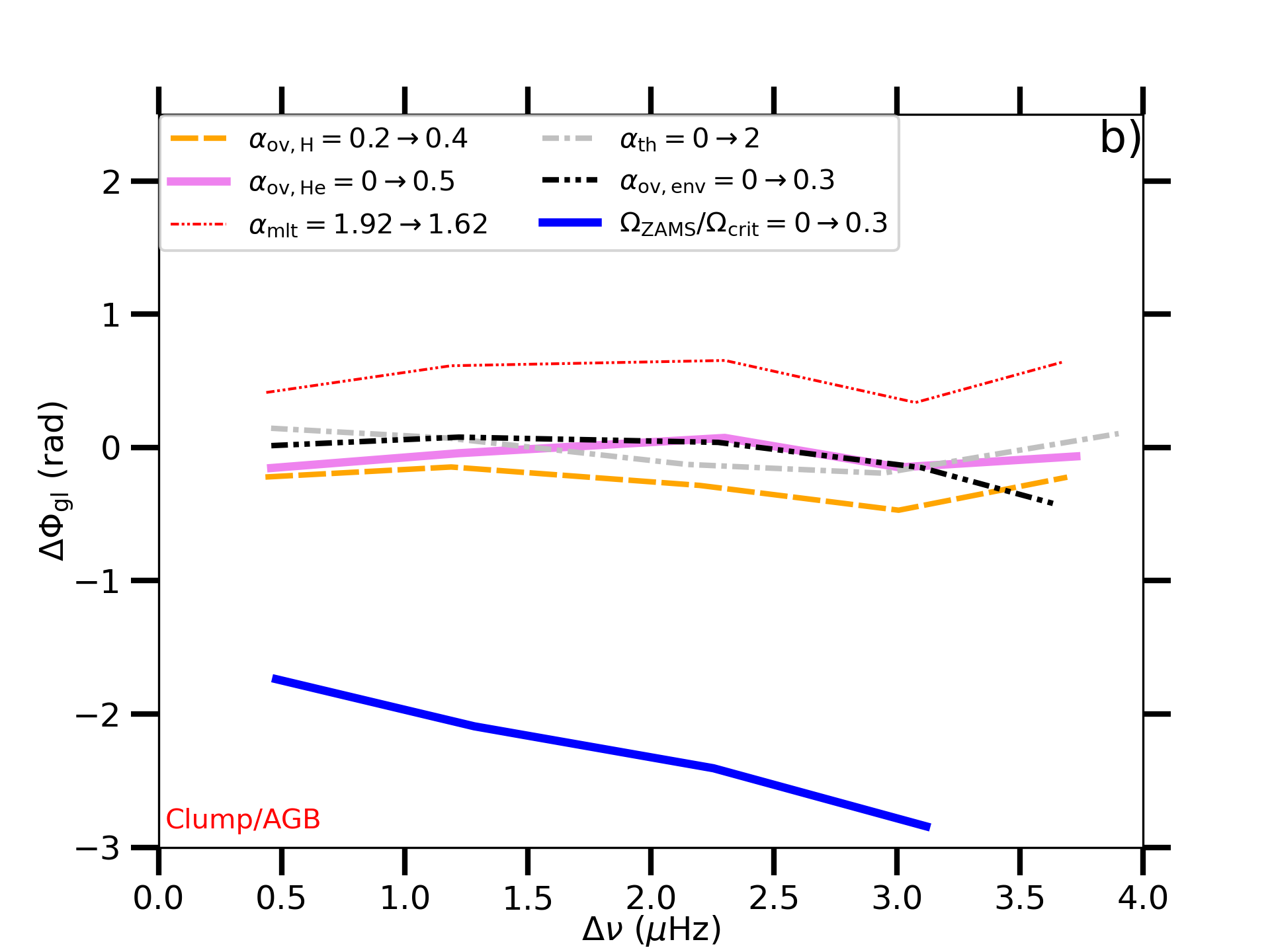}}
		}
	\end{minipage}
	\caption{Dependence of the modulation phase $\Phigl$ on input physics as a function of $\Dnu$. The relative difference $\Delta\Phigl = \Phi\ind{gl,mod} - \Phi\ind{gl,ref}$, where $\Phi\ind{gl,ref}$ and $\Phi\ind{gl,mod}$ are the modulation phases found in the reference and modified models, is shown for $0.5\,\mu$Hz wide $\Dnu$ bins between $0.1\, \mu$Hz and $4.0\, \mu$Hz (except for the first $\Dnu$ bin that is $[0.1, 0.5]\, \mu$Hz). The reference model is defined with $1.75\, M_{\odot}$, solar metallicity and the input physics are those indicated in Table.~\ref{Table:reference_model}. The settings of the modified models are similar to those of the reference models, except one parameter that is changed as indicated on the labels. \emph{a)} and \emph{b)} are obtained on the RGB and clump stage/AGB, respectively.
	}
	\label{fig:relative_diff_Phigl}
\end{figure*}

However, the phase difference is still noticeable in \Kepler\ observations for $M \geq 1.5\, M_{\odot}$. We need to understand what other input physics would allow such difference at high mass. For this, we explored the influence of rotation on the modulation phase $\Phigl$ of the glitch signature, for the $1.75\,M_{\odot}$ model only since additional refined input such as magnetic braking must be taken into account when $M \leq 1.2\,M_{\odot}$. We implemented rotation and rotational-induced mixing as 1D diffusive processes in the shellular approximation \citep[e.g.][]{1997A&A...321..465M}, exactly as presented in \citet{2022A&A...668A.115D}. Rotation is taken into account from the zero-age main sequence (ZAMS) up to the early-AGB, with a rotation rate $\Omega\ind{ZAMS}/\Omega\ind{crit} = 0.3$ at the ZAMS, where $\Omega\ind{crit}$ is the surface angular break-up velocity. Such high rotation rate is typical in B stars \citep{2010ApJ...722..605H} and is large enough to change the envelope composition. In Fig.~\ref{fig:relative_diff_Phigl}, we see that rotation significantly modifies $\Phigl$, with an average shift of $-2\,$rad on the RGB (blue thick curve). Remarkably, the phase shift is larger during the clump/AGB evolution, of about $-0.5\,$rad relatively to that on the RGB when $\Dnu \geq 2.3\,\mu$Hz. This could explain the phase difference observed between RGB and clump/AGB stars at high mass, which is found to be $-1.0\,$rad according to observations \citep{2021A&A...650A.115D}. \\

Given the strong effects of rotation on the glitch signature, we investigated other input physics that would generate efficient mixing in stellar interiors, both for the $1.0\,M_{\odot}$ (Fig.~\ref{fig:relative_diff_params_glitch}) and $1.75\,M_{\odot}$ (Fig.~\ref{fig:relative_diff_Phigl}) models. Envelope overshooting, which affects the composition of the envelope from the main sequence, has weak effects on $\Phigl$, as shown in Figs~\ref{fig:relative_diff_Phigl}, \ref{fig:relative_diff_params_glitch} (black thick dash dot dotted line, the changes induced by considering envelope overshooting in the models barely exceed $0.2\,$rad). So, no negative phase difference between He- and H-shell burning stars could be obtained. Similar conclusions can be drawn when considering extra mixing in the core. Core overshooting during the main sequence (orange dashed line) induces a similar phase shift, $\Delta\Phigl = \Phi\ind{gl,\alpha\ind{ov,H} = 0.4} - \Phi\ind{gl,\alpha\ind{ov,H} = 0.2}$, in both the H-shell and He-burning phases. This shift is more significant than that observed for a $1\,M_{\odot}$ model, as expected (Fig.~\ref{fig:relative_diff_params_glitch}). By contrast, overshooting during the core-He burning phase (pink thick line) only weakly affects $\Phigl$ (Fig.~\ref{fig:relative_diff_Phigl}). Other mixing above the H-burning shell can be caused by thermohaline mixing \citep{2010A&A...521A...9C, 2010A&A...522A..10C, 2011A&A...536A..28L, 2012A&A...543A.108L}. We do not notice any significant impact on $\Phigl$ when adding thermohaline mixing to the models. For $M \geq 1.5\, M_{\odot}$, the thermohaline mixing layer located above the H-burning shell is not able to join the H-burning shell with the convective envelope during the He-core burning phase. In such case, the transfer of H-burning products into the stellar envelope by thermohaline mixing is inefficient. 

Finally, changing the mixing-length parameter $\alphaMLT$ affects $\Phigl$ equally in the shell-H and the He-burning phases, the $\Delta\Phigl = \Phi\ind{gl,\alphaMLT = 1.62} - \Phi\ind{gl,\alphaMLT = 1.92}$ in the shell-H burning phase is similar to that on the He-burning phase (see red thin dash dot dotted line in Fig.~\ref{fig:relative_diff_Phigl}, \ref{fig:relative_diff_params_glitch}\emph{e}+\emph{f}). Further investigation is needed to quantify the importance of these mixing mechanisms in causing the phase shift of the glitch signature between H-shell and He-burning stars at high mass.\\

\subsection{Validity of the asymptotic approach at low $\Dnu$}

In stellar models, we were able to infer the frequencies of the pressure radial, dipole, and quadrupole modes in an efficient way down to $\Dnu \sim 0.09\, \mu$Hz (equivalently $\numax \sim 0.5\, \mu$Hz and $R = 120\, R_{\odot}$) with \adipls\ (see Fig.~\ref{fig:model_frequencies_ADIPLS}). This offers the opportunity to test the relevance of the asymptotic expansion at low $\Dnu$.
In this approach, deviations from the asymptotic pattern caused by any sharp variation feature must be small compared to the asymptotic leading-order term, so that these deviations can be treated as perturbations to the asymptotic relation. Here, we treat the ionisation-induced dip of $\Gamma_{1}$ as a structural perturbation to a reference model in absence of the effects of helium ionisation on the stellar structure \citep{2002ESASP.485...65G, 2021A&A...655A..85H}. As a consequence, the glitch signature in $\Dnu$ (Eq.~\ref{eq:fit_glitch_signature_Dreau_2021}) is assumed to be a perturbation to the asymptotic expansion \citep{2007MNRAS.375..861H, 2022A&A...663A..60H}. Accordingly, we expect the amplitude of the glitch signature to be small with respect to $\Dnu$. In our analysis, we identified the asymptotic expansion to be quantitatively inaccurate when $\Dnu \leq 0.5\, \mu$Hz. The reason for this breakdown is twofold. Firstly, the amplitude $\Amplgl$ of the glitch modulation becomes larger than $0.1$ when $\Dnu \leq 0.5\, \mu$Hz (Fig.~\ref{fig:glitch_structure_parameters}\emph{a}), and eventually reaches $0.5$ at the RGB tip, corresponding to a modulation amplitude equal to half the value of $\Dnu$\footnote{We remind that $\Amplgl$ is a dimensionless parameter expressed in fraction of $\Dnu$, as depicted by Eq.~\ref{eq:fit_glitch_signature_Dreau_2021}.}. At the same time, the amplitude of the dip, $\HHeII$, in the $\Gamma_{1}$ profile is larger and its extent, $\DHeII$, is narrower, resulting in a sharper variation of $\Gamma_{1}$ (see Fig.~\ref{fig:glitch_structure_parameters}\emph{c}+\emph{d}). Secondly, the glitch signature has a significant effect on the local measurement of $\varepsilon$. This signature noticeably affects the profile of $\varepsilon$ around $\Dnu \sim 0.7\,\mu$Hz (see Fig.~\ref{fig:seismic_parameters_asymptotic_pattern}\emph{a}). 
In this case, the perturbation approach cannot be adapted at low $\Dnu$. This conclusion is supported by observational results. The amplitude of the glitch modulation $\delta_{n,\ell}^{\mathrm{gl}}$ introduced in the local large separation $\Dnu_{n,\ell}$ at $\Dnu = 0.8\, \mu$Hz in \Kepler\ observations is $\sim 0.07\, \Dnu$ on the RGB and $\sim 0.08\, \Dnu$ on the AGB \citep{2021A&A...650A.115D}. \\

In addition to the large glitch amplitude at low $\Dnu$, the asymptotic expansion of the mode frequencies cannot be valid below $\Dnu \leq 0.5\, \mu$Hz (equivalently $\numax \leq 2\, \mu$Hz) because the assumption $n \gg \ell$ is not fulfilled. In \citet{2021A&A...650A.115D}, we encountered difficulties to match the observed and the template oscillation spectra based on the asymptotic relation given by Eq.~\ref{eq:nunl_asympt} when $\Dnu \leq 0.5\, \mu$Hz. At the optimal value of $\Dnu$ that maximises the cross correlation between the observed and template oscillation spectra, the observed and modeled modes do not overlap for all radial orders $n$. This reflects that the asymptotic expansion is not suitable to accurately reproduce the oscillation spectrum of red giants when $\Dnu \leq 0.5\,\mu$Hz\footnote{We note that the frequency resolution of \Kepler\ observations is insufficient for a comprehensive seismic study of red giants with $\Dnu \leq 0.5\, \mu$Hz, still the frequency of the modes can be correctly inferred \citep{2013A&A...559A.137M, 2020MNRAS.493.1388Y}.}. 

Eventually, both the asymptotic expansion and the perturbation approach of the signature of HeII are not accurate when $\Dnu \leq 0.5\, \mu$Hz. The efficiency of the classification method based on the signature of the HeII zone may not only be affected by the insufficient frequency resolution at low $\Dnu$, but also by the inadequate scheme to derive p-mode frequencies. Accordingly, a better suited framework to interpret the oscillation spectrum of high-luminosity red giants with $\Dnu \leq 0.5\, \mu$Hz is necessary. 

\section{Conclusion}
\label{sec:conclusion}

The unprecedented high-precision photometric data collected by \Kepler\ give access to seismic parameters, which are of major importance to constrain physical mechanisms in stellar interiors from early to late evolutionary stages. In this work, we have physically interpreted the observed seismic parameters of high-luminosity red giants on the RGB and AGB, in terms of structure parameters by means of stellar models. We computed a grid of stellar models with the stellar evolution code \mesa\ including different input physics such as overshooting, thermohaline mixing and mass loss. Then, we computed the non-radial mode frequencies in red-giant evolved models ($\Dnu \leq 4\, \mu$Hz) with the stellar oscillation code \adipls\ by adopting the squared Brunt-V\"ais\"al\"a frequency $\NBV^{2}$ equal to zero in the core \citep{2018MNRAS.478.4697B}. This method suppresses the g modes in the core, removing the need to compute mixed modes in red-giant models, which would otherwise require tens of thousands of meshpoints to resolve their eigenmodes. However, unlike the original approach, it retains an inconsistent $\Gamma_{1}$ profile. We verified that this method, inspired by \citet{2018MNRAS.478.4697B} and originally applied to non-radial mode frequencies in RGB stars with $\Dnu \leq 4.5\,\mu$Hz, is also valid for non-radial modes in clump and AGB stars. In these cases, it introduces a possible bias of $0.02\Dnu$ and deviations of $0.03\Dnu$–$0.04\Dnu$ for dipole modes, and a deviation of $0.02\Dnu$ for quadrupole modes. By combining these non-radial modes with the radial modes obtained without modifying the Brunt-V\"ais\"al\"a frequency down to $\numax \sim 0.1\, \mu$Hz (equivalently, $\Dnu \sim 0.06\, \mu$Hz), we computed the seismic parameters of the asymptotic pattern and the glitch signature. Thus, we could explore their dependence on the input physics and structure parameters returned by \mesa. The most impactful parameters are those modifying the physical conditions and composition of the envelope. The stellar mass $M$ and metallicity [Fe/H] are the predominant parameters that influence the seismic parameters. The mixing-length parameter $\alphaMLT$, the mass-loss rate on the RGB $\eta\ind{R}$, and the initial helium abundance $Y_{0}$ have moderate impact on the asymptotic and glitch parameters, while the presence of extra-mixing regions such as overshooting and thermohaline mixing have minor influence.\\
The strength of the $\Gamma_{1}$ variation in the HeII ionisation region is correlated with the average temperature and density in this zone. Among these two parameters, the density difference is predominant when comparing RGB and AGB stars. This leads to a distinct modification of the helium ionisation zone, which in turn explains why the amplitude of the glitch signature is stronger in AGB stars than in RGB stars. This result extends previous findings established between RGB and clump stars \citep{2014MNRAS.445.3685C}. \\
We verified that the asymptotic approach is not valid when $\Dnu \leq 0.5\, \mu$Hz. Below this limit, the amplitude of the glitch signature is larger than $0.1$ in units of $\Dnu$, showing that the glitch signature cannot be treated as a perturbation to the asymptotic relation. Moreover, the asymptotic relation does not suitably reproduce the observed pattern of modes when $\Dnu \leq 0.5\, \mu$Hz since the non-radial modes get closer to the neighbouring radial mode, forming a tightly packed triplet of modes. Accordingly, a suitable expression that includes the substantial glitch signature is required to describe the p-mode pattern towards low $\Dnu$. \\
The HeII zone is located closer to the surface in high-mass stars and progressively shifts inward as $\Dnu$ decreases. In fact, these stars have larger effective temperature at fixed $\Dnu$, meaning that the temperature threshold for helium ionisation is reached closer to the surface. At low mass ($M \leq 1.5\, M_{\odot}$), we recovered the phase difference between RGB and clump/AGB stars in the glitch signature, which allows us to identify the evolutionary status of red giants \citep{2015A&A...579A..84V, 2021A&A...650A.115D}. We identified the mass loss as the main cause of this phase difference, as the seismic parameters are mostly sensitive to the stellar mass. At high mass ($M \geq 1.5 \, M_{\odot}$), mass loss has minimal effect, but we found evidence that this phase difference can be retrieved by taking rotational-induced mixing into account. Henceforth, a combination of these input physics could allow us to reproduce this phase difference, which is found to be $\sim -1.0\,$rad between clump/AGB and RGB stars. \\

An accurate expression for the p-mode frequencies when $\Dnu \leq 0.5\, \mu$Hz would not only make the mode identification easier in observations, but it would also improve the efficiency of the classification methods. Indeed, the identification of evolutionary stages for red giants, which is based on the glitch signature \citep{2012A&A...541A..51K, 2015A&A...579A..84V}, assumes that the asymptotic relation is valid. On top of being affected by the frequency resolution, the uncertainties on the evolutionary stages are subject to the deviations from the asymptotic regime, which would explain the increasing disagreements between different identification methods towards low $\Dnu$. Eventually, our investigation into the physical mechanisms producing the phase difference between RGB and clump/AGB stars in the glitch signature when $M \geq 1.5\, M_{\odot}$ could be expanded. Our exploratory tests suggest that this phase difference may be recovered by considering any additional input physics that have the potential to change the physical conditions and composition of the envelope, where the HeII zone lies.

\begin{acknowledgements}

The authors warmly thank the anonymous referee for their valuable comments, which helped improve the depth of the discussion and the overall clarity of the manuscript. The authors also express their sincere thanks to the \Kepler\ team for their tireless efforts in making high-precision data accessible, which have been instrumental in making these results possible. G.D. gratefully acknowledges support from the Australian Research Council through Discovery Project DP190100666. D.S. is supported by the Australian Research Council (DP190100666).

\end{acknowledgements}

\bibliographystyle{Outils_latex/aa}
\bibliography{Outils_latex/aa54004-25}

\begin{thebibliography}{96}
\expandafter\ifx\csname natexlab\endcsname\relax\def\natexlab#1{#1}\fi

\bibitem[{{Baglin} {et~al.}(2006){Baglin}, {Auvergne}, {Barge}, {Deleuil},
  {Catala}, {Michel}, {Weiss}, \& {COROT Team}}]{2006ESASP1306...33B}
{Baglin}, A., {Auvergne}, M., {Barge}, P., {et~al.} 2006, in ESA Special
  Publication, Vol. 1306, The CoRoT Mission Pre-Launch Status - Stellar
  Seismology and Planet Finding, ed. M.~{Fridlund}, A.~{Baglin}, J.~{Lochard},
  \& L.~{Conroy}, 33

\bibitem[{{Ball} {et~al.}(2018){Ball}, {Theme{\ss}l}, \&
  {Hekker}}]{2018MNRAS.478.4697B}
{Ball}, W.~H., {Theme{\ss}l}, N., \& {Hekker}, S. 2018, \mnras, 478, 4697

\bibitem[{{Baudin} {et~al.}(2012){Baudin}, {Barban}, {Goupil}, {Samadi},
  {Lebreton}, {Bruntt}, {Morel}, {Lef{\`e}vre}, {Michel}, {Mosser}, {Carrier},
  {De Ridder}, {Hatzes}, {Hekker}, {Kallinger}, {Auvergne}, {Baglin}, \&
  {Catala}}]{2012A&A...538A..73B}
{Baudin}, F., {Barban}, C., {Goupil}, M.~J., {et~al.} 2012, \aap, 538, A73

\bibitem[{{Bedding} {et~al.}(2010){Bedding}, {Huber}, {Stello}, {Elsworth},
  {Hekker}, {Kallinger}, {Mathur}, {Mosser}, {Preston}, {Ballot}, {Barban},
  {Broomhall}, {Buzasi}, {Chaplin}, {Garc{\'{\i}}a}, {Gruberbauer}, {Hale}, {De
  Ridder}, {Frandsen}, {Borucki}, {Brown}, {Christensen-Dalsgaard},
  {Gilliland}, {Jenkins}, {Kjeldsen}, {Koch}, {Belkacem}, {Bildsten}, {Bruntt},
  {Campante}, {Deheuvels}, {Derekas}, {Dupret}, {Goupil}, {Hatzes}, {Houdek},
  {Ireland}, {Jiang}, {Karoff}, {Kiss}, {Lebreton}, {Miglio}, {Montalb{\'a}n},
  {Noels}, {Roxburgh}, {Sangaralingam}, {Stevens}, {Suran}, {Tarrant}, \&
  {Weiss}}]{2010ApJ...713L.176B}
{Bedding}, T.~R., {Huber}, D., {Stello}, D., {et~al.} 2010, \apjl, 713, L176

\bibitem[{{Bedding} {et~al.}(2011){Bedding}, {Mosser}, {Huber},
  {Montalb{\'a}n}, {Beck}, {Christensen-Dalsgaard}, {Elsworth},
  {Garc{\'{\i}}a}, {Miglio}, {Stello}, {White}, {De Ridder}, {Hekker}, {Aerts},
  {Barban}, {Belkacem}, {Broomhall}, {Brown}, {Buzasi}, {Carrier}, {Chaplin},
  {di Mauro}, {Dupret}, {Frandsen}, {Gilliland}, {Goupil}, {Jenkins},
  {Kallinger}, {Kawaler}, {Kjeldsen}, {Mathur}, {Noels}, {Aguirre}, \&
  {Ventura}}]{2011Natur.471..608B}
{Bedding}, T.~R., {Mosser}, B., {Huber}, D., {et~al.} 2011, \nat, 471, 608

\bibitem[{{Blocker}(1995)}]{1995A&A...297..727B}
{Blocker}, T. 1995, \aap, 297, 727

\bibitem[{{Borucki} {et~al.}(2010){Borucki}, {Koch}, {Basri}, {Batalha},
  {Brown}, {Caldwell}, {Caldwell}, {Christensen-Dalsgaard}, {Cochran},
  {DeVore}, {Dunham}, {Dupree}, {Gautier}, {Geary}, {Gilliland}, {Gould},
  {Howell}, {Jenkins}, {Kondo}, {Latham}, {Marcy}, {Meibom}, {Kjeldsen},
  {Lissauer}, {Monet}, {Morrison}, {Sasselov}, {Tarter}, {Boss}, {Brownlee},
  {Owen}, {Buzasi}, {Charbonneau}, {Doyle}, {Fortney}, {Ford}, {Holman},
  {Seager}, {Steffen}, {Welsh}, {Rowe}, {Anderson}, {Buchhave}, {Ciardi},
  {Walkowicz}, {Sherry}, {Horch}, {Isaacson}, {Everett}, {Fischer}, {Torres},
  {Johnson}, {Endl}, {MacQueen}, {Bryson}, {Dotson}, {Haas}, {Kolodziejczak},
  {Van Cleve}, {Chandrasekaran}, {Twicken}, {Quintana}, {Clarke}, {Allen},
  {Li}, {Wu}, {Tenenbaum}, {Verner}, {Bruhweiler}, {Barnes}, \&
  {Prsa}}]{2010Sci...327..977B}
{Borucki}, W.~J., {Koch}, D., {Basri}, G., {et~al.} 2010, Science, 327, 977

\bibitem[{{Bossini} {et~al.}(2015){Bossini}, {Miglio}, {Salaris},
  {Pietrinferni}, {Montalb{\'a}n}, {Bressan}, {Noels}, {Cassisi}, {Girardi}, \&
  {Marigo}}]{2015MNRAS.453.2290B}
{Bossini}, D., {Miglio}, A., {Salaris}, M., {et~al.} 2015, \mnras, 453, 2290

\bibitem[{{Bossini} {et~al.}(2017){Bossini}, {Miglio}, {Salaris}, {Vrard},
  {Cassisi}, {Mosser}, {Montalb{\'a}n}, {Girardi}, {Noels}, {Bressan},
  {Pietrinferni}, \& {Tayar}}]{2017MNRAS.469.4718B}
{Bossini}, D., {Miglio}, A., {Salaris}, M., {et~al.} 2017, \mnras, 469, 4718

\bibitem[{{Broomhall} {et~al.}(2014){Broomhall}, {Miglio}, {Montalb{\'a}n},
  {Eggenberger}, {Chaplin}, {Elsworth}, {Scuflaire}, {Ventura}, \&
  {Verner}}]{2014MNRAS.440.1828B}
{Broomhall}, A.-M., {Miglio}, A., {Montalb{\'a}n}, J., {et~al.} 2014, \mnras,
  440, 1828

\bibitem[{{Cantiello} \& {Langer}(2010)}]{2010A&A...521A...9C}
{Cantiello}, M. \& {Langer}, N. 2010, \aap, 521, A9

\bibitem[{{Charbonnel} \& {Lagarde}(2010)}]{2010A&A...522A..10C}
{Charbonnel}, C. \& {Lagarde}, N. 2010, \aap, 522, A10

\bibitem[{{Charbonnel} \& {Zahn}(2007)}]{2007A&A...467L..15C}
{Charbonnel}, C. \& {Zahn}, J.~P. 2007, \aap, 467, L15

\bibitem[{{Choi} {et~al.}(2016){Choi}, {Dotter}, {Conroy}, {Cantiello},
  {Paxton}, \& {Johnson}}]{2016ApJ...823..102C}
{Choi}, J., {Dotter}, A., {Conroy}, C., {et~al.} 2016, \apj, 823, 102

\bibitem[{{Christensen-Dalsgaard}(2008)}]{2008Ap&SS.316..113C}
{Christensen-Dalsgaard}, J. 2008, \apss, 316, 113

\bibitem[{{Christensen-Dalsgaard} {et~al.}(2001){Christensen-Dalsgaard},
  {Kjeldsen}, \& {Mattei}}]{2001ApJ...562L.141C}
{Christensen-Dalsgaard}, J., {Kjeldsen}, H., \& {Mattei}, J.~A. 2001, \apjl,
  562, L141

\bibitem[{{Christensen-Dalsgaard} {et~al.}(2014){Christensen-Dalsgaard}, {Silva
  Aguirre}, {Elsworth}, \& {Hekker}}]{2014MNRAS.445.3685C}
{Christensen-Dalsgaard}, J., {Silva Aguirre}, V., {Elsworth}, Y., \& {Hekker},
  S. 2014, \mnras, 445, 3685

\bibitem[{{Corsaro} {et~al.}(2015){Corsaro}, {De Ridder}, \&
  {Garc{\'\i}a}}]{2015A&A...578A..76C}
{Corsaro}, E., {De Ridder}, J., \& {Garc{\'\i}a}, R.~A. 2015, \aap, 578, A76

\bibitem[{{Cowling}(1941)}]{1941MNRAS.101..367C}
{Cowling}, T.~G. 1941, \mnras, 101, 367

\bibitem[{{Cunha} \& {Metcalfe}(2007)}]{2007ApJ...666..413C}
{Cunha}, M.~S. \& {Metcalfe}, T.~S. 2007, \apj, 666, 413

\bibitem[{{Deheuvels} {et~al.}(2016){Deheuvels}, {Brand{\~a}o}, {Silva
  Aguirre}, {Ballot}, {Michel}, {Cunha}, {Lebreton}, \&
  {Appourchaux}}]{2016A&A...589A..93D}
{Deheuvels}, S., {Brand{\~a}o}, I., {Silva Aguirre}, V., {et~al.} 2016, \aap,
  589, A93

\bibitem[{{Dr{\'e}au} {et~al.}(2022){Dr{\'e}au}, {Lebreton}, {Mosser},
  {Bossini}, \& {Yu}}]{2022A&A...668A.115D}
{Dr{\'e}au}, G., {Lebreton}, Y., {Mosser}, B., {Bossini}, D., \& {Yu}, J. 2022,
  \aap, 668, A115

\bibitem[{{Dr{\'e}au} {et~al.}(2021){Dr{\'e}au}, {Mosser}, {Lebreton}, {Gehan},
  \& {Kallinger}}]{2021A&A...650A.115D}
{Dr{\'e}au}, G., {Mosser}, B., {Lebreton}, Y., {Gehan}, C., \& {Kallinger}, T.
  2021, \aap, 650, A115

\bibitem[{{Dziembowski} {et~al.}(2001){Dziembowski}, {Gough}, {Houdek}, \&
  {Sienkiewicz}}]{2001MNRAS.328..601D}
{Dziembowski}, W.~A., {Gough}, D.~O., {Houdek}, G., \& {Sienkiewicz}, R. 2001,
  \mnras, 328, 601

\bibitem[{{Eggleton} {et~al.}(2006){Eggleton}, {Dearborn}, \&
  {Lattanzio}}]{2006Sci...314.1580E}
{Eggleton}, P.~P., {Dearborn}, D. S.~P., \& {Lattanzio}, J.~C. 2006, Science,
  314, 1580

\bibitem[{{Eggleton} {et~al.}(2008){Eggleton}, {Dearborn}, \&
  {Lattanzio}}]{2008ApJ...677..581E}
{Eggleton}, P.~P., {Dearborn}, D. S.~P., \& {Lattanzio}, J.~C. 2008, \apj, 677,
  581

\bibitem[{{Farnir} {et~al.}(2019){Farnir}, {Dupret}, {Salmon}, {Noels}, \&
  {Buldgen}}]{2019A&A...622A..98F}
{Farnir}, M., {Dupret}, M.~A., {Salmon}, S.~J.~A.~J., {Noels}, A., \&
  {Buldgen}, G. 2019, \aap, 622, A98

\bibitem[{{Gilliland} {et~al.}(2010){Gilliland}, {Brown},
  {Christensen-Dalsgaard}, {Kjeldsen}, {Aerts}, {Appourchaux}, {Basu},
  {Bedding}, {Chaplin}, {Cunha}, {De Cat}, {De Ridder}, {Guzik}, {Handler},
  {Kawaler}, {Kiss}, {Kolenberg}, {Kurtz}, {Metcalfe}, {Monteiro}, {Szab{\'o}},
  {Arentoft}, {Balona}, {Debosscher}, {Elsworth}, {Quirion}, {Stello},
  {Su{\'a}rez}, {Borucki}, {Jenkins}, {Koch}, {Kondo}, {Latham}, {Rowe}, \&
  {Steffen}}]{2010PASP..122..131G}
{Gilliland}, R.~L., {Brown}, T.~M., {Christensen-Dalsgaard}, J., {et~al.} 2010,
  \pasp, 122, 131

\bibitem[{{Gough}(1986)}]{1986HiA.....7..283G}
{Gough}, D.~O. 1986, Highlights of Astronomy, 7, 283

\bibitem[{{Gough}(1990)}]{1990LNP...367..283G}
{Gough}, D.~O. 1990, {Comments on Helioseismic Inference}, ed. Y.~{Osaki} \&
  H.~{Shibahashi}, Vol. 367, 283

\bibitem[{{Gough}(2002)}]{2002ESASP.485...65G}
{Gough}, D.~O. 2002, in ESA Special Publication, Vol. 485, Stellar Structure
  and Habitable Planet Finding, ed. B.~{Battrick}, F.~{Favata}, I.~W.
  {Roxburgh}, \& D.~{Galadi}, 65--73

\bibitem[{{Henyey} {et~al.}(1965){Henyey}, {Vardya}, \&
  {Bodenheimer}}]{1965ApJ...142..841H}
{Henyey}, L., {Vardya}, M.~S., \& {Bodenheimer}, P. 1965, \apj, 142, 841

\bibitem[{{Hon} {et~al.}(2017){Hon}, {Stello}, \& {Yu}}]{2017MNRAS.469.4578H}
{Hon}, M., {Stello}, D., \& {Yu}, J. 2017, \mnras, 469, 4578

\bibitem[{{Hon} {et~al.}(2018){Hon}, {Stello}, \& {Yu}}]{2018MNRAS.476.3233H}
{Hon}, M., {Stello}, D., \& {Yu}, J. 2018, \mnras, 476, 3233

\bibitem[{{Houdayer} {et~al.}(2022){Houdayer}, {Reese}, \&
  {Goupil}}]{2022A&A...663A..60H}
{Houdayer}, P.~S., {Reese}, D.~R., \& {Goupil}, M.-J. 2022, \aap, 663, A60

\bibitem[{{Houdayer} {et~al.}(2021){Houdayer}, {Reese}, {Goupil}, \&
  {Lebreton}}]{2021A&A...655A..85H}
{Houdayer}, P.~S., {Reese}, D.~R., {Goupil}, M.-J., \& {Lebreton}, Y. 2021,
  \aap, 655, A85

\bibitem[{{Houdek} \& {Gough}(2007)}]{2007MNRAS.375..861H}
{Houdek}, G. \& {Gough}, D.~O. 2007, \mnras, 375, 861

\bibitem[{{Howell} {et~al.}(2014){Howell}, {Sobeck}, {Haas}, {Still},
  {Barclay}, {Mullally}, {Troeltzsch}, {Aigrain}, {Bryson}, {Caldwell},
  {Chaplin}, {Cochran}, {Huber}, {Marcy}, {Miglio}, {Najita}, {Smith},
  {Twicken}, \& {Fortney}}]{2014PASP..126..398H}
{Howell}, S.~B., {Sobeck}, C., {Haas}, M., {et~al.} 2014, \pasp, 126, 398

\bibitem[{{Huang} {et~al.}(2010){Huang}, {Gies}, \&
  {McSwain}}]{2010ApJ...722..605H}
{Huang}, W., {Gies}, D.~R., \& {McSwain}, M.~V. 2010, \apj, 722, 605

\bibitem[{{Huber} {et~al.}(2011){Huber}, {Bedding}, {Stello}, {Hekker},
  {Mathur}, {Mosser}, {Verner}, {Bonanno}, {Buzasi}, {Campante}, {Elsworth},
  {Hale}, {Kallinger}, {Silva Aguirre}, {Chaplin}, {De Ridder},
  {Garc{\'{\i}}a}, {Appourchaux}, {Frandsen}, {Houdek}, {Molenda-{\.Z}akowicz},
  {Monteiro}, {Christensen-Dalsgaard}, {Gilliland}, {Kawaler}, {Kjeldsen},
  {Broomhall}, {Corsaro}, {Salabert}, {Sanderfer}, {Seader}, \&
  {Smith}}]{2011ApJ...743..143H}
{Huber}, D., {Bedding}, T.~R., {Stello}, D., {et~al.} 2011, \apj, 743, 143

\bibitem[{{Huber} {et~al.}(2010){Huber}, {Bedding}, {Stello}, {Mosser},
  {Mathur}, {Kallinger}, {Hekker}, {Elsworth}, {Buzasi}, {De Ridder},
  {Gilliland}, {Kjeldsen}, {Chaplin}, {Garc{\'{\i}}a}, {Hale}, {Preston},
  {White}, {Borucki}, {Christensen-Dalsgaard}, {Clarke}, {Jenkins}, \&
  {Koch}}]{2010ApJ...723.1607H}
{Huber}, D., {Bedding}, T.~R., {Stello}, D., {et~al.} 2010, \apj, 723, 1607

\bibitem[{{Iglesias} \& {Rogers}(1996)}]{1996ApJ...464..943I}
{Iglesias}, C.~A. \& {Rogers}, F.~J. 1996, \apj, 464, 943

\bibitem[{{Kallinger} {et~al.}(2018){Kallinger}, {Beck}, {Stello}, \&
  {Garcia}}]{2018A&A...616A.104K}
{Kallinger}, T., {Beck}, P.~G., {Stello}, D., \& {Garcia}, R.~A. 2018, \aap,
  616, A104

\bibitem[{{Kallinger} {et~al.}(2012){Kallinger}, {Hekker}, {Mosser}, {De
  Ridder}, {Bedding}, {Elsworth}, {Gruberbauer}, {Guenther}, {Stello}, {Basu},
  {Garc{\'{\i}}a}, {Chaplin}, {Mullally}, {Still}, \&
  {Thompson}}]{2012A&A...541A..51K}
{Kallinger}, T., {Hekker}, S., {Mosser}, B., {et~al.} 2012, \aap, 541, A51

\bibitem[{{Khan} {et~al.}(2018){Khan}, {Hall}, {Miglio}, {Davies}, {Mosser},
  {Girardi}, \& {Montalb{\'a}n}}]{2018ApJ...859..156K}
{Khan}, S., {Hall}, O.~J., {Miglio}, A., {et~al.} 2018, \apj, 859, 156

\bibitem[{{Khan} {et~al.}(2019){Khan}, {Miglio}, {Mosser}, {Arenou},
  {Belkacem}, {Brown}, {Katz}, {Casagrande}, {Chaplin}, {Davies}, {Rendle},
  {Rodrigues}, {Bossini}, {Cantat-Gaudin}, {Elsworth}, {Girardi}, {North}, \&
  {Vallenari}}]{2019A&A...628A..35K}
{Khan}, S., {Miglio}, A., {Mosser}, B., {et~al.} 2019, \aap, 628, A35

\bibitem[{{Kippenhahn} {et~al.}(1980){Kippenhahn}, {Ruschenplatt}, \&
  {Thomas}}]{1980A&A....91..175K}
{Kippenhahn}, R., {Ruschenplatt}, G., \& {Thomas}, H.~C. 1980, \aap, 91, 175

\bibitem[{{Kjeldsen} \& {Bedding}(1995)}]{1995A&A...293...87K}
{Kjeldsen}, H. \& {Bedding}, T.~R. 1995, \aap, 293, 87

\bibitem[{{Lagarde} {et~al.}(2011){Lagarde}, {Charbonnel}, {Decressin}, \&
  {Hagelberg}}]{2011A&A...536A..28L}
{Lagarde}, N., {Charbonnel}, C., {Decressin}, T., \& {Hagelberg}, J. 2011,
  \aap, 536, A28

\bibitem[{{Lagarde} {et~al.}(2012){Lagarde}, {Decressin}, {Charbonnel},
  {Eggenberger}, {Ekstr{\"o}m}, \& {Palacios}}]{2012A&A...543A.108L}
{Lagarde}, N., {Decressin}, T., {Charbonnel}, C., {et~al.} 2012, \aap, 543,
  A108

\bibitem[{{Langer} {et~al.}(1985){Langer}, {El Eid}, \&
  {Fricke}}]{1985A&A...145..179L}
{Langer}, N., {El Eid}, M.~F., \& {Fricke}, K.~J. 1985, \aap, 145, 179

\bibitem[{{Lebreton} \& {Goupil}(2012)}]{2012A&A...544L..13L}
{Lebreton}, Y. \& {Goupil}, M.~J. 2012, \aap, 544, L13

\bibitem[{{Lopes} {et~al.}(1997){Lopes}, {Turck-Chieze}, {Michel}, \&
  {Goupil}}]{1997ApJ...480..794L}
{Lopes}, I., {Turck-Chieze}, S., {Michel}, E., \& {Goupil}, M.-J. 1997, \apj,
  480, 794

\bibitem[{{Maeder}(1975)}]{1975A&A....40..303M}
{Maeder}, A. 1975, \aap, 40, 303

\bibitem[{{Marigo} \& {Aringer}(2009)}]{2009A&A...508.1539M}
{Marigo}, P. \& {Aringer}, B. 2009, \aap, 508, 1539

\bibitem[{{Mazumdar} {et~al.}(2012){Mazumdar}, {Michel}, {Antia}, \&
  {Deheuvels}}]{2012A&A...540A..31M}
{Mazumdar}, A., {Michel}, E., {Antia}, H.~M., \& {Deheuvels}, S. 2012, \aap,
  540, A31

\bibitem[{{Mazumdar} {et~al.}(2014){Mazumdar}, {Monteiro}, {Ballot}, {Antia},
  {Basu}, {Houdek}, {Mathur}, {Cunha}, {Silva Aguirre}, {Garcia}, {Salabert},
  {Verner}, {Christensen-Dalsgaard}, {Metcalfe}, {Sanderfer}, {Seader},
  {Smith}, \& {Chaplin}}]{2014ApJ...782...18M}
{Mazumdar}, A., {Monteiro}, M. J. P. F.~G., {Ballot}, J., {et~al.} 2014, \apj,
  782, 18

\bibitem[{{Meynet} \& {Maeder}(1997)}]{1997A&A...321..465M}
{Meynet}, G. \& {Maeder}, A. 1997, \aap, 321, 465

\bibitem[{{Michaud} {et~al.}(2010){Michaud}, {Richer}, \&
  {Richard}}]{2010A&A...510A.104M}
{Michaud}, G., {Richer}, J., \& {Richard}, O. 2010, \aap, 510, A104

\bibitem[{{Miglio} {et~al.}(2012){Miglio}, {Brogaard}, {Stello}, {Chaplin},
  {D'Antona}, {Montalb{\'a}n}, {Basu}, {Bressan}, {Grundahl}, {Pinsonneault},
  {Serenelli}, {Elsworth}, {Hekker}, {Kallinger}, {Mosser}, {Ventura},
  {Bonanno}, {Noels}, {Silva Aguirre}, {Szabo}, {Li}, {McCauliff}, {Middour},
  \& {Kjeldsen}}]{2012MNRAS.419.2077M}
{Miglio}, A., {Brogaard}, K., {Stello}, D., {et~al.} 2012, \mnras, 419, 2077

\bibitem[{{Miglio} {et~al.}(2021){Miglio}, {Chiappini}, {Mackereth}, {Davies},
  {Brogaard}, {Casagrande}, {Chaplin}, {Girardi}, {Kawata}, {Khan}, {Izzard},
  {Montalb{\'a}n}, {Mosser}, {Vincenzo}, {Bossini}, {Noels}, {Rodrigues},
  {Valentini}, \& {Mandel}}]{2021A&A...645A..85M}
{Miglio}, A., {Chiappini}, C., {Mackereth}, J.~T., {et~al.} 2021, \aap, 645,
  A85

\bibitem[{{Miglio} {et~al.}(2010){Miglio}, {Montalb{\'a}n}, {Carrier}, {De
  Ridder}, {Mosser}, {Eggenberger}, {Scuflaire}, {Ventura}, {D'Antona},
  {Noels}, \& {Baglin}}]{2010A&A...520L...6M}
{Miglio}, A., {Montalb{\'a}n}, J., {Carrier}, F., {et~al.} 2010, \aap, 520, L6

\bibitem[{{Montalb{\'a}n} {et~al.}(2010){Montalb{\'a}n}, {Miglio}, {Noels},
  {Scuflaire}, \& {Ventura}}]{2010ApJ...721L.182M}
{Montalb{\'a}n}, J., {Miglio}, A., {Noels}, A., {Scuflaire}, R., \& {Ventura},
  P. 2010, \apjl, 721, L182

\bibitem[{{Montalb{\'a}n} {et~al.}(2012){Montalb{\'a}n}, {Miglio}, {Noels},
  {Scuflaire}, {Ventura}, \& {D'Antona}}]{2012ASSP...26...23M}
{Montalb{\'a}n}, J., {Miglio}, A., {Noels}, A., {et~al.} 2012, in Red Giants as
  Probes of the Structure and Evolution of the Milky Way, ed. A.~{Miglio},
  J.~{Montalban}, \& A.~{Noels}, Astrophysics and Space Science Proceedings, 23

\bibitem[{{Monteiro} {et~al.}(1994){Monteiro}, {Christensen-Dalsgaard}, \&
  {Thompson}}]{1994A&A...283..247M}
{Monteiro}, M.~J.~P.~F.~G., {Christensen-Dalsgaard}, J., \& {Thompson}, M.~J.
  1994, \aap, 283, 247

\bibitem[{{Monteiro} \& {Thompson}(2005)}]{2005MNRAS.361.1187M}
{Monteiro}, M.~J.~P.~F.~G. \& {Thompson}, M.~J. 2005, \mnras, 361, 1187

\bibitem[{{Mosser} {et~al.}(2011){Mosser}, {Belkacem}, {Goupil}, {Michel},
  {Elsworth}, {Barban}, {Kallinger}, {Hekker}, {De Ridder}, {Samadi}, {Baudin},
  {Pinheiro}, {Auvergne}, {Baglin}, \& {Catala}}]{2011A&A...525L...9M}
{Mosser}, B., {Belkacem}, K., {Goupil}, M., {et~al.} 2011, \aap, 525, L9

\bibitem[{{Mosser} {et~al.}(2013{\natexlab{a}}){Mosser}, {Dziembowski},
  {Belkacem}, {Goupil}, {Michel}, {Samadi}, {Soszy{\'n}ski}, {Vrard},
  {Elsworth}, {Hekker}, \& {Mathur}}]{2013A&A...559A.137M}
{Mosser}, B., {Dziembowski}, W.~A., {Belkacem}, K., {et~al.}
  2013{\natexlab{a}}, \aap, 559, A137

\bibitem[{{Mosser} {et~al.}(2013{\natexlab{b}}){Mosser}, {Michel}, {Belkacem},
  {Goupil}, {Baglin}, {Barban}, {Provost}, {Samadi}, {Auvergne}, \&
  {Catala}}]{2013A&A...550A.126M}
{Mosser}, B., {Michel}, E., {Belkacem}, K., {et~al.} 2013{\natexlab{b}}, \aap,
  550, A126

\bibitem[{{Mosser} {et~al.}(2019){Mosser}, {Michel}, {Samadi}, {Miglio},
  {Davies}, {Girardi}, \& {Goupil}}]{2019A&A...622A..76M}
{Mosser}, B., {Michel}, E., {Samadi}, R., {et~al.} 2019, \aap, 622, A76

\bibitem[{{Paxton} {et~al.}(2011){Paxton}, {Bildsten}, {Dotter}, {Herwig},
  {Lesaffre}, \& {Timmes}}]{2011ApJS..192....3P}
{Paxton}, B., {Bildsten}, L., {Dotter}, A., {et~al.} 2011, \apjs, 192, 3

\bibitem[{{Paxton} {et~al.}(2013){Paxton}, {Cantiello}, {Arras}, {Bildsten},
  {Brown}, {Dotter}, {Mankovich}, {Montgomery}, {Stello}, {Timmes}, \&
  {Townsend}}]{2013ApJS..208....4P}
{Paxton}, B., {Cantiello}, M., {Arras}, P., {et~al.} 2013, \apjs, 208, 4

\bibitem[{{Paxton} {et~al.}(2015){Paxton}, {Marchant}, {Schwab}, {Bauer},
  {Bildsten}, {Cantiello}, {Dessart}, {Farmer}, {Hu}, {Langer}, {Townsend},
  {Townsley}, \& {Timmes}}]{2015ApJS..220...15P}
{Paxton}, B., {Marchant}, P., {Schwab}, J., {et~al.} 2015, \apjs, 220, 15

\bibitem[{{Paxton} {et~al.}(2018){Paxton}, {Schwab}, {Bauer}, {Bildsten},
  {Blinnikov}, {Duffell}, {Farmer}, {Goldberg}, {Marchant}, {Sorokina},
  {Thoul}, {Townsend}, \& {Timmes}}]{2018ApJS..234...34P}
{Paxton}, B., {Schwab}, J., {Bauer}, E.~B., {et~al.} 2018, \apjs, 234, 34

\bibitem[{{Paxton} {et~al.}(2019){Paxton}, {Smolec}, {Schwab}, {Gautschy},
  {Bildsten}, {Cantiello}, {Dotter}, {Farmer}, {Goldberg}, {Jermyn}, {Kanbur},
  {Marchant}, {Thoul}, {Townsend}, {Wolf}, {Zhang}, \&
  {Timmes}}]{2019ApJS..243...10P}
{Paxton}, B., {Smolec}, R., {Schwab}, J., {et~al.} 2019, \apjs, 243, 10

\bibitem[{{Pinsonneault} {et~al.}(2018){Pinsonneault}, {Elsworth}, {Tayar},
  {Serenelli}, {Stello}, {Zinn}, {Mathur}, {Garc{\'\i}a}, {Johnson}, {Hekker},
  {Huber}, {Kallinger}, {M{\'e}sz{\'a}ros}, {Mosser}, {Stassun}, {Girardi},
  {Rodrigues}, {Silva Aguirre}, {An}, {Basu}, {Chaplin}, {Corsaro}, {Cunha},
  {Garc{\'\i}a-Hern{\'a}ndez}, {Holtzman}, {J{\"o}nsson}, {Shetrone}, {Smith},
  {Sobeck}, {Stringfellow}, {Zamora}, {Beers}, {Fern{\'a}ndez-Trincado},
  {Frinchaboy}, {Hearty}, \& {Nitschelm}}]{2018ApJS..239...32P}
{Pinsonneault}, M.~H., {Elsworth}, Y.~P., {Tayar}, J., {et~al.} 2018, \apjs,
  239, 32

\bibitem[{{Reimers}(1975)}]{1975MSRSL...8..369R}
{Reimers}, D. 1975, Memoires of the Societe Royale des Sciences de Liege, 8,
  369

\bibitem[{{Ricker} {et~al.}(2015){Ricker}, {Winn}, {Vanderspek}, {Latham},
  {Bakos}, {Bean}, {Berta-Thompson}, {Brown}, {Buchhave}, {Butler}, {Butler},
  {Chaplin}, {Charbonneau}, {Christensen-Dalsgaard}, {Clampin}, {Deming},
  {Doty}, {De Lee}, {Dressing}, {Dunham}, {Endl}, {Fressin}, {Ge}, {Henning},
  {Holman}, {Howard}, {Ida}, {Jenkins}, {Jernigan}, {Johnson}, {Kaltenegger},
  {Kawai}, {Kjeldsen}, {Laughlin}, {Levine}, {Lin}, {Lissauer}, {MacQueen},
  {Marcy}, {McCullough}, {Morton}, {Narita}, {Paegert}, {Palle}, {Pepe},
  {Pepper}, {Quirrenbach}, {Rinehart}, {Sasselov}, {Sato}, {Seager},
  {Sozzetti}, {Stassun}, {Sullivan}, {Szentgyorgyi}, {Torres}, {Udry}, \&
  {Villasenor}}]{2015JATIS...1a4003R}
{Ricker}, G.~R., {Winn}, J.~N., {Vanderspek}, R., {et~al.} 2015, Journal of
  Astronomical Telescopes, Instruments, and Systems, 1, 014003

\bibitem[{{Roxburgh} \& {Vorontsov}(2003)}]{2003A&A...411..215R}
{Roxburgh}, I.~W. \& {Vorontsov}, S.~V. 2003, \aap, 411, 215

\bibitem[{{Scherrer} {et~al.}(1983){Scherrer}, {Wilcox},
  {Christensen-Dalsgaard}, \& {Gough}}]{1983SoPh...82...75S}
{Scherrer}, P.~H., {Wilcox}, J.~M., {Christensen-Dalsgaard}, J., \& {Gough},
  D.~O. 1983, \solphys, 82, 75

\bibitem[{{Shibahashi} \& {Osaki}(1981)}]{1981PASJ...33..713S}
{Shibahashi}, H. \& {Osaki}, Y. 1981, \pasj, 33, 713

\bibitem[{{Stello} {et~al.}(2014){Stello}, {Compton}, {Bedding},
  {Christensen-Dalsgaard}, {Kiss}, {Kjeldsen}, {Bellamy}, {Garc{\'{\i}}a}, \&
  {Mathur}}]{2014ApJ...788L..10S}
{Stello}, D., {Compton}, D.~L., {Bedding}, T.~R., {et~al.} 2014, \apjl, 788,
  L10

\bibitem[{{Tassoul}(1980)}]{1980ApJS...43..469T}
{Tassoul}, M. 1980, \apjs, 43, 469

\bibitem[{{Trabucchi} {et~al.}(2017){Trabucchi}, {Wood}, {Montalb{\'a}n},
  {Marigo}, {Pastorelli}, \& {Girardi}}]{2017ApJ...847..139T}
{Trabucchi}, M., {Wood}, P.~R., {Montalb{\'a}n}, J., {et~al.} 2017, \apj, 847,
  139

\bibitem[{{Ulrich}(1972)}]{1972ApJ...172..165U}
{Ulrich}, R.~K. 1972, \apj, 172, 165

\bibitem[{{Unno} {et~al.}(1989){Unno}, {Osaki}, {Ando}, {Saio}, \&
  {Shibahashi}}]{1989nos..book.....U}
{Unno}, W., {Osaki}, Y., {Ando}, H., {Saio}, H., \& {Shibahashi}, H. 1989,
  {Nonradial oscillations of stars}, ed. {Unno, W., Osaki, Y., Ando, H., Saio,
  H., \& Shibahashi, H.}

\bibitem[{{Verma} {et~al.}(2014){Verma}, {Faria}, {Antia}, {Basu}, {Mazumdar},
  {Monteiro}, {Appourchaux}, {Chaplin}, {Garc{\'{\i}}a}, \&
  {Metcalfe}}]{2014ApJ...790..138V}
{Verma}, K., {Faria}, J.~P., {Antia}, H.~M., {et~al.} 2014, \apj, 790, 138

\bibitem[{{Verma} {et~al.}(2019){Verma}, {Raodeo}, {Basu}, {Silva Aguirre},
  {Mazumdar}, {Mosumgaard}, {Lund}, \& {Ranadive}}]{2019MNRAS.483.4678V}
{Verma}, K., {Raodeo}, K., {Basu}, S., {et~al.} 2019, \mnras, 483, 4678

\bibitem[{{Vorontsov}(1988)}]{1988IAUS..123..151V}
{Vorontsov}, S.~V. 1988, in IAU Symposium, Vol. 123, Advances in Helio- and
  Asteroseismology, ed. J.~{Christensen-Dalsgaard} \& S.~{Frandsen}, 151

\bibitem[{{Vorontsov} {et~al.}(1992){Vorontsov}, {Baturin}, \&
  {Pamiatnykh}}]{1992MNRAS.257...32V}
{Vorontsov}, S.~V., {Baturin}, V.~A., \& {Pamiatnykh}, A.~A. 1992, \mnras, 257,
  32

\bibitem[{{Vrard} {et~al.}(2015){Vrard}, {Mosser}, {Barban}, {Belkacem},
  {Elsworth}, {Kallinger}, {Hekker}, {Samadi}, \& {Beck}}]{2015A&A...579A..84V}
{Vrard}, M., {Mosser}, B., {Barban}, C., {et~al.} 2015, \aap, 579, A84

\bibitem[{{Vrard} {et~al.}(2016){Vrard}, {Mosser}, \&
  {Samadi}}]{2016A&A...588A..87V}
{Vrard}, M., {Mosser}, B., \& {Samadi}, R. 2016, \aap, 588, A87

\bibitem[{{White} {et~al.}(2011){White}, {Bedding}, {Stello},
  {Christensen-Dalsgaard}, {Huber}, \& {Kjeldsen}}]{2011ApJ...743..161W}
{White}, T.~R., {Bedding}, T.~R., {Stello}, D., {et~al.} 2011, \apj, 743, 161

\bibitem[{{Yu} {et~al.}(2020){Yu}, {Bedding}, {Stello}, {Huber}, {Compton},
  {Gizon}, \& {Hekker}}]{2020MNRAS.493.1388Y}
{Yu}, J., {Bedding}, T.~R., {Stello}, D., {et~al.} 2020, \mnras, 493, 1388

\bibitem[{{Yu} {et~al.}(2021){Yu}, {Hekker}, {Bedding}, {Stello}, {Huber},
  {Gizon}, {Khanna}, \& {Bi}}]{2021MNRAS.501.5135Y}
{Yu}, J., {Hekker}, S., {Bedding}, T.~R., {et~al.} 2021, \mnras, 501, 5135

\bibitem[{{Zahn}(1991)}]{1991A&A...252..179Z}
{Zahn}, J.~P. 1991, \aap, 252, 179

\end{thebibliography}

\newpage 

\begin{appendix}

\section{Mass dependence of the glitch modulation period $\Ggl$}
\label{appendix:mass_dependence_Ggl}

\begin{figure}[htbp]
    \centering
	\includegraphics[width=1.0\linewidth]{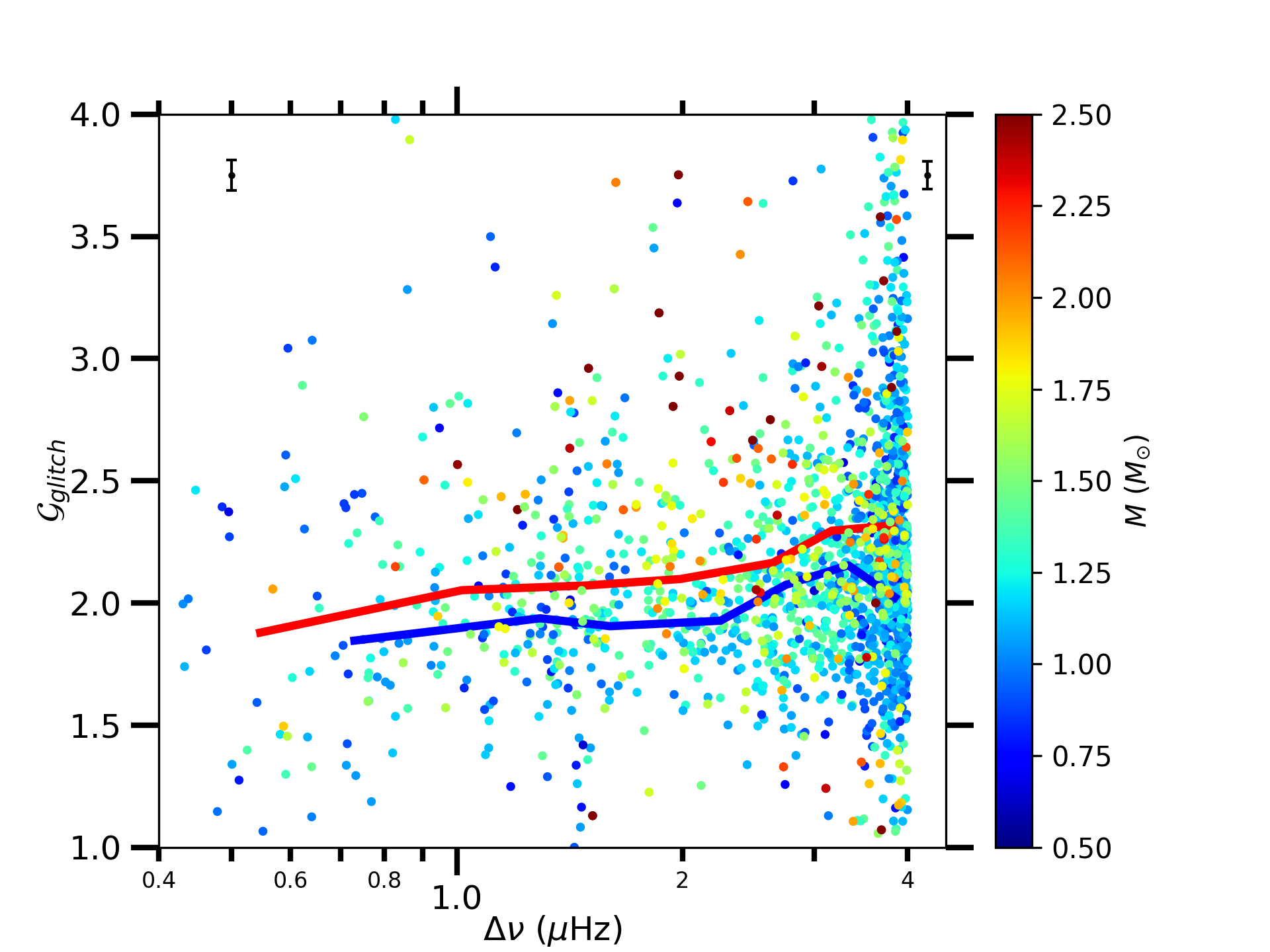}
	\caption{Period of the glitch modulation from the \Kepler\ data used in \citet{2021A&A...650A.115D} as a function of $\Dnu$, where the stellar mass is colour-coded. Mean error bars estimated for $\Dnu$ below or above $1\, \mu$Hz are represented. The thick solid lines are the median values in bins of $0.5\, \mu$Hz $\Dnu$, regardless the evolutionary stage, shown in blue for low-mass stars ($M \leq 1.2\, M_{\odot}$) and in red for high-mass stars ($M \geq 1.2\, M_{\odot}$).
	}
	\label{fig:G_gl_dependence_on_mass_obs}
\end{figure}

Stellar models highlight a noticeable mass dependence on the modulation period $\Ggl$ (Fig.~\ref{fig:G_gl_Phi_gl_dependence_on_mass_models}\emph{a}+\emph{b}). Then, we inspect how this glitch parameter varies with mass in \Kepler\ observations. This part supplements the observational results presented in \citet{2021A&A...650A.115D}. To summarise, the frequency large separation $\Dnu$ is obtained as the value that optimises the cross correlation between the observed oscillation spectrum and the template spectrum based on the asymptotic relation of p-mode frequencies in red giants (Eq.~\ref{eq:nunl_asympt}). The way the modulation period $\Ggl$ is extracted is similar to that described in Sect.~\ref{subsec:method_seismic_param_asympt}, except that the fit of the glitch signature is limited to the observed mode frequencies. Stellar masses are assessed from the semi-empirical asteroseismic scaling relation presented in \cite{1995A&A...293...87K}. Deviations from the asteroseismic scaling relations are corrected by a factor that is adjusted star by star \citep{2018ApJS..239...32P}. When this correcting factor is not available, which concerns about 10\% of our \Kepler\ targets, we estimated the mass with the semi-empirical relation without any correction factor. In Fig.~\ref{fig:G_gl_dependence_on_mass_obs}, we show the mass dependence of the modulation period $\Ggl$ as a function of $\Dnu$, from the data set used in \citet{2021A&A...650A.115D}, including $\sim$ 2,100 RGB, clump and AGB stars. We note that low-mass stars ($M \leq 1.2\, M_{\odot}$) tend to have a lower modulation period $\Ggl$ compared to their higher-mass counterparts ($M \geq 1.2\, M_{\odot}$) at fixed $\Dnu$. This means that the helium ionisation occurs closer to the surface when the stellar mass increases. These observations are in line with the results obtained with stellar models, presented in Sect.~\ref{subsec:glitch_period_result}. The effective temperature $\Teff$ is larger in high-mass stars, which makes the physical conditions for helium ionisation reached closer to the surface, hence a larger modulation period $\Ggl$.

\section{Effects of input physics on seismic parameters}
\label{appendix:effects_input_physics}

We aim at investigating the effects of stellar parameters and input physics on seismic parameters. To this end, we generated a grid of stellar models based on the reference model of mass $1\, M_{\odot}$ and solar metallicity presented in Table.~\ref{Table:reference_model}. We changed the input physics relatively to the reference model and compared the seismic parameters between the reference and modified models in several bins of $\Dnu$. In order to smooth the evolutionary tracks, we computed the median values of seismic parameters every three consecutive models and kept the average values in $0.5\,\mu$Hz wide $\Dnu$ bins. These averaged values obtained in the modified models, noted $X\ind{mod}$, are then compared to those of the reference model, noted $X\ind{ref}$, by means of the relative difference

\begin{equation}
    \label{eq:relative_difference}
    \Delta X = \frac{X\ind{mod} - X\ind{ref}}{X\ind{ref}}.
\end{equation}
Here, $X$ can be the acoustic offset $\eps$, the dimensionless small frequency separations $d_{01}$ and $d_{02}$, the modulation amplitude $\Amplgl$ and period $\Ggl$. For the modulation phase $\Phigl$, we directly compare the reference and modified models with the absolute difference $\Delta\Phigl = \Phi\ind{gl,mod} - \Phi\ind{gl,ref}$. \\

From Figs~\ref{fig:relative_diff_params_aympt};~\ref{fig:relative_diff_params_glitch}; and \ref{fig:relative_diff_params_HeII}, we can infer the impact of input physics on seismic parameters and HeII ionisation zone characteristics as a function of the evolutionary stage. 

\subsection{Mass, mass loss and metallicity}

The main parameters affecting the seismic quantities of Eqs.~\ref{eq:nunl_asympt},\ref{eq:fit_glitch_signature_Dreau_2021} are the stellar mass (when varied from $1\, M_{\odot}$ to $1.75\, M_{\odot}$) and metallicity (from $0\,$dex to $-1\,$dex), with relative differences reaching more than 20\% for some seismic parameters. Given the strong dependence of the seismic parameters with the stellar mass $M$, we expect the mass loss rate $\eta\ind{R}$ would have an effect too. Because most of the mass loss occurs near the luminosity-tip of the RGB, we investigate the effect only after He-burning starts. Modifying the mass loss rate, $\eta\ind{R}$, produces systematic shifts in both the seismic parameters from the asymptotic pattern (Eq.~\ref{eq:nunl_asympt}) and the glitch signature (Eq.~\ref{eq:fit_glitch_signature_Dreau_2021}) with respect to the reference value, with relative differences that could reach $\sim 5\%$. The most prominent effect of $\eta\ind{R}$ is on the modulation phase $\Phigl$, which likely accounts for the phase difference observed between RGB and clump/AGB stars, as discussed in Sect.~\ref{subsec:phase_diff_physical_origin}. Decreasing the mass loss parameter from $\eta\ind{R} = 0.3$ to $0.1$ changes the mass difference between the initial and clump mass from $0.15$ to $0.05\, M_{\odot}$. Hence, the impact of mass loss on the seismic parameters remain small compared to the effect of changing the initial mass from, say, $1$ to $1.75\, M_{\odot}$.

\subsection{Mixing length parameter}

Varying the mixing-length parameter $\alphaMLT$ from $1.92$ to $1.62$ induces systematic shifts in both the asymptotic and the glitch signature parameters, with relative differences reaching up to $\sim 5\%$ compared to the reference model. This adjustment particularly affects the acoustic offset $\varepsilon$ and the modulation phase $\Phigl$. It is expected, because $\varepsilon$ is mostly affected by the envelope structure \citep{2014MNRAS.445.3685C}, which the mixing length parameter affects (such as the effective temperature $\Teff$ and radius $R$.) This also justifies the effects of $\alphaMLT$ on the modulation phase $\Phigl$ since a change of $\varepsilon$ leads to a change of $\Dnu$ through Eq.~\ref{eq:link_deps_dDnu}, especially the modulation phase \citep{2015A&A...579A..84V}.

\subsection{Initial helium abundance}

While a change of initial helium abundance $Y_{0}$ from $0.25$ to $0.30$ only weakly affects $\dol$, it clearly impacts the glitch parameters, and $\varepsilon$. This is expected because the glitch signature explicitly depends on the helium abundance \citep{2007MNRAS.375..861H, 2021A&A...655A..85H, 2022A&A...663A..60H}, which gives the opportunity to estimate the helium abundance in cool red-giant stars \citep{2014MNRAS.440.1828B, 2014ApJ...790..138V, 2019MNRAS.483.4678V}. 
A change of 0.05 in $Y_{0}$ induces an average modification of $\sim 10\%$ of the modulation amplitude $\Amplgl$ (Fig.~\ref{fig:relative_diff_params_glitch}\emph{a}+\emph{b}). \\
The noisy profiles of the modulation amplitude $\Amplgl$ could be caused by the method with which the p-mode frequencies are extracted, that is by setting $N\ind{BV}^{2} = 0$ in the core. Indeed, Fig.~\ref{fig:test_computation_pure_p_modes_with_NBV0}\emph{b}+\emph{d} shows that a deviation of $3\%-4\%$ of $\Dnu$ from the exact pure p-mode frequency is possibly introduced in mode frequencies, which is significant relatively to the amplitude of the glitch signature. At low $\Dnu$, we suspect the $\Amplgl$ to be affected by large uncertainties due to the computation of the glitch signature with Eq.~\ref{eq:glitch_contribution_Vrard_2015}. The expected value of $\Dnu_{n}$ according to Eq.~\ref{eq:nunl_asympt} is certainly not the best reference to isolate the glitch signature at low $\Dnu$.

\subsection{Overshooting and thermohaline convection}

\begin{figure*}[htbp]
    \centering
	\begin{minipage}{1.0\linewidth}  
		\rotatebox{0}{\includegraphics[width=0.50\linewidth]{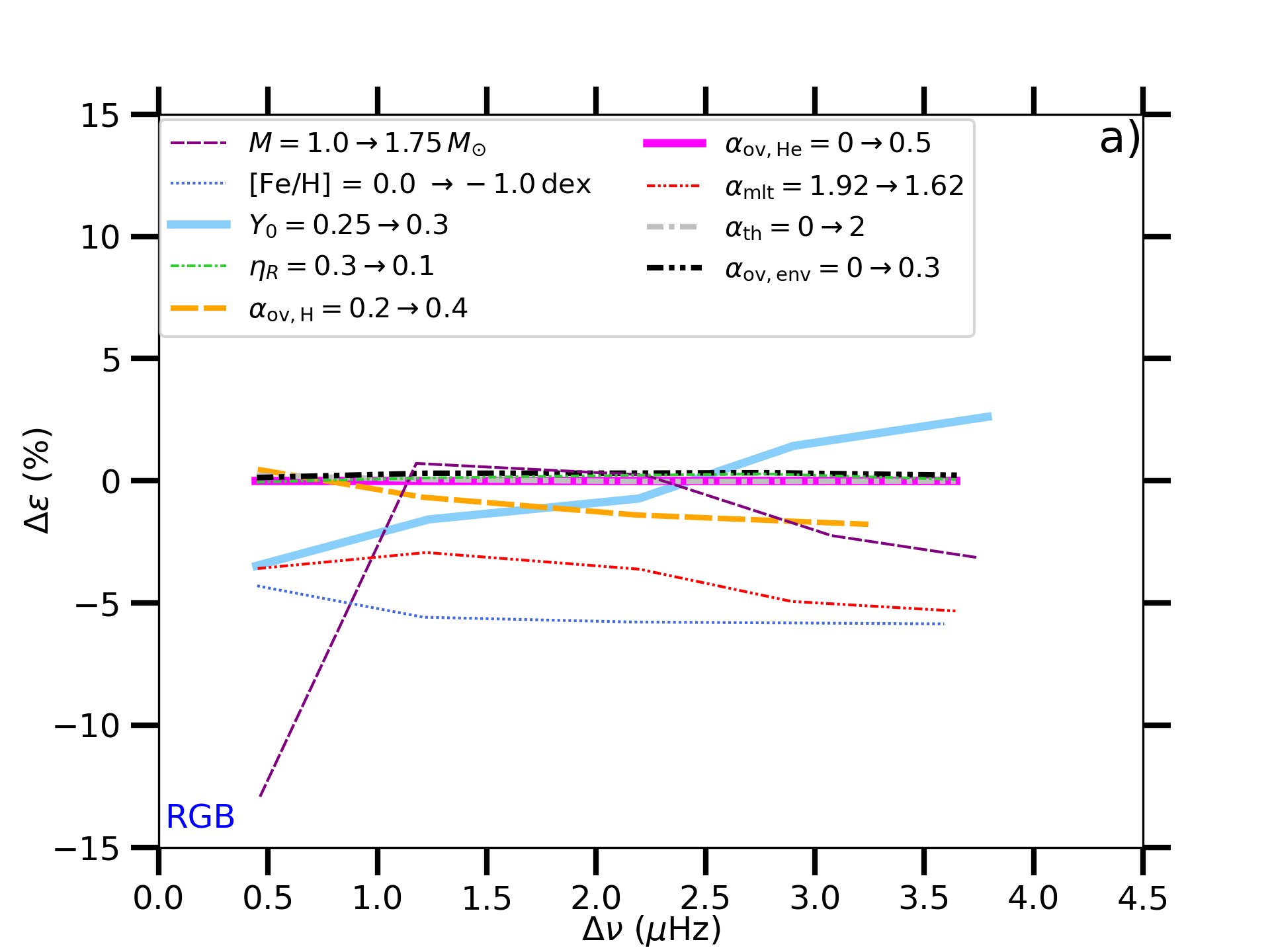}
		\rotatebox{0}{\includegraphics[width=0.50\linewidth]{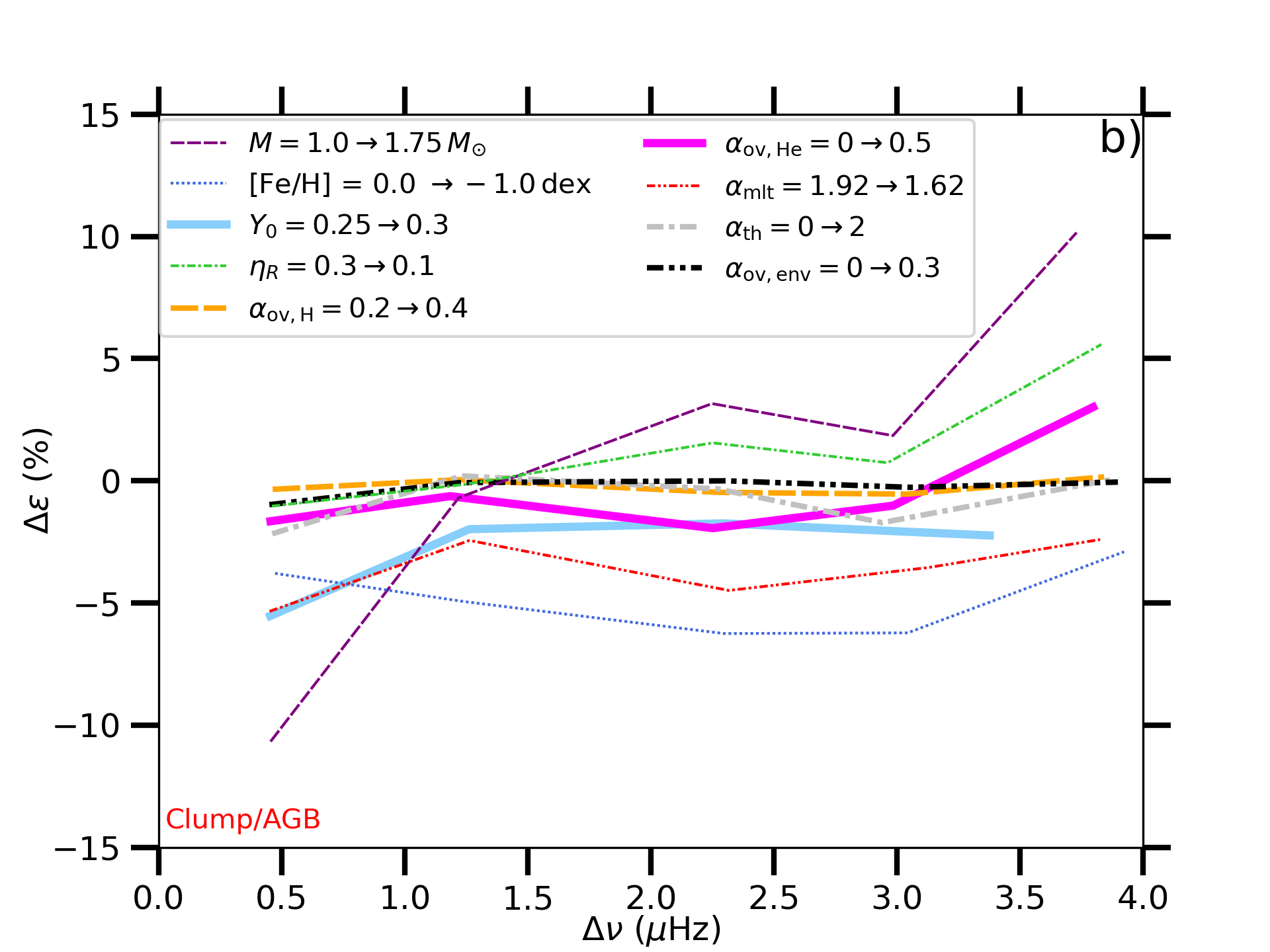}}
		}
	\end{minipage}
	\begin{minipage}{1.0\linewidth}  
		\rotatebox{0}{\includegraphics[width=0.50\linewidth]{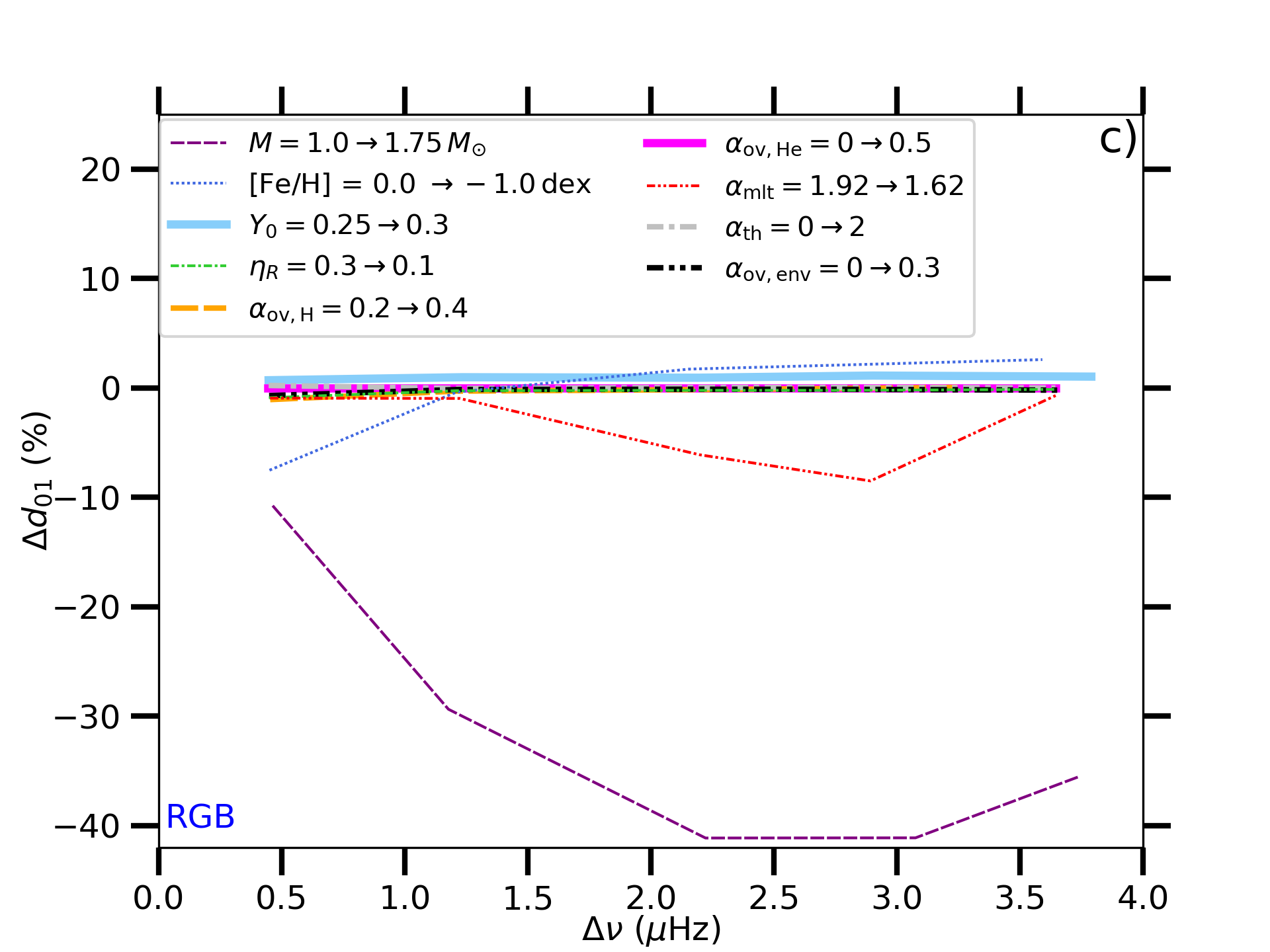}
		\rotatebox{0}{\includegraphics[width=0.50\linewidth]{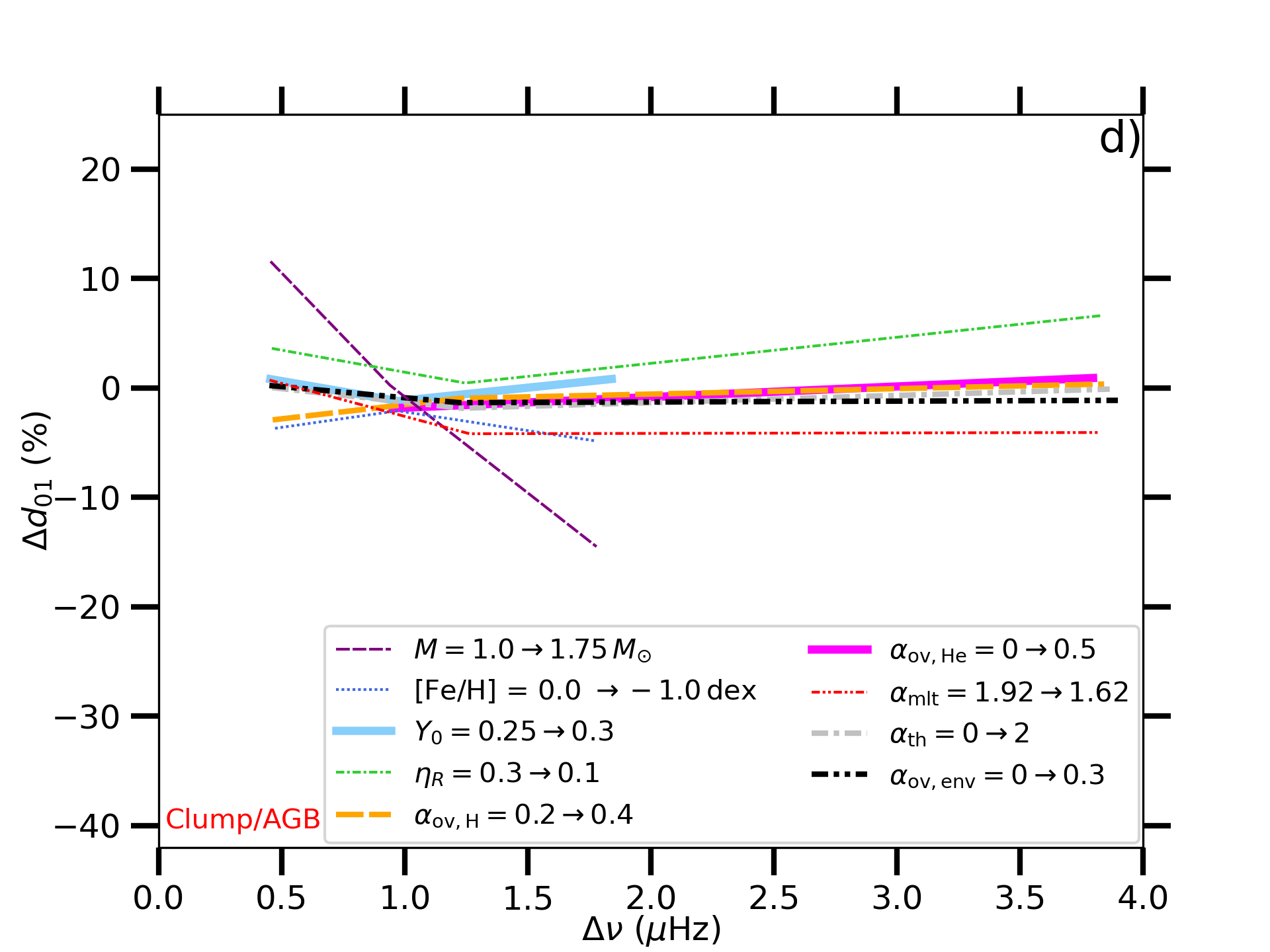}}
		}
	\end{minipage}
	\begin{minipage}{1.0\linewidth}  
		\rotatebox{0}{\includegraphics[width=0.50\linewidth]{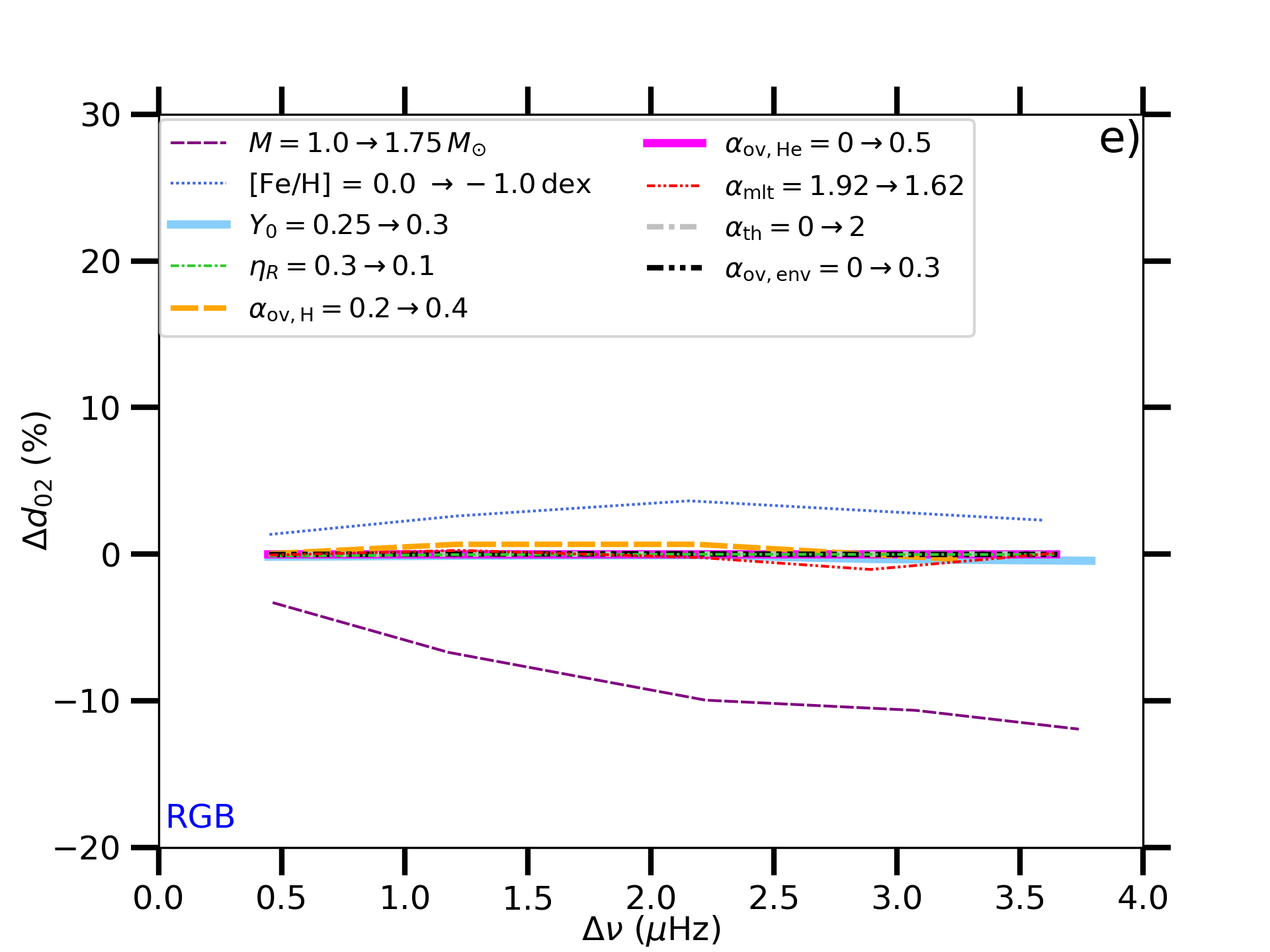}
		\rotatebox{0}{\includegraphics[width=0.50\linewidth]{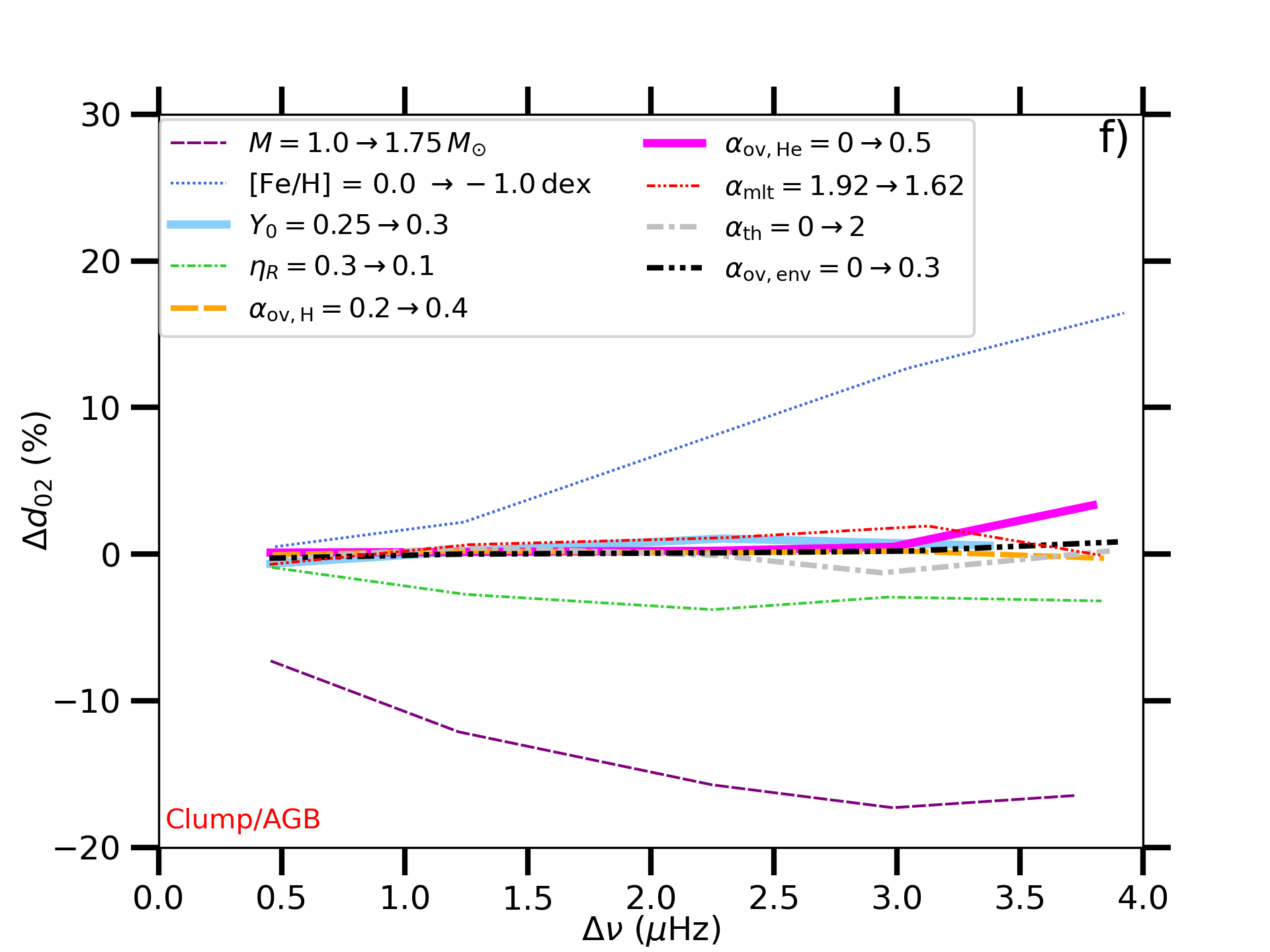}}
		}
	\end{minipage}
	\caption{Sensitivity of the seismic parameters in the asymptotic relation (Eq.~\ref{eq:nunl_asympt}) with respect to input physics of \mesa\ models. The relative difference of seismic parameters given by Eq.~\ref{eq:relative_difference} between the reference and modified models is shown for $0.5\,\mu$Hz-bins for $\Dnu$ in the range $[0.1,4.0]\, \mu$Hz (except for the first $\Dnu$ bin that is $[0.1, 0.5]\, \mu$Hz). The reference model is defined with $1\, M_{\odot}$, solar metallicity and the input physics are those indicated in Table.~\ref{Table:reference_model}. The settings of the modified models are similar to those of the reference models, except one parameter that is changed as indicated by the labels. The left and right columns are obtained on the RGB and clump stage/AGB, respectively. In panel \emph{d)}, some of the lines are shorter than the full range because the code failed to return the frequencies of $\ell = 1$ modes when setting $\NBV^{2} = 0$ in the core of clump models ($\Dnu \gtrsim 3\,\mu$Hz).}
    
	\label{fig:relative_diff_params_aympt}
\end{figure*}

\begin{figure*}[htbp]
    \centering
	\begin{minipage}{1.0\linewidth}  
		\rotatebox{0}{\includegraphics[width=0.50\linewidth]{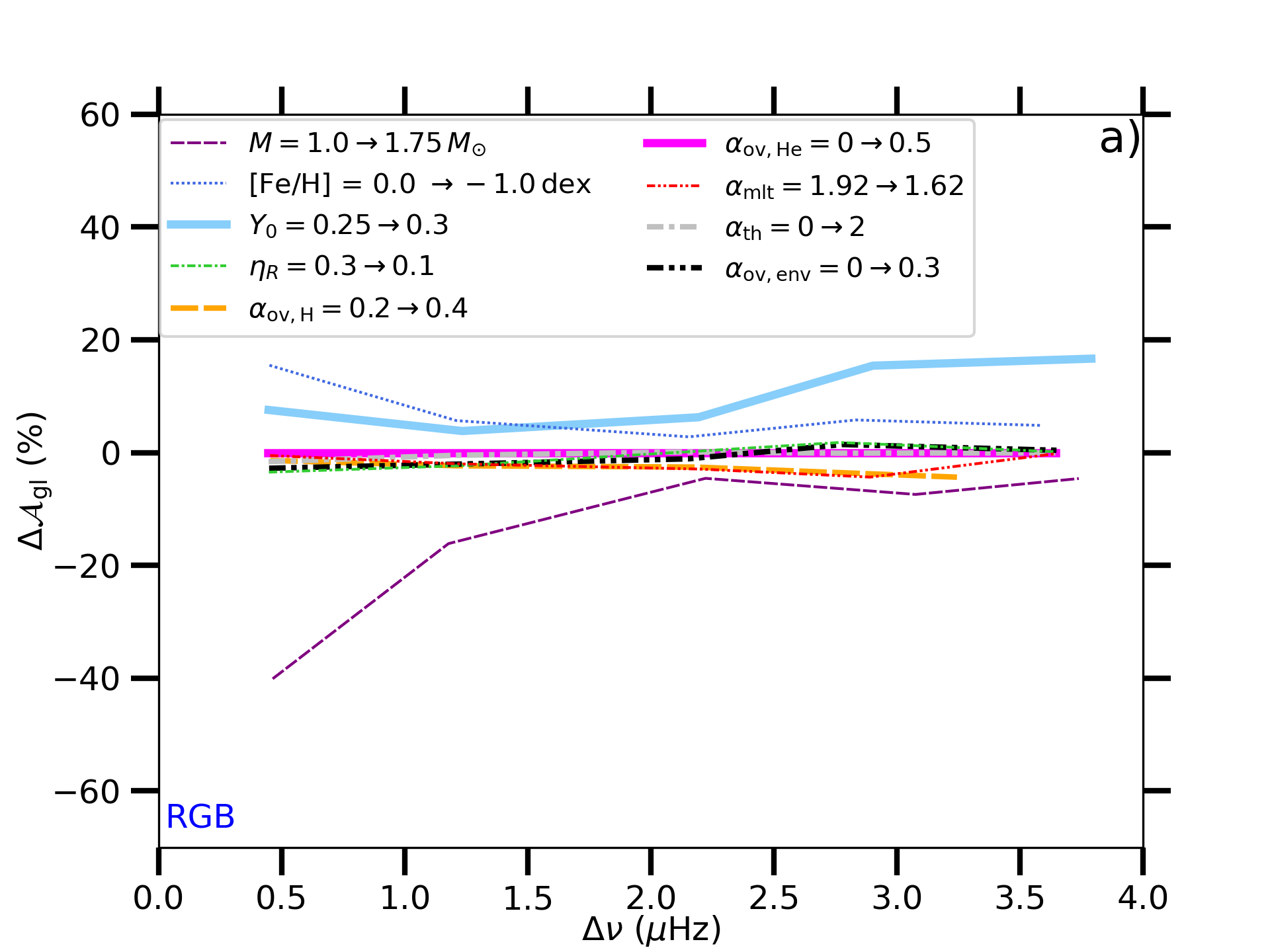}
		\rotatebox{0}{\includegraphics[width=0.50\linewidth]{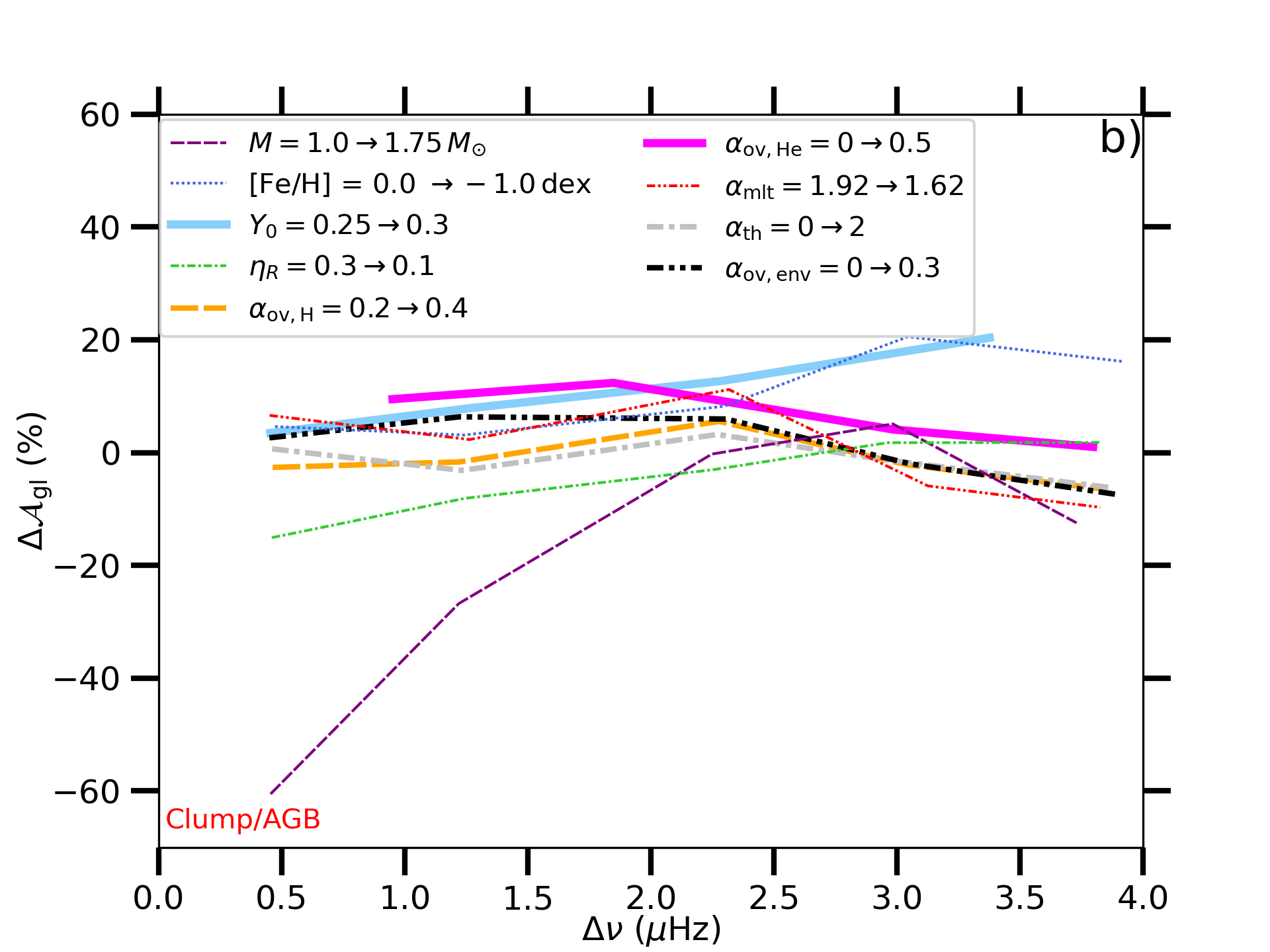}}
		}
	\end{minipage}
	\begin{minipage}{1.0\linewidth}  
		\rotatebox{0}{\includegraphics[width=0.50\linewidth]{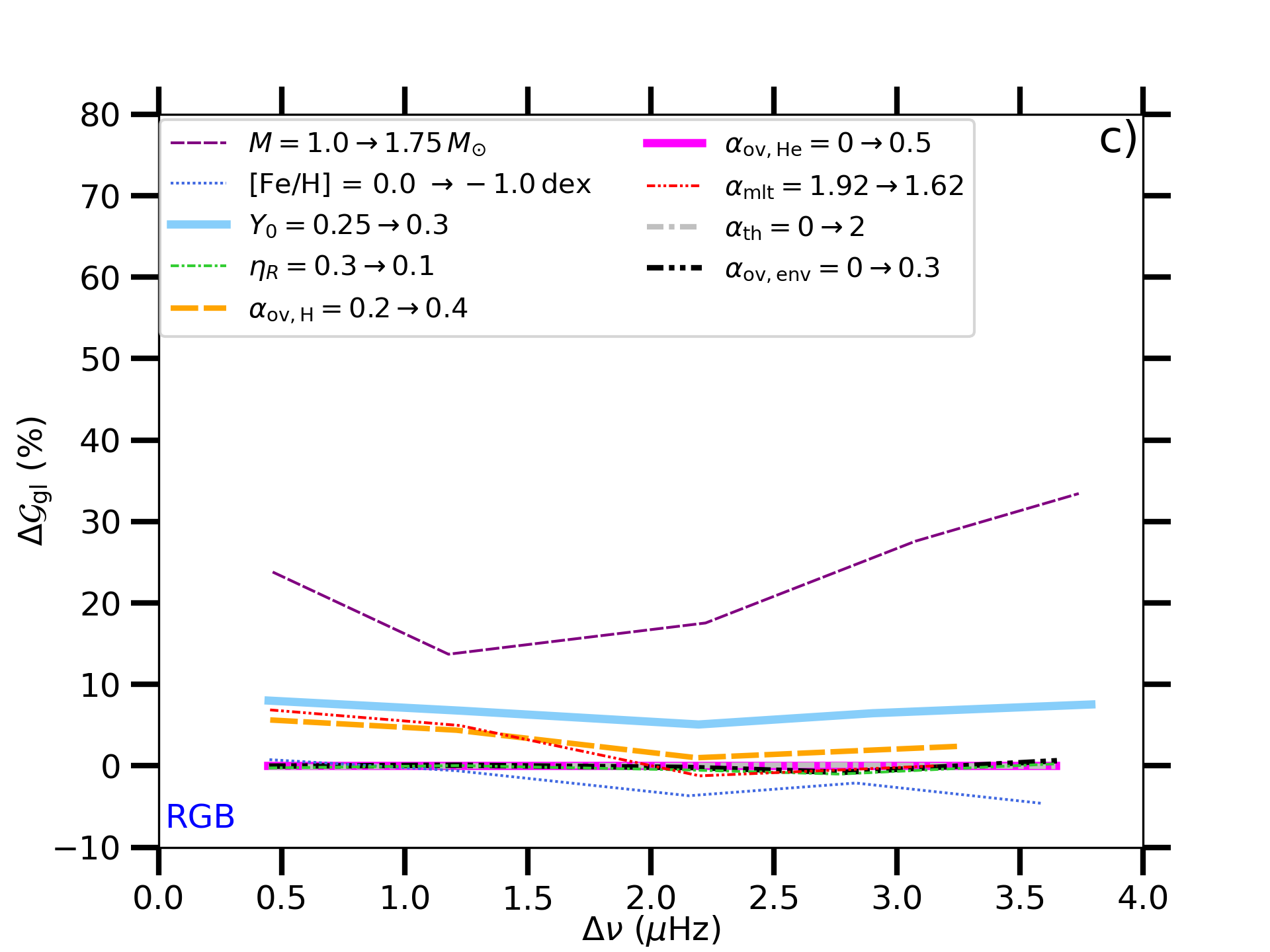}
		\rotatebox{0}{\includegraphics[width=0.50\linewidth]{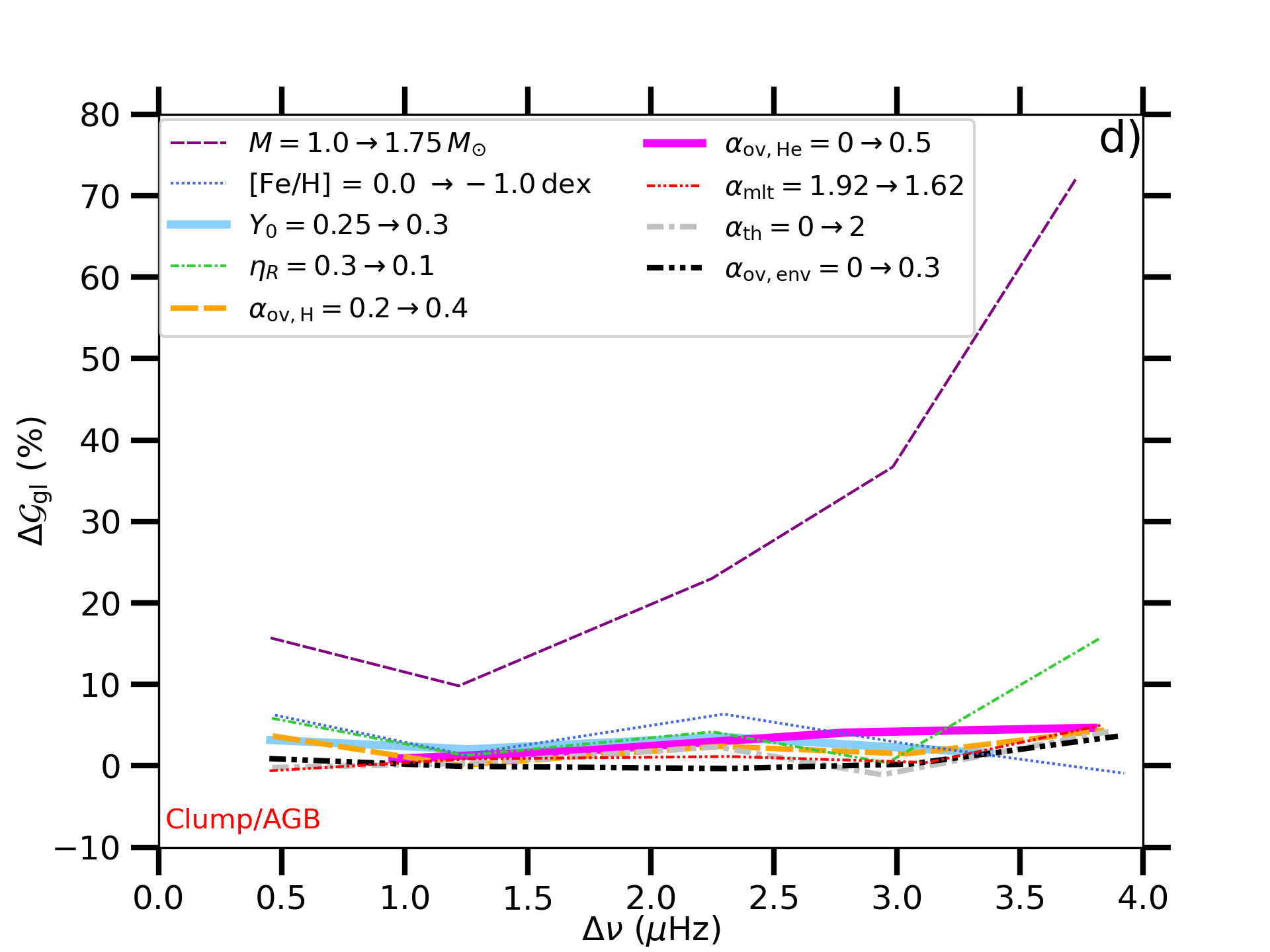}}
		}
	\end{minipage}
	\begin{minipage}{1.0\linewidth}  
		\rotatebox{0}{\includegraphics[width=0.50\linewidth]{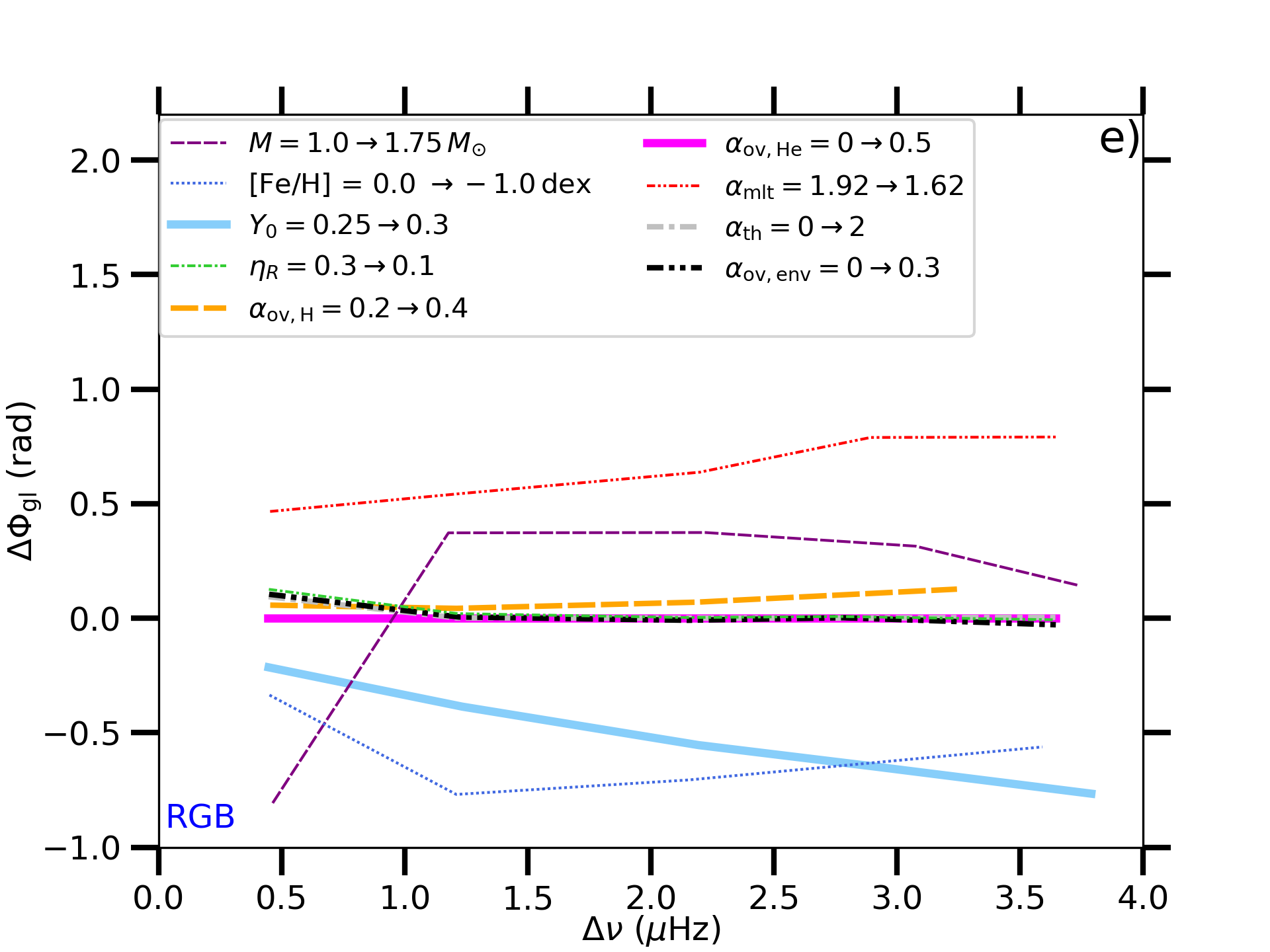}
		\rotatebox{0}{\includegraphics[width=0.50\linewidth]{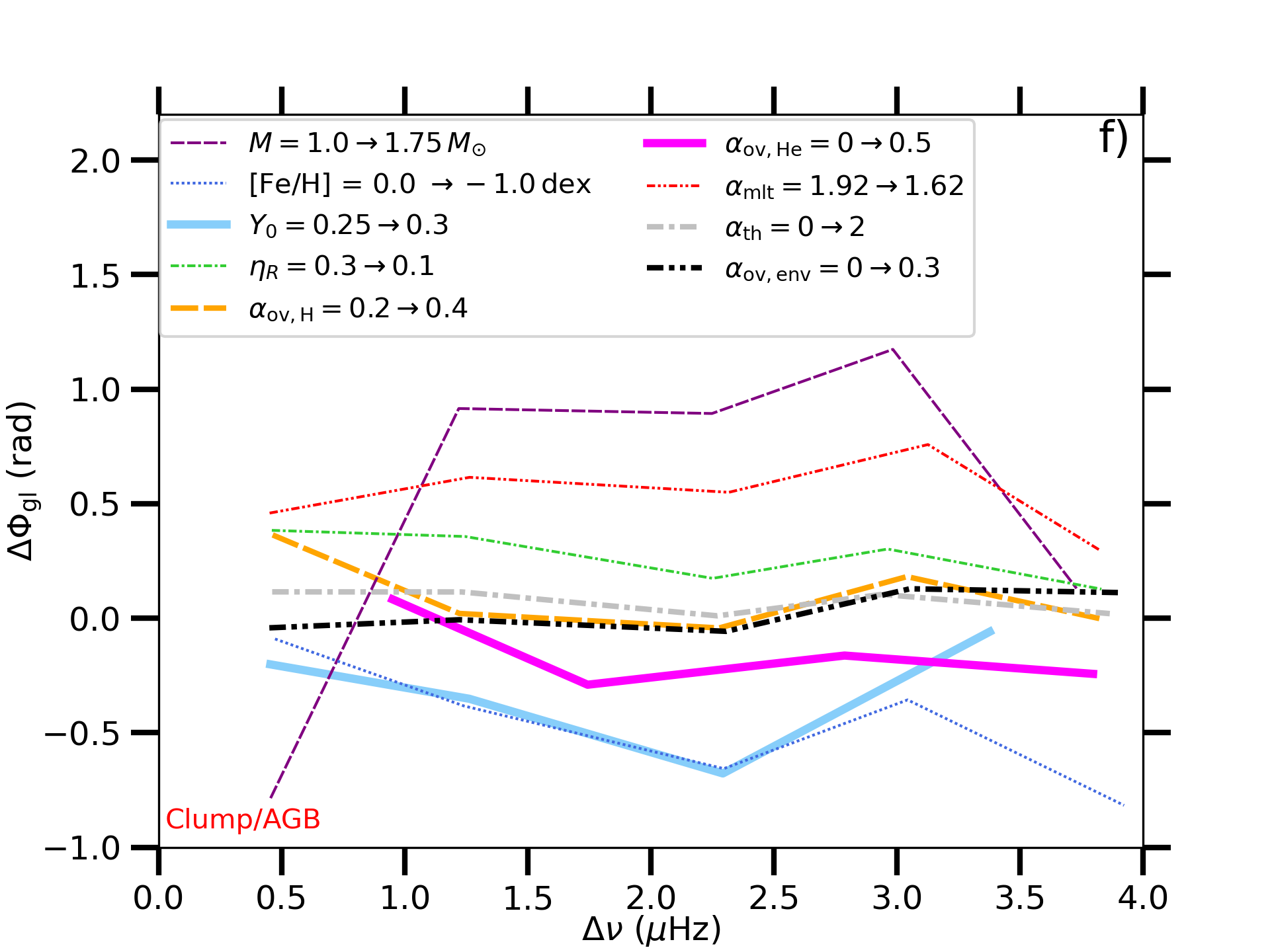}}
		}
	\end{minipage}
	\caption{Sensitivity of the glitch parameters in the damped oscillator model (Eq.~\ref{eq:fit_glitch_signature_Dreau_2021}) with respect to input physics of \mesa\ models. Same label as in Fig.~\ref{fig:relative_diff_params_aympt}, except that the modulation phase differences $\Delta\Phigl$ between the reference and modified models are given by $\Delta\Phigl = \Phi\ind{gl,mod} - \Phi\ind{gl,ref}$, where $\Phi\ind{gl,ref}$ and $\Phi\ind{gl,mod}$ are the modulation phases found in the reference and modified models, respectively.
	}
	\label{fig:relative_diff_params_glitch}
\end{figure*}

\begin{figure*}[htbp]
    \centering
	\begin{minipage}{1.0\linewidth}  
		\rotatebox{0}{\includegraphics[width=0.50\linewidth]{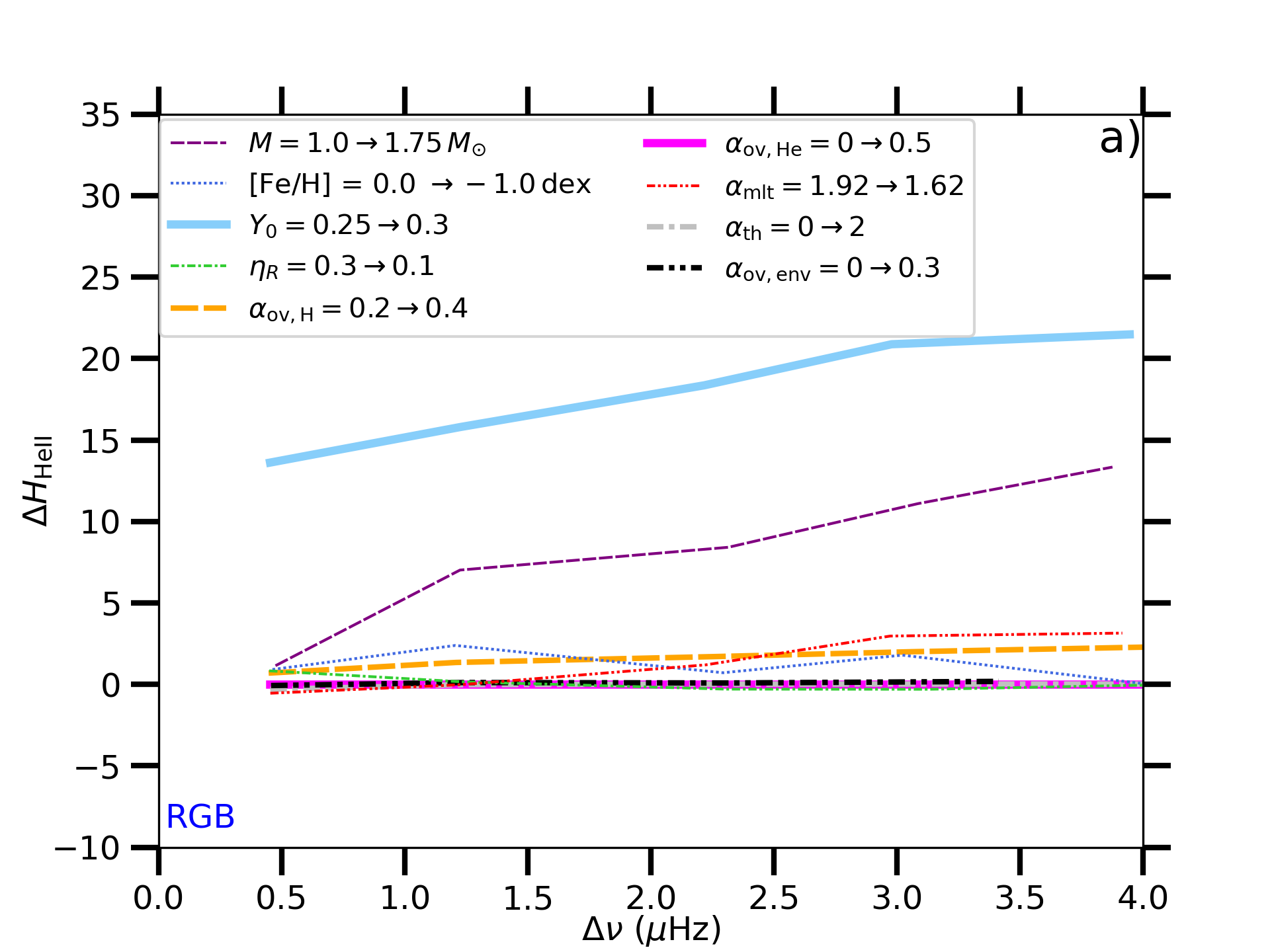}
		\rotatebox{0}{\includegraphics[width=0.50\linewidth]{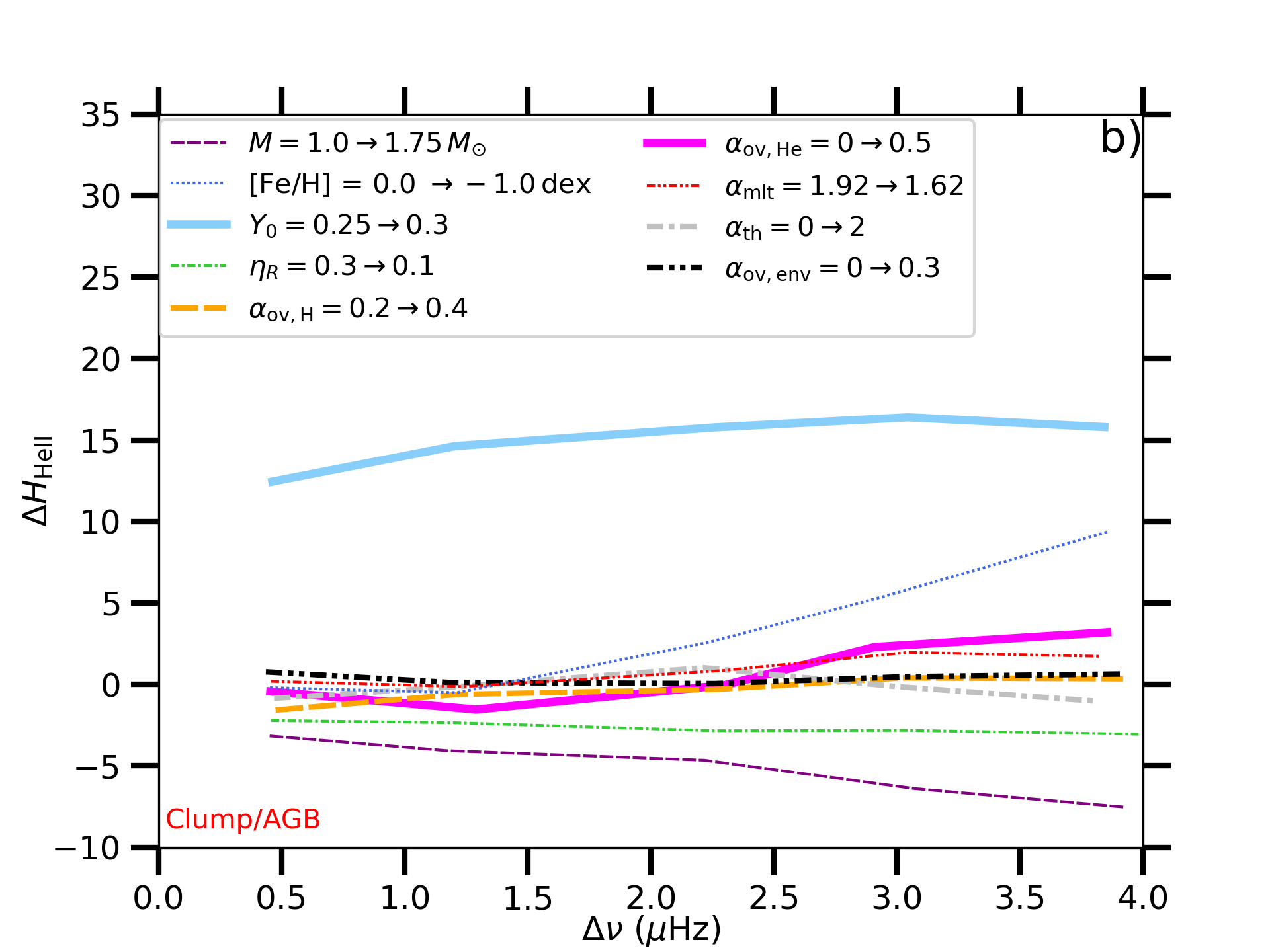}}
		}
	\end{minipage}
	\begin{minipage}{1.0\linewidth}  
		\rotatebox{0}{\includegraphics[width=0.50\linewidth]{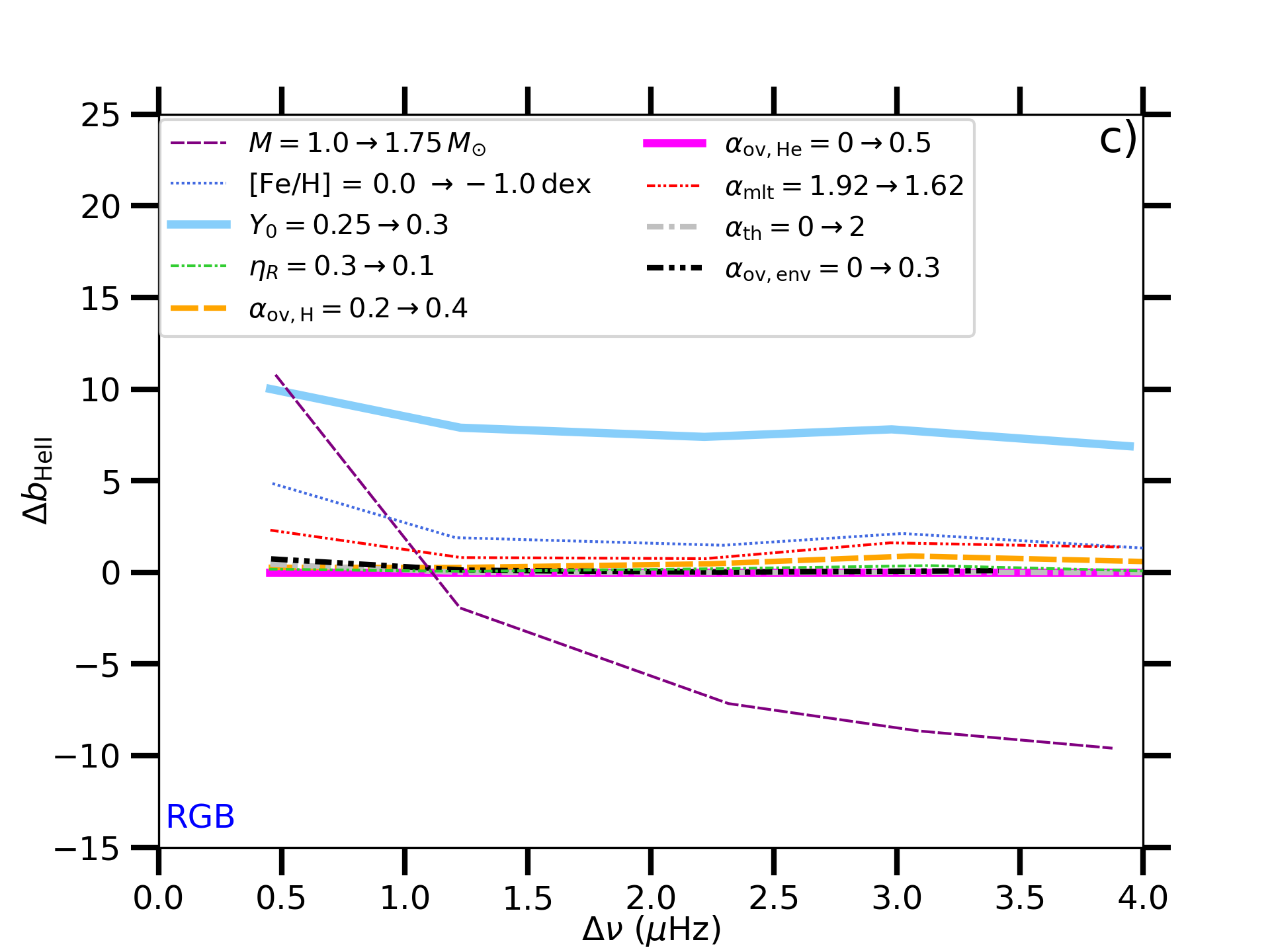}
		\rotatebox{0}{\includegraphics[width=0.50\linewidth]{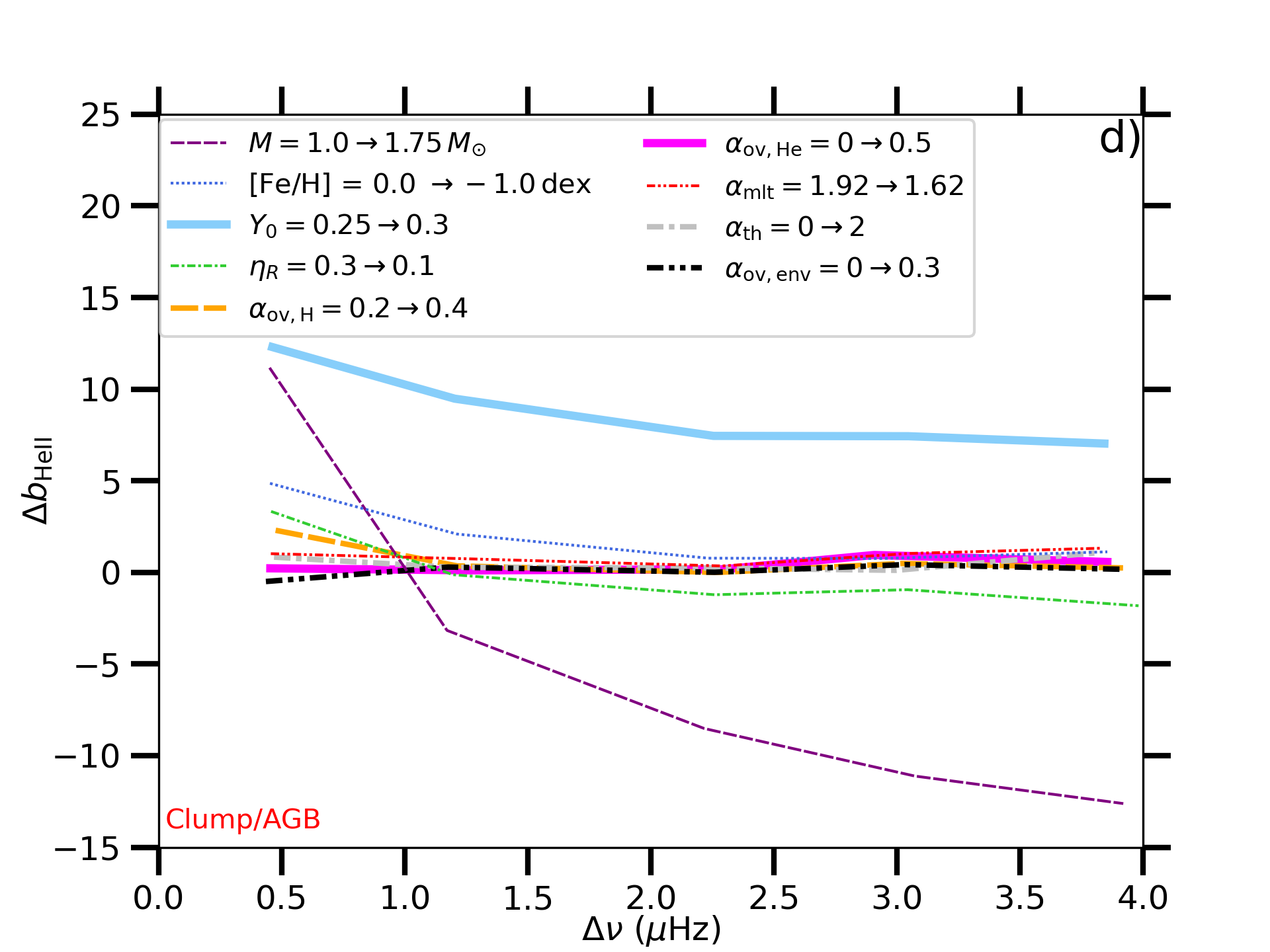}}
		}
	\end{minipage}
	\begin{minipage}{1.0\linewidth}  
		\rotatebox{0}{\includegraphics[width=0.50\linewidth]{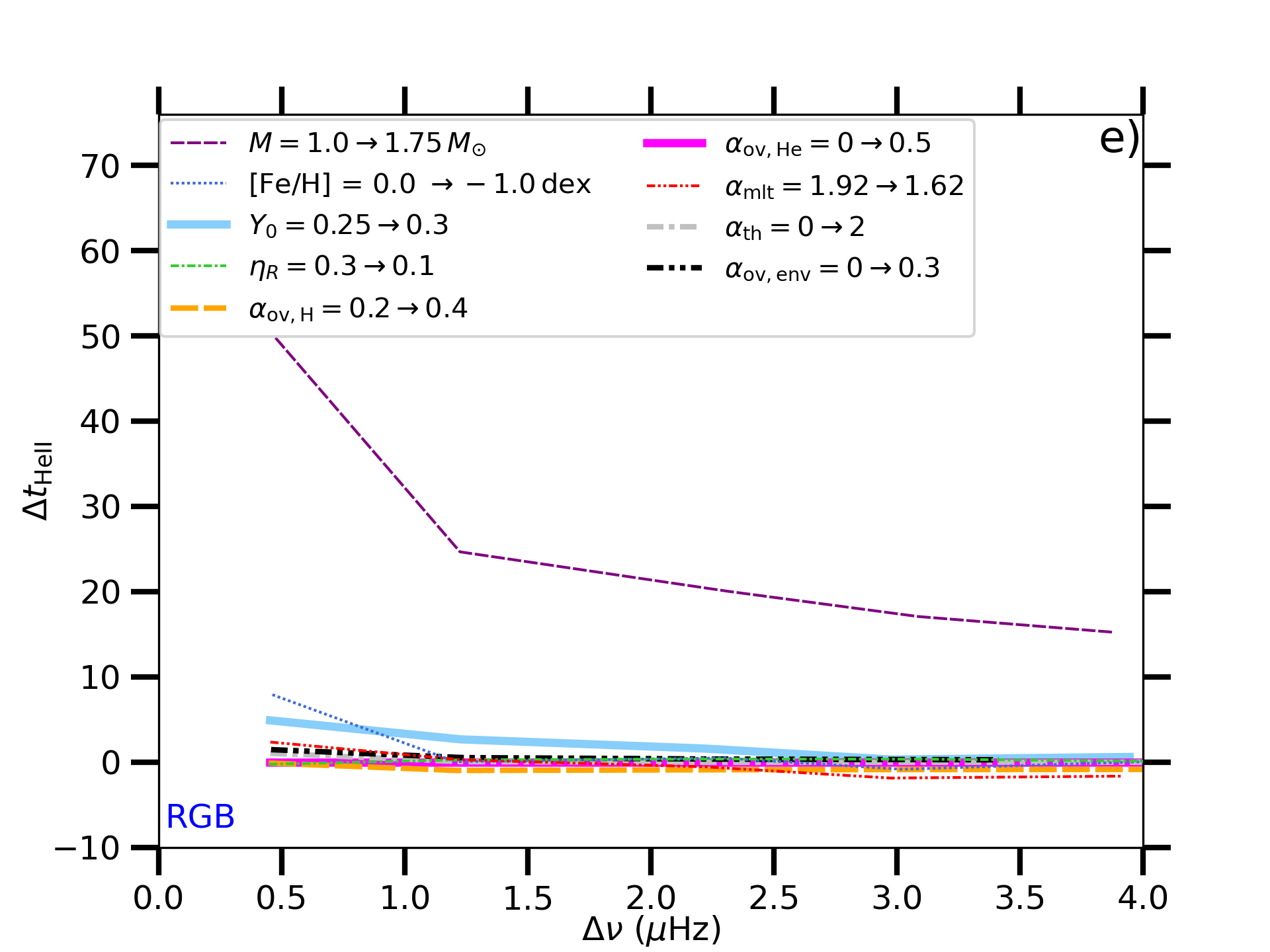}
		\rotatebox{0}{\includegraphics[width=0.50\linewidth]{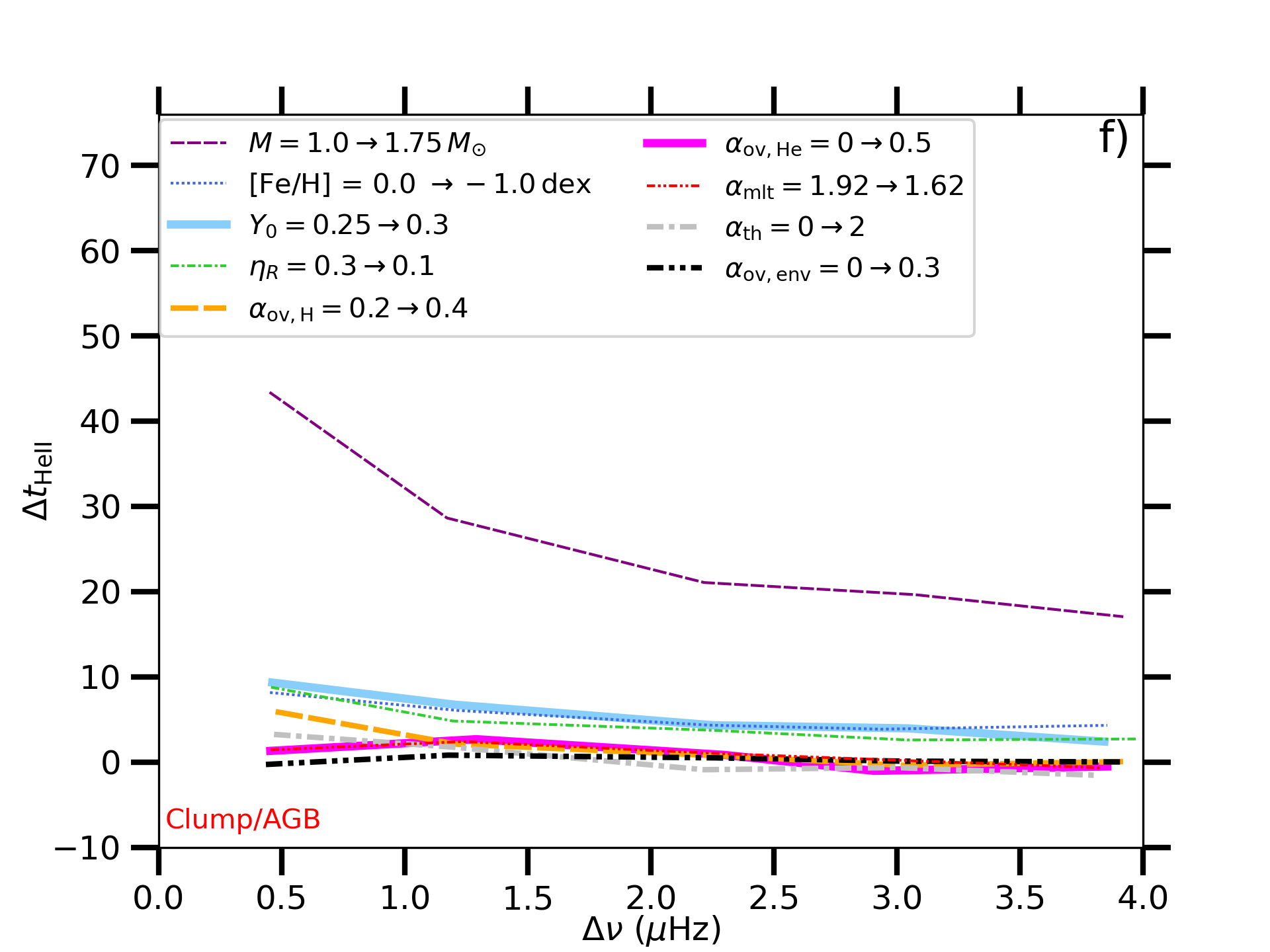}}
		}
	\end{minipage}
	\caption{Sensitivity of the HeII zone parameters introduced in Eq.~\ref{eq:fit_Gamma_1_around_HeII} with respect to input physics of \mesa\ models. Same label as in Fig.~\ref{fig:relative_diff_params_aympt}.
	}
	\label{fig:relative_diff_params_HeII}
\end{figure*}

Processes such as core and envelope overshooting with typical efficiency values ranging from $\alpha =0.2$ to $0.5$ (see labels of Fig. \ref{fig:relative_diff_params_aympt}), as well as thermohaline mixing only marginally affect the seismic parameters derived from asymptotic expressions. The effects do not exceed $3\%$ of the reference values of $\varepsilon$ and $\dol$, which is below the typical uncertainty derived from observations. We expect core-H overshooting during the main sequence not to have a significant impact because the convective core is very small when $M = 1.0\, M_{\odot}$. We note that thermohaline convection with an efficiency parameter of $\alpha_{\mathrm{th}} = 2$, representing a rather inefficient case, connects the H-burning shell to the convective envelope only during the high-luminosity stages of the RGB. This likely accounts for the absence of significant effects when including thermohaline convection in our models.

\end{appendix}

\end{document}